\documentclass[journal]{IEEEtran}%
\pdfoutput=1
\usepackage{graphics,amsmath,amsfonts,amscd,revsymb,latexsym,
enumerate,multirow,epsfig}
\usepackage{amsmath}
\usepackage{amsfonts}
\usepackage{amssymb}
\usepackage{graphicx}%
\providecommand{\U}[1]{\protect\rule{.1in}{.1in}}
\newtheorem{theorem}{Theorem}

\newtheorem{corollary}{Corollary}

\newtheorem{definition}{Definition}

\newtheorem{lemma}{Lemma}

\newtheorem{proposition}{Proposition}

\def\bi{\begin{itemize}}
\def\ei{\end{itemize}}
\def\be{\begin{equation}}
\def\ee{\end{equation}}
\def\bea{\begin{eqnarray}}
\def\eea{\end{eqnarray}}
\def\ben{\begin{eqnarray*}}
\def\een{\end{eqnarray*}}

\def\>{\rangle}
\def\<{\langle}

\newcommand{\1} I

 \DeclareMathOperator{\tr}{Tr}

\def\*{\star}

\def\0{{\mathbf{0}}}
\def\1{{\mathbf{1}}}
\def\2{{\mathbf{2}}}
\def\3{{\mathbf{3}}}
\def\4{{\mathbf{4}}}
\def\5{{\mathbf{5}}}
\def\6{{\mathbf{6}}}
\def\7{{\mathbf{7}}}
\def\8{{\mathbf{8}}}
\def\9{{\mathbf{9}}}

\begin{document}

\title{Entanglement-assisted communication of classical and quantum information}
\author{Min-Hsiu Hsieh and Mark M. Wilde\thanks{Min-Hsiu Hsieh is with the ERATO-SORST
Quantum Computation and Information Project, Japan Science and Technology
Agency, 5-28-3, Hongo, Bunkyo-ku, Tokyo, Japan 113-0033. Mark M. Wilde was
originally a visiting researcher with the Centre for Quantum Technologies,
National University of Singapore, 3 Science Drive 2, Singapore 117543 at the
beginning of this project, and he is now a postdoctoral fellow with the School
of Computer Science, McGill University, Montreal, Canada H3A 2A7 (E-mail:
minhsiuh@gmail.com and mwilde@gmail.com)}}
\maketitle

\begin{abstract}
We consider the problem of transmitting classical and quantum information
reliably over an entanglement-assisted quantum channel. Our main result is a
capacity theorem that gives a three-dimensional achievable rate region. Points
in the region are \textit{rate triples}, consisting of the classical
communication rate, the quantum communication rate, and the entanglement
consumption rate of a particular coding scheme. The crucial protocol in
achieving the boundary points of the capacity region is a protocol that we
name the \textit{classically-enhanced father protocol}. The
classically-enhanced father protocol is more general than other protocols in
the family tree of quantum Shannon theoretic protocols, in the sense that
several previously known quantum protocols are now child protocols of it. The
classically-enhanced father protocol also shows an improvement over a
time-sharing strategy for the case of a qubit dephasing channel---this result
justifies the need for simultaneous coding of classical and quantum
information over an entanglement-assisted quantum channel. Our capacity
theorem is of a multi-letter nature (requiring a limit over many uses of the
channel), but it reduces to a single-letter characterization for at least
three channels:\ the completely depolarizing channel, the quantum erasure
channel, and the qubit dephasing channel.

\end{abstract}


\section{Introduction}

The communication of information over a noisy quantum channel is a fundamental
task in quantum communication theory. A sender may wish to transmit classical
information, quantum information, or both. The Holevo-Schumacher-Westmoreland
(HSW) coding theorem gives an achievable rate at which a sender can transmit
\textit{classical} data to a receiver if she transmits the classical
information over a noisy quantum channel \cite{ieee1998holevo,PhysRevA.56.131}%
. The HSW theorem generalizes Shannon's classical channel coding theorem
\cite{bell1948shannon} to the quantum setting. The Lloyd-Shor-Devetak (LSD)
coding theorem gives an achievable rate at which a sender can transmit
\textit{quantum} data to a receiver through a quantum channel
\cite{PhysRevA.55.1613,capacity2002shor,ieee2005dev}. Devetak and Shor
followed up on these results by determining achievable rates at which a sender
can simultaneously transmit both classical and quantum information over a
quantum channel \cite{cmp2005dev}. The na\"{\i}ve scheme is to employ a
time-sharing strategy, where a sender uses an HSW code for a fraction of the
transmitted qubits and an LSD code for the other fraction. The Devetak-Shor
coding strategy outperforms the na\"{\i}ve time-sharing strategy, at least
when the noisy channel is the qubit dephasing channel
\cite{book2000mikeandike}. This result demonstrates the need to consider
non-trivial coding schemes when communicating more than one resource.

A sender can exploit a quantum channel alone, as in the above examples, or she
can exploit assisting resources as well. Examples of such assisting resources
are a static resource shared with the receiver, as in the case of common
randomness, secret key, or entanglement, or a dynamic resource connecting the
sender to the receiver, as in the case of a noiseless classical or quantum
side channel.

Assisting a quantum channel with noiseless resources sometimes improves
communication rates. The simplest and most striking example of this phenomenon
occurs when a noiseless ebit assists a noiseless qubit channel. The
super-dense coding protocol outlines a simple method to transmit two classical
bits over a noiseless qubit channel assisted by an ebit
\cite{PhysRevLett.69.2881}. This protocol beats the Holevo bound
\cite{book2000mikeandike}, which limits an unassisted noiseless qubit channel
to transmit no more than one classical bit. The super-dense coding protocol
then led Bennett \textit{et al.} to explore if one could improve the classical
capacity of a noisy quantum channel by assisting it with unlimited
entanglement \cite{PhysRevLett.83.3081,ieee2002bennett}. They confirmed their
intuition by proving a channel coding theorem that gives an
entanglement-assisted classical transmission rate higher than that without
assistance. Shor then refined this result by determining trade-offs between
the classical communication rate and the entanglement consumption rate
\cite{arx2004shor}.

Quantum information theorists have since organized protocols that exploit the
different resources of quantum communication, classical communication, and
entanglement into a family tree
\cite{PhysRevLett.93.230504,arx2005dev,arx2006anura,arx2008oppenheim}. One
member of the family tree is the \textit{father protocol}
\cite{PhysRevLett.93.230504,arx2005dev}. The father protocol is so named
because it generates several \textquotedblleft child\textquotedblright%
\ protocols using the theory of resource inequalities
\cite{PhysRevLett.93.230504,arx2005dev}. Devetak \textit{et al.} exploited the
father protocol to demonstrate trade-offs between the quantum communication
rate and the entanglement consumption rate over an entanglement-assisted
quantum channel \cite{PhysRevLett.93.230504}.

An important natural question, in light of the aforementioned trade-off
solutions for two of the three noiseless resources, is then how one might
combine all three different resources. Previous work has addressed trade-offs
for the task of remotely preparing quantum states with the aid of classical
communication, quantum communication, and entanglement
\cite{PhysRevA.68.062319}, but no one has yet considered the triple trade-offs
for channel coding.

In this article, we conduct an investigation of the trade-offs for channel
coding both quantum and classical information over a quantum channel assisted
by noiseless entanglement. We prove the entanglement-assisted classical and
quantum capacity theorem, that gives achievable rates for this task. We extend
the family tree of quantum Shannon theory by developing the
\textit{classically-enhanced father protocol}.\footnote{As a side note, we
mention that former articles discuss the possibility of this protocol but
never fully developed it \cite{itit2008hsieh,arx2005dev}. In addition, the
current authors have both constructed \textquotedblleft classically-enhanced
father\textquotedblright\ error-correcting coding schemes for block codes
\cite{kremsky:012341} and for convolutional codes \cite{arx2008wildeUQCC}.}
This protocol is more general than any of the existing protocols in the tree
and achieves rates in the three-dimensional capacity region. We dub this
protocol the \textquotedblleft classically-enhanced
father\ protocol\textquotedblright\ because it is an extension of the father
protocol, and it generates five child\ protocols in the sense of
Refs.~\cite{PhysRevLett.93.230504,arx2005dev}. Two of its child protocols are
classically-enhanced quantum communication \cite{cmp2005dev} and
entanglement-assisted classical communication
\cite{PhysRevLett.83.3081,ieee2002bennett,arx2004shor} (we detail the others
in Section~\ref{sec:children}). We also demonstrate that isometric encodings
are sufficient for achieving our rate formulas, resolving an open problem from
Ref.~\cite{arx2005dev}.

A benefit of the classically-enhanced father protocol is that it inspires the
design of classically-enhanced entanglement-assisted quantum error-correcting
codes \cite{kremsky:012341,arx2008wildeUQCC}. We give evidence in
Section~\ref{sec:optimal-codes} that it is possible to reach the achievable
rates without encoding classical information into the entanglement shared
between the sender and receiver.

We structure this article as follows. In the next section, we give some
definitions and establish notation used in the remainder of the article.
Section~\ref{sec:description} provides a description of a general protocol for
communication of classical and quantum information with the assistance of
entanglement. We then state the main capacity theorem, Theorem~\ref{gf}, in
Section~\ref{sec:main-theorem} and show how the classical capacity theorem
\cite{ieee1998holevo,PhysRevA.56.131}, the quantum capacity theorem
\cite{PhysRevA.55.1613,capacity2002shor,ieee2005dev}, the classically-enhanced
quantum capacity region \cite{cmp2005dev}, the father capacity region
\cite{arx2005dev}, and the entanglement-assisted classical capacity region
\cite{arx2004shor}\ are all special cases of the entanglement-assisted
classical and quantum capacity region. We prove the converse of
Theorem~\ref{gf} in Section~\ref{sec:converse} and prove the direct-coding
part of Theorem~\ref{gf} in Section~\ref{sec:direct-coding}.
Section~\ref{sec:children}\ discusses the child protocols that the
classically-enhanced father protocol generates. We then give three example
channels, the completely depolarizing channel, the quantum erasure channel,
and the qubit dephasing channel, that admit a single-letter solution for the
capacity region (meaning that we have a complete understanding of the capacity
region for these channels). We also show that the classically-enhanced father
protocol gives an improvement over a time-sharing strategy when the noisy
channel is the qubit dephasing channel. We end by summarizing our results and
by posing several open questions.

\section{Definitions and Notation}

The ensemble $\left\{  p\left(  x\right)  ,\psi_{x}^{ABE}\right\}
_{x\in\mathcal{X}}$, where each state $\psi_{x}^{ABE}$ is a pure tripartite
state, is essential in the ensuing analysis of this article. The
\textit{coherent information} $I\left(  A\rangle B\right)  _{\psi_{x}}$\ of
each state $\psi_{x}^{ABE}$ in the ensemble is as follows:%
\[
I\left(  A\rangle B\right)  _{\psi_{x}}\equiv H\left(  B\right)  _{\psi_{x}%
}-H\left(  AB\right)  _{\psi_{x}},
\]
where $H\left(  B\right)  _{\psi_{x}}$ is the von Neumann entropy of the
reduction of the state $\psi_{x}^{ABE}$ to the system $B$ with a similar
definition for $H\left(  AB\right)  _{\psi_{x}}$. The \textit{quantum mutual
information} $I\left(  A;B\right)  _{\psi_{x}}$ of each state $\psi_{x}^{ABE}$
is as follows:%
\[
I\left(  A;B\right)  _{\psi_{x}}\equiv H\left(  A\right)  _{\psi_{x}}+I\left(
A\rangle B\right)  _{\psi_{x}}.
\]
We can classically correlate states in some system $X$ with each state
$\psi_{x}^{ABE}$ to produce an augmented ensemble%
\[
\left\{  p\left(  x\right)  ,\left\vert x\right\rangle \left\langle
x\right\vert ^{X}\otimes\psi_{x}^{ABE}\right\}  _{x\in\mathcal{X}},
\]
where the set $\left\{  \left\vert x\right\rangle \right\}  _{x\in\mathcal{X}%
}$ is some preferred orthonormal basis for the auxiliary system $X$. The
expected density operator of this augmented ensemble is the following
classical-quantum state:%
\[
\sigma^{XABE}\equiv\sum_{x\in\mathcal{X}}p\left(  x\right)  \left\vert
x\right\rangle \left\langle x\right\vert ^{X}\otimes\psi_{x}^{ABE}.
\]
The \textit{Holevo information} of the classical variable $X$ with the quantum
system $B$ is $I\left(  X;B\right)  _{\sigma}$. For the special case of a
classical system $X$, taking the expectation of the above entropic quantities
with respect to the density $p\left(  x\right)  $ gives the respective
conditional entropy $H\left(  A|X\right)  _{\sigma}$, conditional coherent
information $I\left(  A\rangle B|X\right)  _{\sigma}$, and conditional mutual
information $I\left(  A;B|X\right)  _{\sigma}$:%
\begin{align*}
H\left(  A|X\right)  _{\sigma}  &  \equiv\sum_{x\in\mathcal{X}}p\left(
x\right)  H\left(  A\right)  _{\psi_{x}},\\
I\left(  A\rangle B|X\right)  _{\sigma}  &  \equiv\sum_{x\in\mathcal{X}%
}p\left(  x\right)  I\left(  A\rangle B\right)  _{\psi_{x}},\\
I\left(  A;B|X\right)  _{\sigma}  &  \equiv\sum_{x\in\mathcal{X}}p\left(
x\right)  I\left(  A;B\right)  _{\psi_{x}}.
\end{align*}
One can easily prove that $I\left(  A\rangle B|X\right)  _{\sigma}=I\left(
A\rangle BX\right)  _{\sigma}$. We use the notation $I\left(  A\rangle
BX\right)  _{\sigma}$ for conditional coherent information in what follows.
The above definitions lead to the following useful identities:%
\begin{align}
H\left(  A|X\right)  _{\sigma}  &  =\frac{1}{2}I\left(  A;B|X\right)
_{\sigma}+\frac{1}{2}I\left(  A;E|X\right)  _{\sigma}%
,\label{eq:entropy-identity}\\
I\left(  A\rangle BX\right)  _{\sigma}  &  =\frac{1}{2}I\left(  A;B|X\right)
_{\sigma}-\frac{1}{2}I\left(  A;E|X\right)  _{\sigma}.
\label{eq:coherent-identity}%
\end{align}
Proving the above identities is a simple matter of noting that the von Neumann
entropy is equal for the reduced systems of a pure bipartite state. Adding the
above identities gives the following one:%
\begin{equation}
H\left(  A|X\right)  _{\sigma}+I\left(  A\rangle BX\right)  _{\sigma}=I\left(
A;B|X\right)  _{\sigma}. \label{eq:ent-coh-mut-identity}%
\end{equation}
The chain rule for quantum mutual information proves to be useful as well:%
\begin{equation}
I(AX;B)_{\sigma}=I(A;B|X)_{\sigma}+I(X;B)_{\sigma}. \label{eq:chain-rule}%
\end{equation}
All of the above information quantities possess operational interpretations in
the theorems in this article.
%(consider the expressions in Theorem~\ref{gf} and
%(\ref{eq:CEFR})).

A noisy quantum channel $\mathcal{N}^{A^{\prime}\rightarrow B}$ acts as a
completely-positive trace-preserving (CPTP) map. It takes a quantum system
$A^{\prime}$ as an input and produces a noisy output quantum system $B$.

A \textit{conditional quantum encoder} $\mathcal{E}^{MA\rightarrow B}$, or
\textit{conditional quantum channel }\cite{thesis2005yard}, is a collection
$\left\{  \mathcal{E}_{m}^{A\rightarrow B}\right\}  _{m}$\ of CPTP maps. Its
inputs are a classical system $M$ and a quantum system $A$ and its output is a
quantum system $B$. A classical-quantum state $\rho^{MA}$, where%
\[
\rho^{MA}\equiv\sum_{m}p\left(  m\right)  \left\vert m\right\rangle
\left\langle m\right\vert ^{M}\otimes\rho_{m}^{A},
\]
can act as an input to the conditional quantum encoder $\mathcal{E}%
^{MA\rightarrow B}$. The action of the conditional quantum encoder
$\mathcal{E}^{MA\rightarrow B}$ on the classical-quantum state $\rho^{MA}$ is
as follows:%
\begin{align*}
&  \mathcal{E}^{MA\rightarrow B}\left(  \rho^{MA}\right) \\
&  =\text{Tr}_{M}\left\{  \sum_{m}p\left(  m\right)  \left\vert m\right\rangle
\left\langle m\right\vert ^{M}\otimes\mathcal{E}_{m}^{A\rightarrow B}\left(
\rho_{m}^{A}\right)  \right\}  .
\end{align*}
It is actually possible to write \textit{any} quantum channel as a conditional
quantum encoder when its input is a classical-quantum state
\cite{thesis2005yard}. In this article, a conditional quantum encoder
functions as the sender Alice's encoder of classical and quantum information.

A \textit{quantum instrument }$\mathcal{D}^{A\rightarrow BM}$ is a CPTP\ map
whose input is a quantum system $A$ and whose outputs are a quantum system $B$
and a classical system $M$ \cite{arx2005dev,thesis2005yard}. A collection
$\left\{  \mathcal{D}_{m}^{A\rightarrow B}\right\}  _{m}$\ of
completely-positive trace-reducing maps specifies the instrument
$\mathcal{D}^{A\rightarrow BM}$. The action of the instrument $\mathcal{D}%
^{A\rightarrow BM}$ on an arbitrary input state $\rho$ is as follows:%
\begin{equation}
\mathcal{D}^{A\rightarrow BM}\left(  \rho^{A}\right)  =\sum_{m}\mathcal{D}%
_{m}^{A\rightarrow B}\left(  \rho^{A}\right)  \otimes\left\vert m\right\rangle
\left\langle m\right\vert ^{M}. \label{eq:instrument}%
\end{equation}
Tracing out the classical register $M$ gives the induced quantum operation
$\mathcal{D}^{A\rightarrow B}$ where%
\[
\mathcal{D}^{A\rightarrow B}\left(  \rho^{A}\right)  \equiv\sum_{m}%
\mathcal{D}_{m}^{A\rightarrow B}\left(  \rho^{A}\right)  .
\]
This sum map is trace preserving:%
\[
\text{Tr}\left\{  \sum_{m}\mathcal{D}_{m}^{A\rightarrow B}\left(  \rho
^{A}\right)  \right\}  =1.
\]
We can think of the following quantity%
\[
p\left(  m|\rho^{A}\right)  \equiv\text{Tr}\left\{  \mathcal{D}_{m}%
^{A\rightarrow B}\left(  \rho^{A}\right)  \right\}  ,
\]
as a conditional probability $p\left(  m|\rho^{A}\right)  $ of receiving the
classical message $m$ when the state $\rho^{A}$ is input. In this article, a
quantum instrument functions as Bob's decoder of classical and quantum information.

We abbreviate a \textit{capacity region }by the noiseless resources
involved:\ classical communication (C), quantum communication (Q), or
entanglement\ (E), but we abbreviate a \textit{protocol }with a different name
corresponding either to its inventors or an appropriate acronym. For example,
we speak of the C, Q, or CE capacity theorems for classical communication,
quantum communication, and entanglement-assisted classical communication,
respectively, but the corresponding protocols are
Holevo-Schumacher-Westmoreland coding (HSW), Lloyd-Shor-Devetak coding (LSD),
and entanglement-assisted classical coding (EAC).

We note some other points before beginning. The trace norm $\left\Vert
A\right\Vert _{1}$\ of an operator $A$ is as follows:%
\[
\left\Vert A\right\Vert _{1}\equiv\text{Tr}\left\{  \sqrt{A^{\dag}A}\right\}
.
\]
The maximally entangled state on system $T_{A}$ and $T_{B}$ is $\Phi
^{T_{A}T_{B}}$. The omission of a superscript implies a reduced state, e.g.,
the state $\Phi^{T_{A}}$ is the reduced state of $\Phi^{T_{A}T_{B}}$ on
$T_{A}$. Yard's thesis \cite{thesis2005yard} provides a good introduction to
quantum Shannon theory, and we point the reader there for properties such as
strong subadditivity \cite{LR73}\ and the quantum data processing inequality
\cite{PhysRevA.54.2629}.

\section{A General Protocol for Entanglement-Assisted Communication of
Classical and Quantum Information}

\label{sec:description}We begin by defining a general protocol for
entanglement-assisted communication of classical and quantum information
(EACQ)\ for a noisy quantum channel connecting a sender Alice to a receiver
Bob. Alice would like to communicate two items to Bob:

\begin{enumerate}
\item An arbitrary quantum state $\rho^{A_{1}}$\ in a system $A_{1}$ with
dimension $2^{nQ}$.

\item One of $2^{nC}$ classical messages.
\end{enumerate}

\begin{figure}
[ptb]
\begin{center}
\includegraphics[
natheight=5.826200in,
natwidth=8.406000in,
width=3.2716in
]%
{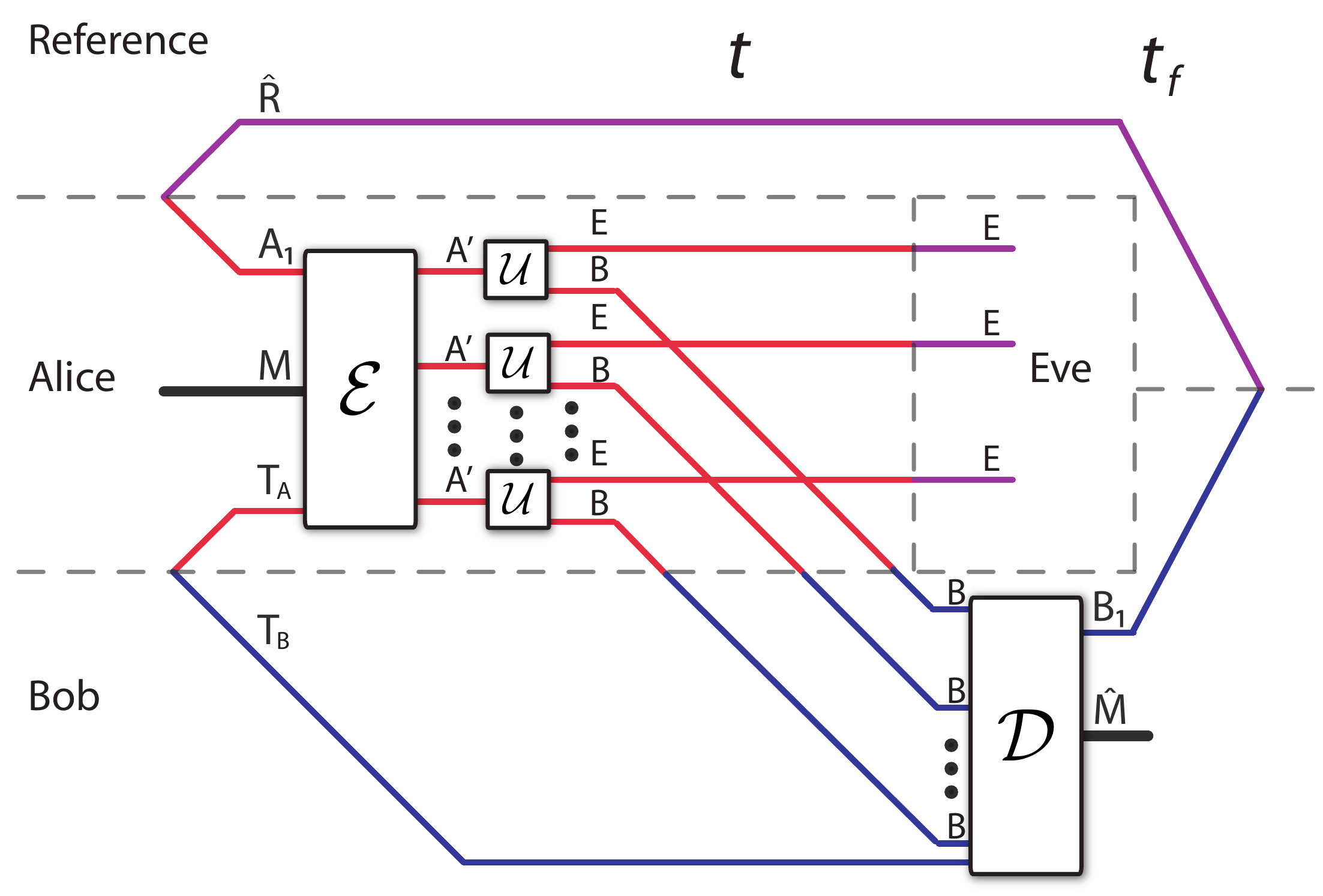}%
\caption{(Color online)\ The above figure depicts a general EACQ protocol. A
sender Alice would like to communicate the quantum information in system
$A_{1}$ and the classical information in system $M$. Her system $T_{A}$
represents shared maximal entanglement with the receiver's system $T_{B}$.
Alice encodes her information and uses the noisy channel a large number of
times. The environment Eve obtains part of the output and the receiver Bob
obtains the other part. Bob combines his received systems with his half of the
entanglement and performs a decoding operation to recover both the classical
and quantum information.}%
\label{fig:GFP}%
\end{center}
\end{figure}

Alice and Bob also share entanglement in the form of a maximally entangled
state $\Phi^{T_{A}T_{B}}$ prior to communication. Alice possesses the system
$T_{A}$, Bob possesses the system $T_{B}$, and the dimension of each system is
$2^{nE}$. We can think of this state as possessing $nE$ ebits of entanglement
because it is equivalent by local isometries to $nE$ \textquotedblleft gold
standard\textquotedblright\ ebits in the state $|\Phi^{+}\rangle^{AB}%
\equiv(\left\vert 00\right\rangle ^{AB}+\left\vert 11\right\rangle
^{AB})/\sqrt{2}$. Alice performs a conditional quantum encoder $\mathcal{E}%
^{MA_{1}T_{A}\rightarrow A^{\prime n}}$ that encodes both her quantum systems
$A_{1}$ and $T_{A}$ and the classical message in system $M$. The encoding
operation $\mathcal{E}^{MA_{1}T_{A}\rightarrow A^{\prime n}}$ prepares a
system $A^{\prime n}$ for input to a noisy quantum channel $\mathcal{N}%
^{A^{\prime n}\rightarrow B^{n}}$. The channel $\mathcal{N}^{A^{\prime
n}\rightarrow B^{n}}$ represents $n$ independent uses of the noisy quantum
channel $\mathcal{N}^{A^{\prime}\rightarrow B}$:%
\[
\mathcal{N}^{A^{\prime n}\rightarrow B^{n}}\equiv\left(  \mathcal{N}%
^{A^{\prime}\rightarrow B}\right)  ^{\otimes n}.
\]
She then sends her state through the quantum channel $\mathcal{N}^{A^{\prime
n}\rightarrow B^{n}}$. Bob receives the system $B^{n}$ and performs a decoding
instrument $\mathcal{D}^{B^{n}T_{B}\rightarrow B_{1}B_{E}\hat{M}}$ on the
channel output $B^{n}$ and his half of the entanglement $T_{B}$. The
instrument $\mathcal{D}^{B^{n}T_{B}\rightarrow B_{1}B_{E}\hat{M}}$ produces a
system $B_{1}$ with the quantum information that Alice sent, a classical
register $\hat{M}$ containing Alice's classical message, and another system
$B_{E}$ that does not contain any useful information. Bob should be able to
identify the classical message with high probability and recover the state
$\rho^{A_{1}}$ with high fidelity. Figure~\ref{fig:GFP} provides a detailed
illustration of this protocol.%
%TCIMACRO{\FRAME{ftbpFU}{3.2716in}{2.271in}{0pt}{\Qcb{(Color online)\ The above
%figure depicts a general EACQ protocol. A sender Alice would like to
%communicate the quantum information in system $A_{1}$ and the classical
%information in system $M$. Her system $T_{A}$ represents shared maximal
%entanglement with the receiver's system $T_{B}$. Alice encodes her information
%and uses the noisy channel a large number of times. The environment Eve
%obtains part of the output and the receiver Bob obtains the other part. Bob
%combines his received systems with his half of the entanglement and performs a
%decoding operation to recover both the classical and quantum information.}%
%}{\Qlb{fig:GFP}}{gfp-figure.pdf}{\special{ language "Scientific Word";
%type "GRAPHIC";  maintain-aspect-ratio TRUE;  display "USEDEF";
%valid_file "F";  width 3.2716in;  height 2.271in;  depth 0pt;
%original-width 8.406in;  original-height 5.8262in;  cropleft "0";
%filename 'class-enhanced-father.pdf';file-properties "XNPEU";}}}%
%BeginExpansion
%EndExpansion

It is useful to consider the isometric extension $U_{\mathcal{N}}^{A^{\prime
}\rightarrow BE}$ of the channel $\mathcal{N}^{A^{\prime}\rightarrow B}$ where
Alice controls the channel input system $A^{\prime}$, Bob has access to the
channel output system $B$, and the environment Eve has access to the system
$E$.\footnote{It should be clear from context when $E$ refers to Eve's system
or when it refers to the entanglement consumption rate.} For an independent
and identically distributed (IID)\ channel $\mathcal{N}^{A^{\prime
n}\rightarrow B^{n}}$ as defined above, we write its isometric extension as
$U_{\mathcal{N}}^{A^{\prime n}\rightarrow B^{n}E^{n}}$. Also, it is useful to
think of Alice's quantum system $\rho^{A_{1}}$ as a restriction of some pure
state $\varphi^{\hat{R}A_{1}}$ where Alice does not have access to the
purification system $\hat{R}$.

We formalize the EACQ quantum information processing task as follows. Define
an $(n,C,Q,E,\epsilon)$ EACQ code by

\begin{itemize}
\item Alice's conditional quantum encoder $\mathcal{E}^{MA_{1}T_{A}\rightarrow
A^{\prime n}}$ with encoding maps $\{\mathcal{E}_{m}^{A_{1}T_{A}\rightarrow
A^{\prime n}}\}_{m\in\lbrack2^{nC}]}$. This encoder encodes both her quantum
information and classical information. Define the following states for each
classical message $m$:%
\begin{equation}
\omega_{m}^{\hat{R}A^{\prime n}T_{B}}\equiv\mathcal{E}_{m}^{A_{1}%
T_{A}\rightarrow A^{\prime n}}(\varphi^{\hat{R}A_{1}}\otimes\Phi^{T_{A}T_{B}%
}), \label{eq:encoded-state}%
\end{equation}
where the dimension of system $A_{1}$ is $2^{nQ}$ and the dimension of system
$T_{A}$ is $2^{nE}$. The density operator that includes the classical register
$M$ and averages over all classical messages is as follows:%
\begin{equation}
\omega^{M\hat{R}A^{\prime n}T_{B}}\equiv\frac{1}{|M|}\sum_{m}\left\vert
m\right\rangle \left\langle m\right\vert ^{M}\otimes\omega_{m}^{\hat
{R}A^{\prime n}T_{B}},
\end{equation}
where $|M|$ is the size of the classical register $M$. The output of the
channel given that Alice sent classical message $m$ is then as follows:%
\[
\omega_{m}^{\hat{R}B^{n}E^{n}T_{B}}\equiv U_{\mathcal{N}}^{A^{\prime
n}\rightarrow B^{n}E^{n}}\left(  \omega_{m}^{\hat{R}A^{\prime n}T_{B}}\right)
.
\]
The average output of the channel is as follows:%
\begin{equation}
\omega^{M\hat{R}B^{n}E^{n}T_{B}}\equiv U_{\mathcal{N}}^{A^{\prime
n}\rightarrow B^{n}E^{n}}\left(  \omega^{M\hat{R}A^{\prime n}T_{B}}\right)  .
\label{eq:state-after-channel}%
\end{equation}

\item Bob's decoding instrument $\mathcal{D}^{B^{n}T_{B}\rightarrow B_{1}%
B_{E}\hat{M}}$, whose action is defined in (\ref{eq:instrument}), is a
collection of completely-positive trace-reducing maps $\{\mathcal{D}%
_{m}^{B^{n}T_{B}\rightarrow B_{1}B_{E}}\}_{m\in\lbrack2^{nC}]}$. The decoding
instrument decodes both the quantum information and classical information that
Alice sends. The density operator corresponding to Bob's output state is as
follows:%
\[
\omega^{M\hat{R}B_{1}B_{E}\hat{M}E^{n}}\equiv\mathcal{D}^{B^{n}T_{B}%
\rightarrow B_{1}B_{E}\hat{M}}\left(  \omega^{M\hat{R}B^{n}E^{n}T_{B}}\right)
.
\]

\end{itemize}

The classical probability of successful transmission of message $m$ is as
follows:%
\[
\Pr\{\hat{M}=m\ |\ M=m\}=\text{Tr}\{(\mathcal{D}_{m}^{B^{n}T_{B}\rightarrow
B_{1}B_{E}})(\omega_{m}^{\hat{R}B^{n}E^{n}T_{B}})\},
\]
where $\hat{M}$ denotes the random variable corresponding to Bob's received
classical message. The final state on the reference system $\hat{R}$ and Bob's
quantum system $B_{1}$ is $\Upsilon^{\hat{R}B_{1}}$ where%
\[
\Upsilon^{\hat{R}B_{1}}\equiv\tr_{\hat{M}B_{E}E^{n}}\{\mathcal{D}^{B^{n}%
T_{B}\rightarrow B_{1}B_{E}\hat{M}}(\omega_{m}^{\hat{R}B^{n}E^{n}T_{B}})\}.
\]

For the $\left(  n,C,Q,E,\epsilon\right)  $ EACQ code to be \textquotedblleft%
$\epsilon$-good,\textquotedblright\ the following two conditions should hold
for all classical messages $m\in\lbrack2^{nC}]$:

\begin{enumerate}
\item Bob decodes any of the classical messages $m$ with high probability%
\begin{equation}
\Pr\{\hat{M}=m\ |\ M=m\}\geq1-\epsilon, \label{eq:good-classical-comm}%
\end{equation}

\item The state $\Upsilon^{\hat{R}B_{1}}$ should be $\epsilon$-close to the
ideal state $\varphi^{\hat{R}B_{1}}\equiv\ $id$^{A_{1}\rightarrow B_{1}%
}(\varphi^{\hat{R}A_{1}})$:%
\begin{equation}
\left\Vert \Upsilon^{\hat{R}B_{1}}-\varphi^{\hat{R}B_{1}}\right\Vert _{1}%
\leq\epsilon, \label{eq:good-quantum-comm}%
\end{equation}
so that Bob recovers the quantum information in system $A_{1}$ with high fidelity.
\end{enumerate}

A rate triple $(C,Q,E)$ is \emph{achievable} if there exists an $\left(
n,C-\delta,Q-\delta,E+\delta,\epsilon\right)  $ EACQ code for any
$\epsilon,\delta>0$ and sufficiently large $n$. The capacity region
$\mathcal{C}(\mathcal{N})$ is a three-dimensional region containing all
achievable rate triples $(C,Q,E)$.

\section{The Entanglement-Assisted Classical and Quantum Capacity Theorem}

\label{sec:main-theorem}We now state our main theorem:\ the
entanglement-assisted classical and quantum capacity (CQE)\ theorem that
involves all three noiseless resources.%
\begin{figure*}
[ptb]
\begin{center}
\includegraphics[
natheight=5.750100in,
natwidth=11.111100in,
width=6.5518in
]%
{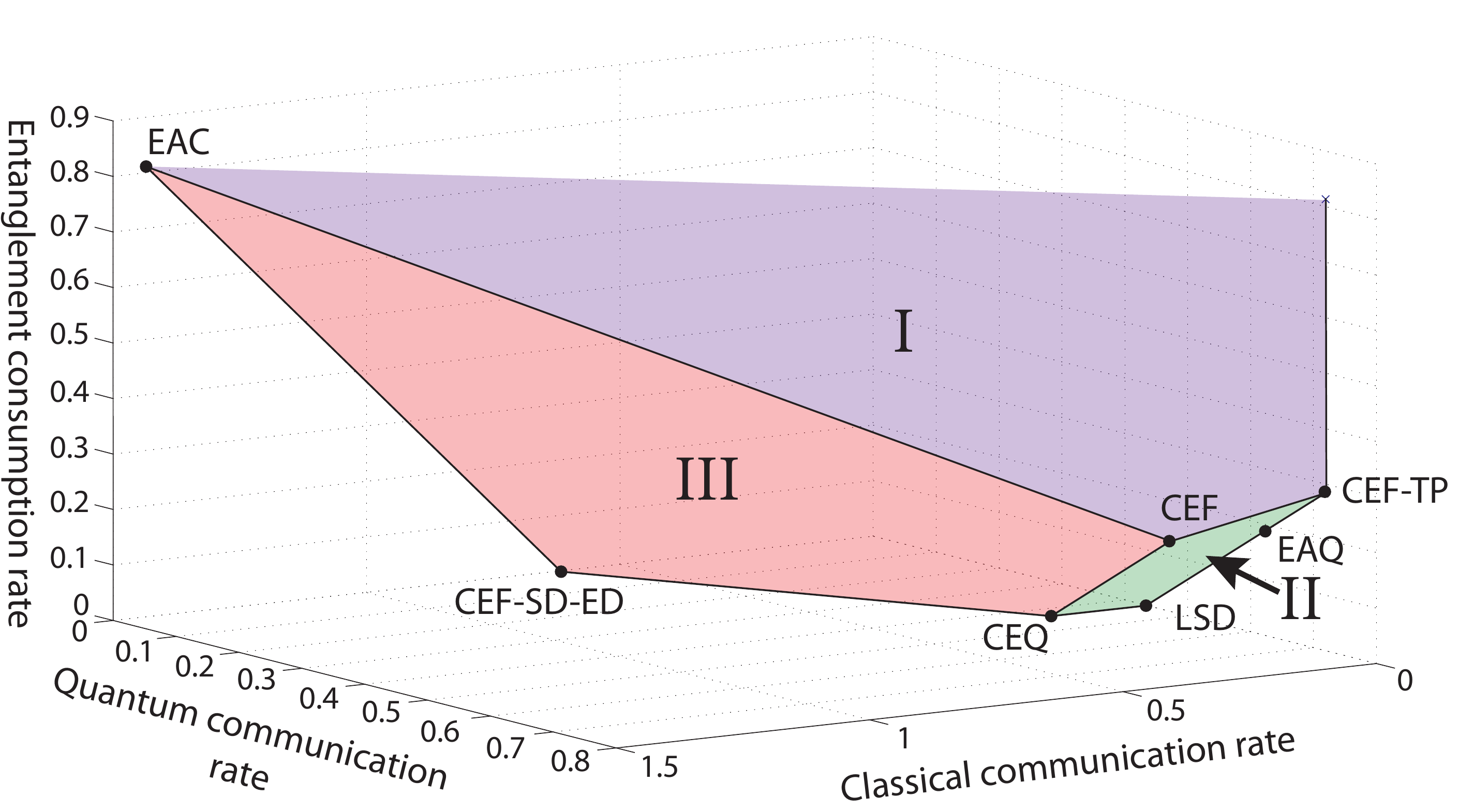}%
\caption{(Color online) An example of the one-shot, one-state achievable
region $\mathcal{C}_{\text{CQE,}\sigma}^{\left(  1\right)  }\left(
\mathcal{N}\right)  $ corresponding to a state $\sigma^{XABE}$ that arises
from a qubit dephasing channel with dephasing parameter $p=0.2$. The state
input to the channel $\mathcal{N}$ is $\sigma^{XAA^{\prime}}$, defined in
(\ref{eq:example-input-state}). The plot features seven achievable corner
points of the one-shot, one-state region. We can achieve the convex hull of
these eight points by time-sharing any two different coding strategies. We can
also achieve any point above an achievable point by consuming more
entanglement than necessary. The seven achievable points correspond to the
father protocol (EAQ) \cite{PhysRevLett.93.230504,arx2005dev}, the
Devetak-Shor protocol for classically-enhanced quantum communication
(CEQ)\ \cite{cmp2005dev}, Shor's protocol for entanglement-assisted classical
communication with limited entanglement (EAC) \cite{arx2004shor}, quantum
communication (LSD)\ \cite{PhysRevA.55.1613,capacity2002shor,ieee2005dev},
combining CEF\ with entanglement distribution and super-dense coding
(CEF-SD-ED) as detailed in Section \ref{sec:children}, the
classically-enhanced father protocol (CEF) outlined in
Section~\ref{sec:direct-coding}, and combining the classically-enhanced father
protocol with teleportation \cite{PhysRevLett.70.1895} (CEF-TP). Observe that
we can obtain EAC\ by combining CEF\ with super-dense coding as detailed in
Section \ref{sec:children}, so that the points CEQ, CEF, EAC, and CEF-SD-ED
all lie in plane III. Observe that we can obtain CEQ from CEF by entanglement
distribution and we can obtain LSD from EAQ and EAQ from CEF-TP,\ both by
entanglement distribution. Thus, the points CEF, CEQ, LSD, EAQ, and
CEF-TP\ all lie in plane II. Finally, observe that we can obtain all corner
points by combining CEF with the unit protocols in (\ref{TP}-\ref{ED}). This
one-shot, one-state achievable region for the state $\sigma^{XABE}$ is tight.
The bounds in (\ref{gf1}-\ref{gf3}) uniquely specify the respective planes
I-III. We obtain the full achievable region by taking the union over all
states $\sigma$ of the one-shot, one-state regions $\mathcal{C}_{\sigma
}^{\left(  1\right)  }\left(  \mathcal{N}\right)  $ and taking the
regularization, as outlined in Theorem~\ref{gf}. The above region is a
translation of the unit resource capacity region to the classically-enhanced
father protocol.}%
\label{fig:one-shot-region-state}%
\end{center}
\end{figure*}
%EndExpansion

\begin{theorem}
\label{gf}The capacity region $\mathcal{C}_{\text{CQE}}(\mathcal{N})$ of an
entanglement-assisted quantum channel $\mathcal{N}$ for simultaneously
transmitting both quantum information and classical information is equal to
the following expression:%
\begin{equation}
\mathcal{C}_{\text{CQE}}(\mathcal{N})=\overline{\bigcup_{k=1}^{\infty}\frac
{1}{k}\mathcal{C}_{\text{CQE}}^{(1)}(\mathcal{N}^{\otimes k})}, \label{gfc}%
\end{equation}
where the overbar indicates the closure of a set. The \textquotedblleft
one-shot\textquotedblright\ region $\mathcal{C}_{\text{CQE}}^{(1)}%
(\mathcal{N})$ is the union of the regions $\mathcal{C}_{\text{CQE},\sigma
}^{(1)}(\mathcal{N})$:%
\[
\mathcal{C}_{\text{CQE}}^{(1)}(\mathcal{N})\equiv\bigcup_{\sigma}%
\mathcal{C}_{\text{CQE},\sigma}^{(1)}(\mathcal{N}),
\]
where $\mathcal{C}_{\text{CQE},\sigma}^{(1)}(\mathcal{N})$ is the set of all
$C,Q,E\geq0$, such that%
\begin{align}
C+2Q  &  \leq I(AX;B)_{\sigma},\label{gf1}\\
Q  &  \leq I(A\rangle BX)_{\sigma}+E,\label{gf2}\\
C+Q  &  \leq I(X;B)_{\sigma}+I(A\rangle BX)_{\sigma}+E. \label{gf3}%
\end{align}
The above entropic quantities are with respect to a \textquotedblleft
one-shot\textquotedblright\ quantum state $\sigma^{XABE}$ where%
\begin{equation}
\sigma^{XABE}\equiv\sum_{x}p(x)\left\vert x\right\rangle \left\langle
x\right\vert ^{X}\otimes U_{\mathcal{N}}^{A^{\prime}\rightarrow BE}(\phi
_{x}^{AA^{\prime}}), \label{eq:maximization-state}%
\end{equation}
the states $\phi_{x}^{AA^{\prime}}$ are pure, and it is sufficient to consider
$\left\vert \mathcal{X}\right\vert \leq\min\left\{  \left\vert A^{\prime
}\right\vert ,\left\vert B\right\vert \right\}  ^{2}+1$ by the method in
Ref.~\cite{ieee2008yard}.
\end{theorem}

The capacity region in Theorem~\ref{gf} is a union of general polyhedra, each
specified by (\ref{gf1}-\ref{gf3}), where the union is over all possible
states of the form (\ref{eq:maximization-state}) and a potentially infinite
number of uses of the channel. Figure~\ref{fig:one-shot-region-state}%
\ illustrates an example of the general polyhedron specified by (\ref{gf1}%
-\ref{gf3}), where the channel is the qubit dephasing channel\footnote{The
action of the qubit dephasing channel with dephasing parameter $p$ on a
density operator $\rho$ is $\rho\rightarrow\left(  1-p\right)  \rho+pZ\rho
Z$.} with dephasing parameter $p=0.2$, and the input state is%
\begin{equation}
\sigma^{XAA^{\prime}}\equiv\frac{1}{2}(\left\vert 0\right\rangle \left\langle
0\right\vert ^{X}\otimes\phi_{0}^{AA^{\prime}}+\left\vert 1\right\rangle
\left\langle 1\right\vert ^{X}\otimes\phi_{1}^{AA^{\prime}}),
\label{eq:example-input-state}%
\end{equation}
where%
\begin{align*}
\left\vert \phi_{0}\right\rangle ^{AA^{\prime}}  &  \equiv\sqrt{1/4}\left\vert
00\right\rangle ^{AA^{\prime}}+\sqrt{3/4}\left\vert 11\right\rangle
^{AA^{\prime}},\\
\left\vert \phi_{1}\right\rangle ^{AA^{\prime}}  &  \equiv\sqrt{3/4}\left\vert
00\right\rangle ^{AA^{\prime}}+\sqrt{1/4}\left\vert 11\right\rangle
^{AA^{\prime}}.
\end{align*}
The state $\sigma^{XABE}$ resulting from the channel is $U_{\mathcal{N}%
}^{A^{\prime}\rightarrow BE}(\sigma^{XAA^{\prime}})$ where $U_{\mathcal{N}}$
is an isometric extension of the qubit dephasing channel. The figure caption
provides a detailed explanation of the one-shot, one-state region
$\mathcal{C}_{\text{CQE},\sigma}^{(1)}$ (note that
Figure~\ref{fig:one-shot-region-state} displays the one-shot, one-state region
and does not display the full capacity region).

The above capacity region has the simple interpretation. In Ref.~\cite{HW09},
we determined a unit resource capacity region. This unit resource region
outlines what is achievable if one does not possess a noisy channel, but only
possesses the three noiseless resources of classical communication, quantum
communication, and entanglement. There, we found that the optimal strategy is
to combine teleportation, super-dense coding, and entanglement distribution.
Interestingly, the above set of inequalities demonstrates that the one-shot,
one-state region is a translation of the unit resource capacity region to the
classically-enhanced father protocol. Indeed, eliminating the entropic
quantities from (\ref{gf1}-\ref{gf3}) reveals that the inequalities are the
same as those that specify the unit resource capacity region.

Proving that Theorem~\ref{gf} holds consists of proving it in two steps,
traditionally called the \textit{direct coding theorem} and the
\textit{converse}. For our case, the \textit{direct coding theorem }proves
that the region corresponding to the right hand side of (\ref{gfc}) is an
achievable rate region. It constructs an EACQ protocol whose rates are in the
region of the right hand side of (\ref{gfc}) and shows that its fidelity of
quantum communication is high and its probability of error of classical
communication is small. The \textit{converse} assumes that a good code with
high fidelity and low probability of error\ exists and shows that the region
on the right hand side of (\ref{gfc}) bounds the achievable rate region. We
prove the converse in Section~\ref{sec:converse}\ and the direct coding
theorem in Section~\ref{sec:direct-coding}.

\subsection{Special Cases of the Capacity Theorem}

We first consider five special cases of the above capacity theorem that arise
when $Q$ and $E$ both vanish, $C$ and $E$ both vanish, or one of $C$, $Q$, or
$E$ vanishes. The first two cases correspond respectively to the
Holevo-Schumacher-Westmoreland coding theorem and the Lloyd-Shor-Devetak
coding theorem. Each of the other special cases traces out a two-dimensional
achievable rate region in the three-dimensional capacity region. The five
coding scenarios are as follows:

\begin{enumerate}
\item Classical communication (C)\ when there is no entanglement assistance or
quantum communication \cite{ieee1998holevo,PhysRevA.56.131}. The achievable
rate region lies on the $(C,0,0)$ ray extending from the origin.

\item Quantum communication (Q)\ when there is no entanglement assistance or
classical communication \cite{PhysRevA.55.1613,capacity2002shor,ieee2005dev}.
The achievable rate region lies on the $(0,Q,0)$ ray extending from the origin.

\item Entanglement-assisted quantum communication (QE) when there is no
classical communication \cite{PhysRevLett.93.230504,arx2005dev}. The
achievable rate region lies in the $\left(  0,Q,E\right)  $ quarter-plane of
the three-dimensional region in (\ref{gfc}).

\item Classically-enhanced quantum communication (CQ)\ when there is no
entanglement assistance \cite{cmp2005dev}. The achievable rate region lies in
the $\left(  C,Q,0\right)  $ quarter-plane of the three-dimensional region in
(\ref{gfc}).

\item Entanglement-assisted classical communication (CE) when there is no
quantum communication \cite{arx2004shor}. The achievable rate region lies in
the $\left(  C,0,E\right)  $ quarter-plane of the three-dimensional region in
(\ref{gfc}).
\end{enumerate}

\subsubsection{Classical Capacity}

The following theorem gives the one-dimensional capacity region $\mathcal{C}%
_{\text{C}}(\mathcal{N})$ of a quantum channel $\mathcal{N}$ for classical
communication \cite{ieee1998holevo,PhysRevA.56.131}.

\begin{theorem}
The classical capacity region $\mathcal{C}_{\text{C}}(\mathcal{N})$ is given
by%
\begin{equation}
\mathcal{C}_{\text{C}}(\mathcal{N})=\overline{\bigcup_{k=1}^{\infty}\frac
{1}{k}\mathcal{C}_{\text{C}}^{(1)}(\mathcal{N}^{\otimes k})}.
\end{equation}
The \textquotedblleft one-shot\textquotedblright\ region $\mathcal{C}%
_{\text{C}}^{(1)}(\mathcal{N})$ is the union of the regions $\mathcal{C}%
_{\text{C},\sigma}^{(1)}(\mathcal{N})$, where $\mathcal{C}_{\text{C},\sigma
}^{(1)}(\mathcal{N})$ is the set of all $C\geq0$, such that
\begin{equation}
C\leq I(X;B)_{\sigma}+I\left(  A\rangle BX\right)  _{\sigma}.
\label{eq:HSW-bound}%
\end{equation}
The entropic quantity is with respect to the state $\sigma^{XABE}$ in
(\ref{eq:maximization-state}).
\end{theorem}

The bound in (\ref{eq:HSW-bound}) is a special case of the bound in
(\ref{gf3}) with $Q=0$ and $E=0$. The above characterization of the classical
capacity\ region may seem slightly different from the original
HSW\ characterization, until we make a few observations. First, we rewrite the
coherent information $I\left(  A\rangle BX\right)  _{\sigma}$ as $H\left(
B|X\right)  _{\sigma}-H\left(  E|X\right)  _{\sigma}$. Then $I(X;B)_{\sigma
}+I\left(  A\rangle BX\right)  _{\sigma}=H\left(  B\right)  _{\sigma}-H\left(
E|X\right)  _{\sigma}$. Next, pure states of the form $\left\vert
\varphi\right\rangle _{x}^{A^{\prime}}$ are sufficient to attain the classical
capacity of a quantum channel \cite{arx2004shor}. We briefly recall this
argument. An ensemble of the following form realizes the classical capacity of
a quantum channel:%
\[
\rho^{XA^{\prime}}\equiv\sum_{x}p_{X}\left(  x\right)  \left\vert
x\right\rangle \left\langle x\right\vert ^{X}\otimes\rho_{x}^{A^{\prime}}.
\]
This ensemble itself is a restriction of the ensemble in
(\ref{eq:maximization-state}) to the systems $X$ and $A^{\prime}$. Each mixed
state $\rho_{x}^{A^{\prime}}$ admits a spectral decomposition of the form
$\rho_{x}^{A^{\prime}}=\sum_{y}p_{Y|X}\left(  y|x\right)  \psi_{x,y}%
^{A^{\prime}}$ where $\psi_{x,y}^{A^{\prime}}$ is a pure state. We can define
an augmented classical-quantum state $\theta^{XYA^{\prime}}$ as follows:%
\[
\theta^{XYA^{\prime}}\equiv\sum_{x,y}p_{Y|X}\left(  y|x\right)  p_{X}\left(
x\right)  \left\vert x\right\rangle \left\langle x\right\vert ^{X}%
\otimes\left\vert y\right\rangle \left\langle y\right\vert ^{Y}\otimes
\psi_{x,y}^{A^{\prime}},
\]
so that Tr$_{Y}\{\theta^{XYA^{\prime}}\}=\rho^{XA^{\prime}}$. Sending the
$A^{\prime}$ system of the states $\rho^{XA^{\prime}}$ and $\theta
^{XYA^{\prime}}$ leads to the respective states $\rho^{XB}$ and $\theta^{XYB}%
$. Then the following equality and inequality hold:%
\begin{align*}
I\left(  X;B\right)  _{\rho}  & =I\left(  X;B\right)  _{\theta}\\
& \leq I\left(  XY;B\right)  _{\theta},
\end{align*}
where the equality holds because Tr$_{Y}\{\theta^{XYA^{\prime}}\}=\rho
^{XA^{\prime}}$ and the inequality follows from quantum data processing.
Redefining the classical variable as the joint random variable $X,Y$ reveals
that it is sufficient to consider pure state ensembles for the classical
capacity. Returning to our main argument, then $H\left(  E|X\right)  _{\sigma
}=H\left(  B|X\right)  _{\sigma}$ so that $I(X;B)_{\sigma}+I\left(  A\rangle
BX\right)  _{\sigma}=H\left(  B\right)  _{\sigma}-H\left(  B|X\right)
_{\sigma}=I(X;B)_{\sigma}$ for states of this form. Thus, the expression in
(\ref{eq:HSW-bound}) can never exceed the classical capacity and finds its
maximum exactly at the Holevo information.

\subsubsection{Quantum Capacity}

The following theorem gives the one-dimensional quantum capacity region
$\mathcal{C}_{\text{Q}}(\mathcal{N})$ of a quantum channel $\mathcal{N}$
\cite{PhysRevA.55.1613,capacity2002shor,ieee2005dev}.

\begin{theorem}
The quantum capacity region $\mathcal{C}_{\text{Q}}(\mathcal{N})$ is given by%
\begin{equation}
\mathcal{C}_{\text{Q}}(\mathcal{N})=\overline{\bigcup_{k=1}^{\infty}\frac
{1}{k}\mathcal{C}_{\text{Q}}^{(1)}(\mathcal{N}^{\otimes k})}.
\end{equation}
The \textquotedblleft one-shot\textquotedblright\ region $\mathcal{C}%
_{\text{Q}}^{(1)}(\mathcal{N})$ is the union of the regions $\mathcal{C}%
_{\text{Q},\sigma}^{(1)}(\mathcal{N})$, where $\mathcal{C}_{\text{Q},\sigma
}^{(1)}(\mathcal{N})$ is the set of all $Q\geq0$, such that
\begin{equation}
Q\leq I(A\rangle BX)_{\sigma}. \label{eq:LSD-bound}%
\end{equation}
The entropic quantity is with respect to the state $\sigma^{XABE}$ in
(\ref{eq:maximization-state}) with the restriction that the density $p(x)$ is degenerate.
\end{theorem}

The bound in (\ref{eq:LSD-bound})\ is a special case of the bound in
(\ref{gf2}) with $E=0$. The other bounds in Theorem~\ref{gf} are looser than
the bound in (\ref{gf2}) when $C,E=0$.

\subsubsection{Entanglement-Assisted Quantum Capacity}

The following theorem gives the two-dimensional entanglement-assisted quantum
capacity region $\mathcal{C}_{\text{QE}}(\mathcal{N})$ of a quantum channel
$\mathcal{N}$ \cite{PhysRevLett.93.230504,arx2005dev}.

\begin{theorem}
The entanglement-assisted quantum capacity region $\mathcal{C}_{\text{QE}%
}(\mathcal{N})$ is given by
\begin{equation}
\mathcal{C}_{\text{QE}}(\mathcal{N})=\overline{\bigcup_{k=1}^{\infty}\frac
{1}{k}\mathcal{C}_{\text{QE}}^{(1)}(\mathcal{N}^{\otimes k})}. \label{eaQ}%
\end{equation}
The \textquotedblleft one-shot\textquotedblright\ region $\mathcal{C}%
_{\text{QE}}^{(1)}(\mathcal{N})$ is the union of the regions $\mathcal{C}%
_{\text{QE},\sigma}^{(1)}(\mathcal{N})$, where $\mathcal{C}_{\text{QE},\sigma
}^{(1)}(\mathcal{N})$ is the set of all $Q,E\geq0$, such that
\begin{align}
Q  &  \leq\frac{1}{2}I(AX;B)_{\sigma},\label{eaQ1}\\
Q  &  \leq E+I(A\rangle BX)_{\sigma}. \label{eaQ2}%
\end{align}
The entropic quantities are with respect to the state $\sigma^{XABE}$ in
(\ref{eq:maximization-state}) with the restriction that the density $p(x)$ is degenerate.
\end{theorem}

The bounds in (\ref{eaQ1}) and (\ref{eaQ2}) are a special case of the
respective bounds in (\ref{gf1}) and (\ref{gf2}) with $C=0$. The other bounds
in Theorem~\ref{gf} are looser than the bounds in (\ref{gf1}) and (\ref{gf2})
when $C=0$. Observe that the region is a union of general pentagons (see the
$QE$-plane in Figure~\ref{fig:one-shot-region-state} for an example of one of
these general pentagons in the union).

\subsubsection{Classically-Enhanced Quantum Capacity}

The following theorem gives the two-dimensional capacity region $\mathcal{C}%
_{\text{CQ}}(\mathcal{N})$ for classically-enhanced quantum communication
through a quantum channel $\mathcal{N}$ \cite{cmp2005dev}.

\begin{theorem}
The classically-enhanced quantum capacity region $\mathcal{C}_{\text{CQ}%
}(\mathcal{N})$ is given by
\begin{equation}
\mathcal{C}_{\text{CQ}}(\mathcal{N})=\overline{\bigcup_{k=1}^{\infty}\frac
{1}{k}\mathcal{C}_{\text{CQ}}^{(1)}(\mathcal{N}^{\otimes k})}.
\end{equation}
The \textquotedblleft one-shot\textquotedblright\ region $\mathcal{C}%
_{\text{CQ}}^{(1)}(\mathcal{N})$ is the union of the regions $\mathcal{C}%
_{\text{CQ},\sigma}^{(1)}(\mathcal{N})$, where $\mathcal{C}_{\text{CQ},\sigma
}^{(1)}(\mathcal{N})$ is the set of all $C,Q\geq0$, such that%
\begin{align}
C+Q  &  \leq I(X;B)_{\sigma}+I(A\rangle BX)_{\sigma},\label{eq:ceq-1}\\
Q  &  \leq I(A\rangle BX)_{\sigma}. \label{eq:ceq-2}%
\end{align}
The entropic quantities are with respect to the state $\sigma^{XABE}$ in
(\ref{eq:maximization-state}).
\end{theorem}

The bounds in (\ref{eq:ceq-1}) and (\ref{eq:ceq-2}) are a special case of the
respective bounds in (\ref{gf2}) and (\ref{gf3}) with $E=0$. Observe that the
region is a union of trapezoids (see the $CQ$-plane in
Figure~\ref{fig:one-shot-region-state} for an example of one of these
rectangles in the union).

The above characterization is a slightly improved characterization of the
Devetak-Shor region from Ref.~\cite{cmp2005dev}. Indeed, the one-shot,
one-state region there was a union of rectangles given by the following set of
inequalities:%
\begin{align}
C  &  \leq I(X;B)_{\sigma},\\
Q  &  \leq I(A\rangle BX)_{\sigma}.
\end{align}
These rectangles are inside the trapezoids above. Though, our characterization
in (\ref{eq:ceq-1}-\ref{eq:ceq-2}) is the same as theirs when we consider the
union over all the one-shot, one-state regions.

\subsubsection{Entanglement-assisted Classical Capacity with Limited
Entanglement}

\begin{theorem}
\label{thm:shors-theorem}The entanglement-assisted classical capacity region
$\mathcal{C}_{\text{CE}}(\mathcal{N})$ of a quantum channel $\mathcal{N}$\ is%
\begin{equation}
\mathcal{C}_{\text{CE}}(\mathcal{N})=\overline{\bigcup_{k=1}^{\infty}\frac
{1}{k}\mathcal{C}_{\text{CE}}^{(1)}(\mathcal{N}^{\otimes k})}. \label{eaC}%
\end{equation}
The \textquotedblleft one-shot\textquotedblright\ region $\mathcal{C}%
_{\text{CE}}^{(1)}(\mathcal{N})$ is the union of the regions $\mathcal{C}%
_{\text{CE},\sigma}^{(1)}(\mathcal{N})$, where $\mathcal{C}_{\text{CE},\sigma
}^{(1)}(\mathcal{N})$ is the set of all $C,E\geq0$, such that
\begin{align}
C  &  \leq I(AX;B)_{\sigma},\label{eac1}\\
C  &  \leq I(X;B)_{\sigma}+I(A\rangle BX)_{\sigma}+E. \label{eac2}%
\end{align}
where the entropic quantities are with respect to the state $\sigma^{XABE}$ in
(\ref{eq:maximization-state}).
\end{theorem}

The bounds in (\ref{eac1}) and (\ref{eac2}) are a special case of the
respective bounds in (\ref{gf1}) and (\ref{gf3}) with $Q=0$. Observe that the
region is a union of general polyhedra (see the CE-plane in
Figure~\ref{fig:one-shot-region-state} for an example of one of these general
polyhedra in the union).

The above characterization of the CE achievable region is again an improvement
over the characterization in
Refs.~\cite{ieee2002bennett,arx2004shor,arx2005dev}. It specifies a union of
general trapezoids. The region in
Refs.~\cite{ieee2002bennett,arx2004shor,arx2005dev} was a union of general
rectangles of the form:
\begin{align}
C &  \leq I(AX;B)_{\sigma},\\
E &  \geq H(A|X)_{\sigma}.
\end{align}
These general rectangles are inside the above general trapezoids (note that
the bounds in (\ref{eac1}-\ref{eac2}) intersect at $E=H(A|X)_{\sigma}$), but
the regions coincide when we take the union over all the one-shot, one-state regions.

\section{The Converse Proof}

\label{sec:converse}Our method for proving the converse of Theorem~\ref{gf} is
to apply standard entropic bounds that are available in
Ref.~\cite{book2000mikeandike}. We first recall the Fannes inequality for
continuity of entropy, the Alicki-Fannes inequality for continuity of coherent
information, and another inequality of the Fannes class for continuity of
quantum mutual information.

\begin{theorem}
[Fannes Inequality \cite{F73}]\label{Thm:F}Suppose two states $\rho^{A}$ and
$\sigma^{A}$ are close:%
\[
\left\Vert \rho^{A}-\sigma^{A}\right\Vert _{1}\leq\epsilon.
\]
Then their respective entropies are close:%
\begin{equation}
\left\vert H(A)_{\rho}-H(A)_{\sigma}\right\vert \leq\epsilon\log\left\vert
A\right\vert +H_{2}(\epsilon).
\end{equation}
$\left\vert A\right\vert $ is the dimension of the system $A$ and
$H_{2}\left(  \epsilon\right)  $ is the binary entropy function that has the
property $\lim_{\epsilon\rightarrow0}H_{2}(\epsilon)=0$.
\end{theorem}

\begin{theorem}
[Alicki-Fannes Inequality \cite{0305-4470-37-5-L01}]\label{Thm:AF}Suppose two
states $\rho^{AB}$ and $\sigma^{AB}$ are close:%
\[
\left\Vert \rho^{AB}-\sigma^{AB}\right\Vert _{1}\leq\epsilon.
\]
Then their respective coherent informations are close:%
\begin{equation}
\left\vert I(A\rangle B)_{\rho}-I(A\rangle B)_{\sigma}\right\vert
\leq4\epsilon\log\left\vert A\right\vert +2H_{2}(\epsilon).
\end{equation}

\end{theorem}

\begin{corollary}
\label{cor:F-MI}Suppose two states $\rho^{AB}$ and $\sigma^{AB}$ are close:%
\[
\left\Vert \rho^{AB}-\sigma^{AB}\right\Vert _{1}\leq\epsilon.
\]
Then their respective quantum mutual informations are close:%
\begin{equation}
\left\vert I(A;B)_{\rho}-I(A;B)_{\sigma}\right\vert \leq5\epsilon
\log\left\vert A\right\vert +3H_{2}(\epsilon).
\end{equation}

\end{corollary}

\begin{IEEEproof}
The proof follows in two steps by applying Theorems~\ref{Thm:F} and
\ref{Thm:AF}. First, monotonicity of the trace distance under the discarding
of subsystems implies that $\left\Vert \rho^{A}-\sigma^{A}\right\Vert _{1}%
\leq\epsilon$. Theorem~\ref{Thm:F} then applies. The corollary then follows
from the equality $I\left(  A;B\right)  =H\left(  A\right)  +I\left(  A\rangle
B\right)  $ and the triangle inequality.
\end{IEEEproof}

\begin{IEEEproof}
[Converse]Section~\ref{sec:description}\ describes the most general EACQ
protocol and this most general case is the one we consider in proving the
converse. Suppose Alice shares the maximally entangled state $\Phi^{\hat
{R}A_{1}}$ with the reference system $\hat{R}$ (the protocol should be able to
transmit the entanglement in state $\Phi^{\hat{R}A_{1}}$ with $\epsilon
$-accuracy if it can approximately transmit the entanglement with system
$\hat{R}$ for any pure state on $\hat{R}$ and $A_{1}$). Alice also shares the
maximally entangled state $\Phi^{T_{A}T_{B}}$ with Bob. Alice combines her
system $A_{1}$ of the quantum state $\Phi^{\hat{R}A_{1}}$ with her system
$T_{A}$ of the state $\Phi^{T_{A}T_{B}}$ and the classical register $M$ that
contains her classical information. The most general encoding operation that
she can perform on her three registers $M$, $A_{1}$, and $T_{A}$ is a
conditional quantum encoder $\mathcal{E}^{MA_{1}T_{A}\rightarrow A^{\prime n}%
}$ consisting of a collection $\{\mathcal{E}_{m}^{A_{1}T_{A}\rightarrow
A^{\prime n}}\}_{m}$\ of CPTP maps. For now, we assume this general form of the encoder
but later show in Appendix~\ref{sec:isometric-encoders} that it is only necessary to consider a collection of isometries. Each element $\mathcal{E}_{m}^{A_{1}%
T_{A}\rightarrow A^{\prime n}}$ of the conditional quantum encoder produces
the following state:%
\[
\omega_{m}^{\hat{R}A^{\prime n}E^{\prime}T_{B}}\equiv U_{\mathcal{E}_{m}%
}^{A_{1}T_{A}\rightarrow A^{\prime n}E^{\prime}}(\Phi^{\hat{R}A_{1}}%
\otimes\Phi^{T_{A}T_{B}}),
\]
where we consider the isometric extension $U_{\mathcal{E}_{m}}^{A_{1}%
T_{A}\rightarrow A^{\prime n}E^{\prime}}$ of each element $\mathcal{E}%
_{m}^{A_{1}T_{A}\rightarrow A^{\prime n}}$. The average density operator over
all classical messages is then as follows:%
\[
\frac{1}{|M|}\sum_{m}\left\vert m\right\rangle \left\langle m\right\vert
^{M}\otimes\omega_{m}^{\hat{R}A^{\prime n}E^{\prime}T_{B}}.
\]
Alice sends the $A^{\prime n}$ system through the noisy channel
$U_{\mathcal{N}}^{A^{\prime n}\rightarrow B^{n}E^{n}}$, producing the
following state:%
\begin{align}
&  \omega^{M\hat{R}B^{n}E^{n}E^{\prime}T_{B}}\label{eq:omega-state}\\
&  \equiv\frac{1}{|M|}\sum_{m}\left\vert m\right\rangle \left\langle
m\right\vert ^{M}\otimes U_{\mathcal{N}}^{A^{\prime n}\rightarrow B^{n}E^{n}%
}(\omega_{m}^{\hat{R}A^{\prime n}E^{\prime}T_{B}}).\nonumber
\end{align}
Define $A\equiv\hat{R}T_{B}$ so that the state in (\ref{eq:omega-state}) is a
particular $n^{\text{th}}$ extension of the state in
(\ref{eq:maximization-state}). The above state is the state at time $t$ in
Figure~\ref{fig:GFP}. Bob receives the above state and performs a decoding
instrument $\mathcal{D}^{B^{n}T_{B}\rightarrow B_{1}B_{E}\hat{M}}$. The
protocol ends at time $t_{f}$. Let $\left(  \omega^{\prime}\right)  ^{M\hat
{R}B_{1}B_{E}\hat{M}E^{n}E^{\prime}}$ be the state at time $t_{f}$ after Bob
processes $\omega^{M\hat{R}B^{n}E^{n}E^{\prime}T_{B}}$ with the decoding
instrument $\mathcal{D}^{B^{n}T_{B}\rightarrow B_{1}B_{E}\hat{M}}$. Suppose
that an $\left(  n,C-\delta,Q-\delta,E+\delta,\epsilon\right)  $ EACQ protocol
as given above exists. We prove that the following bounds apply to the
elements of its rate triple $\left(  C-\delta,Q-\delta,E+\delta\right)  $:%
\begin{align}
C+2Q-\delta &  \leq\frac{1}{n}I(AM;B^{n})_{\omega},\label{cgf1}\\
Q-\delta &  \leq\frac{1}{n}I(A\rangle B^{n}M)_{\omega}+E,\label{cgf2}\\
C+Q-\delta &  \leq\frac{1}{n}\left(  I(M;B^{n})_{\omega}+I(A\rangle
B^{n}M)_{\omega}\right)  +E,\label{cgf3}%
\end{align}
for any $\epsilon,\delta>0$ and all sufficiently large $n$. In the ideal case,
the identity quantum channel acts on system $A_{1}$ to produce the maximally
entangled state $\Phi^{\hat{R}B_{1}}$. So for our case, the following
inequality%
\begin{equation}
\left\Vert \left(  \omega^{\prime}\right)  ^{\hat{R}B_{1}}-\Phi^{\hat{R}B_{1}%
}\right\Vert _{1}\leq\epsilon\label{eq:converse-good-q-comm}%
\end{equation}
holds because the protocol is $\epsilon$-good for quantum communication
according to the criterion in\ (\ref{eq:good-quantum-comm}). Also, in the
ideal case, the identity classical channel acts on system $M$ to produce the
maximally correlated state $\overline{\Phi}^{M\hat{M}}$ where
\begin{equation}
\overline{\Phi}^{M\hat{M}} \equiv \frac{1}{|M|}
\sum_m \vert m \rangle \langle m \vert^M \otimes \vert m \rangle \langle m \vert^{\hat{M}}.
\end{equation}
So for our case, the following inequality%
\begin{equation}
\left\Vert \left(  \omega^{\prime}\right)  ^{M\hat{M}}-\overline{\Phi}%
^{M\hat{M}}\right\Vert _{1}\leq\epsilon\label{eq:converse-good-c-comm}%
\end{equation}
holds because the protocol is $\epsilon$-good for classical communication
according to the criterion in\ (\ref{eq:good-classical-comm}).
We first prove the upper bound in (\ref{cgf1}) on the classical and quantum
rates. Shor's version \cite{arx2004shor} of the entanglement-assisted
classical capacity theorem \cite{PhysRevLett.83.3081,ieee2002bennett}\ states
that the rate $I(AM;B^{n})/n$ is achievable and serves as a multi-letter upper bound. This bound implies
that the unlimited entanglement-assisted quantum capacity is $I(AM;B^{n})/2n$.
If it were not so, then one could convert all of the quantum communication to
classical communication by super-dense coding and beat the rate $I(AM;B^{n}%
)/n$. But this result contradicts the optimality of the unlimited
entanglement-assisted classical capacity. These two results imply the bounds
$C\leq I(AM;B^{n})/n$ and $2Q\leq I(AM;B^{n})/n$. But we can go further and
prove that the sum rate is bounded as well. Suppose there exists a protocol
that beats the sum rate in (\ref{cgf1}). With more entanglement, one could
convert all of the quantum communication to classical communication by
super-dense coding. But this result again contradicts the optimality of the
unlimited entanglement-assisted classical capacity. So the bound
$C+2Q-\delta\leq I(AM;B^{n})_{\omega}/n$ holds.
We next prove the upper bound in (\ref{cgf2}) on the quantum communication
rate:%
\begin{align}
&  n(Q-\delta)\nonumber\\
&  =I(\hat{R}\rangle B_{1})_{\Phi^{\hat{R}B_{1}}}\nonumber\\
&  \leq I(\hat{R}\rangle B_{1})_{\omega^{\prime}}+4nQ\epsilon+H_{2}%
(\epsilon)\nonumber\\
&  \leq I(\hat{R}\rangle B_{1}M)_{\omega^{\prime}}+4nQ\epsilon+H_{2}%
(\epsilon)\nonumber\\
&  \leq I(\hat{R}\rangle B^{n}T_{B}M)_{\omega}+4nQ\epsilon+H_{2}%
(\epsilon)\nonumber\\
&  \leq I(\hat{R}T_{B}\rangle B^{n}M)_{\omega}+H(T_{B}|M)_{\omega}%
+4nQ\epsilon+H_{2}(\epsilon)\nonumber\\
&  \leq I(A\rangle B^{n}M)_{\omega}+nE+4nQ\epsilon+H_{2}(\epsilon).
\label{cqrate1}%
\end{align}
The first equality follows by evaluating the coherent information for the
state $\Phi^{\hat{R}B_{1}}$. The first inequality follows from
(\ref{eq:converse-good-q-comm}) and the Alicki-Fannes inequality in
Theorem~\ref{Thm:AF}. The second inequality is from strong subadditivity, and
the third inequality is quantum data processing. The fourth inequality follows
because $H\left(  T_{B}|B^{n}M\right)  \leq H\left(  T_{B}|M\right)  $
(conditioning reduces entropy). The last inequality follows from the definition
$A\equiv\hat{R}T_{B}$ and the fact that $H(T_{B}|M)_{\omega}\leq nE$. The
inequality in (\ref{cgf2}) follows by redefining $\delta$ as $\delta^{\prime
}\equiv\delta+4Q\epsilon+\frac{H_{2}(\epsilon)}{n}$.
We prove the upper bound in (\ref{cgf3})\ on the classical and quantum rates:%
\begin{align*}
&  n\left(  C+Q-\delta\right) \\
&  =I(M;\hat{M})_{\overline{\Phi}^{M\hat{M}}}+I(\hat{R}\rangle B_{1}%
)_{\Phi^{\hat{R}B_{1}}}\\
&  \leq I(M;\hat{M})_{\omega^{\prime}}+I(\hat{R}\rangle B_{1})_{\omega
^{\prime}}+5nC\epsilon+4nQ\epsilon+5H_{2}(\epsilon)\\
&  \leq I(M;B^{n}T_{B})_{\omega^{\prime}}+I(\hat{R}\rangle B^{n}%
T_{B}M)_{\omega^{\prime}}+n\delta^{\prime}\\
&  =I(M;B^{n})_{\omega^{\prime}}+I\left(  M;T_{B}|B^{n}\right)  _{\omega
^{\prime}}+H\left(  B^{n}T_{B}|M\right)  _{\omega^{\prime}}\\
&  -H(\hat{R}B^{n}T_{B}|M)_{\omega^{\prime}}+n\delta^{\prime}\\
&  =I(M;B^{n})_{\omega^{\prime}}+H\left(  T_{B}|B^{n}\right)  _{\omega
^{\prime}}\\
&  +H\left(  B^{n}|M\right)  _{\omega^{\prime}}-H(\hat{R}B^{n}T_{B}%
|M)_{\omega^{\prime}}+n\delta^{\prime}\\
&  \leq I(M;B^{n})_{\omega^{\prime}}+H\left(  T_{B}\right)  _{\omega^{\prime}%
}+I\left(  \hat{R}T_{B}\rangle B^{n}M\right)  _{\omega^{\prime}}%
+n\delta^{\prime}\\
&  =I(M;B^{n})_{\omega^{\prime}}+I\left(  A\rangle B^{n}M\right)
_{\omega^{\prime}}+nE+n\delta^{\prime}.%
\end{align*}
The first equality follows because the mutual information $I(M;\hat{M})$ of
the maximally correlated state $\overline{\Phi}^{M\hat{M}}$ is equal to $nC$.
The first inequality follows by applying (\ref{eq:converse-good-c-comm}) and
Corollary~\ref{cor:F-MI}\ to the mutual information $I(M;\hat{M})$, and
(\ref{eq:converse-good-q-comm}) and the Alicki-Fannes' inequality to the
coherent information $I(\hat{R}\rangle B_{1})$. The second inequality follows
by applying the quantum data processing inequality and strong subadditivity as
we did in the proof of the previous bound and by defining $\delta^{\prime
}\equiv5C\epsilon+4Q\epsilon+5H_{2}(\epsilon)/n$. The second and third
equalities follow by manipulating entropies. The third inequality follows from
the definition of coherent information and because conditioning does not
increase entropy. The last inequality follows from the definition $A\equiv
\hat{R}T_{B}$ and because $nE$ is the maximal value that $H\left(
T_{B}\right)  $ can take.
\end{IEEEproof}

\section{The Direct Coding Theorem}

\label{sec:direct-coding}In this section, we prove the direct coding theorem
for entanglement-assisted communication of classical and quantum information
by giving a combination of strategies that can achieve the rates in
Theorem~\ref{gf}. The most important development is the introduction of the
classically-enhanced father protocol and its corresponding proof in the next
section. This protocol yields a corner point in the achievable region (see,
for example, the point labeled CEF in Figure~\ref{fig:one-shot-region-state}).
Section~\ref{sec:children}\ shows that combining this protocol with
teleportation, super-dense coding, and entanglement distribution allows us to
obtain all other corner points of the achievable rate region. Thus, this
protocol is the most general one available for the channel coding scenario.

\subsection{The Classically-Enhanced Father Protocol}

We can phrase the classically-enhanced father protocol as a \textit{resource
inequality} (see Ref.~\cite{arx2005dev} for the theory of resource
inequalities):%
\begin{align}
&  \langle\mathcal{N}^{A^{\prime}\rightarrow B}\rangle+\frac{1}{2}I\left(
A;E|X\right)  _{\sigma}\left[  qq\right] \nonumber\\
&  \geq\frac{1}{2}I\left(  A;B|X\right)  _{\sigma}\left[  q\rightarrow
q\right]  +I\left(  X;B\right)  _{\sigma}\left[  c\rightarrow c\right]  .
\label{eq:CEFR}%
\end{align}
The precise statement of the classically-enhanced father resource inequality
is a statement of achievability. For any $\epsilon,\delta>0$ and sufficiently
large $n$, there exists a protocol that consumes $n$ uses of the noisy channel
$\mathcal{N}^{A^{\prime}\rightarrow B}$ and consumes $\approx nI\left(
A;E|X\right)  _{\sigma}/2$ ebits. In doing so, the protocol communicates
$\approx nI\left(  A;B|X\right)  _{\sigma}/2$ qubits with $1-\epsilon$
fidelity and $\approx nI\left(  X;B\right)  _{\sigma}$ classical bits with
$\epsilon$ probability of error. The entropic quantities are with respect to
the state $\sigma^{XABE}$ in (\ref{eq:maximization-state}).

The proof of the achievability of the classically-enhanced father protocol
proceeds in several steps. We first establish some definitions relevant to an
entanglement-assisted quantum\ code, or \textit{father} code for short, and
recall the direct coding theorem for entanglement-assisted quantum
(EAQ)\ communication \cite{PhysRevLett.93.230504,arx2005dev,arx2006anura}. We
then define a \textit{random} father code, give a few relevant definitions and
properties, and prove a version of the EAQ\ coding theorem that applies to
random father codes. In particular, we show random father codes exist whose
expected channel input is close to a product state (similar to result of the
random quantum coding theorem in Appendix D of Ref.~\cite{ieee2005dev}). We
follow this development by showing how to \textquotedblleft
paste\textquotedblright\ random father codes together so that the expected
channel input of the pasted random code is close to a product state containing
a classical message. A \textit{random classically-enhanced father code} is
then a collection of \textquotedblleft pasted\textquotedblright\ father codes.
The closeness of each expected channel input to a product state allows us to
apply the HSW\ coding theorem \cite{ieee1998holevo,PhysRevA.56.131}\ so that
Bob can decode the classical message while causing almost no disturbance to
the encoded quantum information. Based on the classical message, Bob
determines which random father code he should be decoding for. This method of
efficiently coding classical and quantum information is the \textquotedblleft
piggybacking\textquotedblright\ technique introduced in Ref.~\cite{cmp2005dev}
and applied again in Refs.~\cite{ieee2008yard,Yard:07062907}. The final
arguments consist of a series of Shannon-theoretic arguments of
derandomization and expurgation. The result is a \textit{deterministic}
classically-enhanced father code that performs well and achieves the rates in
the capacity region in Theorem~\ref{gf}.

\subsection{Father Codes}

The unencoded state of a father code is as follows%
\begin{equation}
\left\vert \varphi\right\rangle ^{\hat{R}A_{1}}\otimes\left\vert
\Phi\right\rangle ^{T_{A}T_{B}}, \label{eq:unencoded-state}%
\end{equation}
where%
\begin{align*}
\left\vert \varphi\right\rangle ^{\hat{R}A_{1}}  &  \equiv\sum_{k=1}^{2^{nQ}%
}\alpha_{k}\left\vert k\right\rangle ^{\hat{R}}\left\vert k\right\rangle
^{A_{1}},\\
\left\vert \Phi\right\rangle ^{T_{A}T_{B}}  &  \equiv\frac{1}{\sqrt{2^{nE}}%
}\sum_{m=1}^{2^{nE}}\left\vert m\right\rangle ^{T_{A}}\left\vert
m\right\rangle ^{T_{B}}.
\end{align*}
The isometric encoder $\mathcal{E}^{A_{1}T_{A}\rightarrow A^{\prime n}}$ of
the father code maps kets on the systems $A_{1}$ and $T_{A}$ as follows%
\[
\left\vert \phi_{k,m}\right\rangle ^{A^{\prime n}}\equiv\mathcal{E}%
^{A_{1}T_{A}\rightarrow A^{\prime n}}\left(  \left\vert k\right\rangle
^{A_{1}}\left\vert m\right\rangle ^{T_{A}}\right)  ,
\]
where the states $\left\vert \phi_{k,m}\right\rangle ^{A^{\prime n}}$ are
mutually orthogonal. Therefore, the encoder $\mathcal{E}^{A_{1}T_{A}%
\rightarrow A^{\prime n}}$ maps the unencoded state in
(\ref{eq:unencoded-state}) to the following encoded state:%
\[
\mathcal{E}^{A_{1}T_{A}\rightarrow A^{\prime n}}\left(  \left\vert
\varphi\right\rangle ^{\hat{R}A_{1}}\otimes\left\vert \Phi\right\rangle
^{T_{A}T_{B}}\right)  =\sum_{k=1}^{2^{nQ}}\alpha_{k}\left\vert k\right\rangle
^{\hat{R}}\left\vert \phi_{k}\right\rangle ^{A^{\prime n}T_{B}},
\]
where we define the states $\left\vert \phi_{k}\right\rangle ^{A^{\prime
n}T_{B}}$ in the following definition.

\begin{definition}
The set $\mathcal{C}\equiv\{\left\vert \phi_{k}\right\rangle ^{A^{\prime
n}T_{B}}\}_{k}$ is a representation of the father code. The EAQ
\textit{codewords} are as follows:%
\begin{equation}
\left\vert \phi_{k}\right\rangle ^{A^{\prime n}T_{B}}\equiv\frac{1}%
{\sqrt{2^{nE}}}\sum_{m=1}^{2^{nE}}\left\vert \phi_{k,m}\right\rangle
^{A^{\prime n}}\left\vert m\right\rangle ^{T_{B}}.
\label{eq:ent-assist-codewords}%
\end{equation}
The \textit{EAQ code density operator} $\rho^{A^{\prime n}T_{B}}\left(
\mathcal{C}\right)  $\ is a uniform mixture of the EAQ codewords:%
\[
\rho^{A^{\prime n}T_{B}}\left(  \mathcal{C}\right)  \equiv\frac{1}{2^{nQ}}%
\sum_{k=1}^{2^{nQ}}\left\vert \phi_{k}\right\rangle \left\langle \phi
_{k}\right\vert ^{A^{\prime n}T_{B}}.
\]
The \textit{channel input density operator} $\rho^{A^{\prime n}}\left(
\mathcal{C}\right)  $ is the part of the code density operator $\rho
^{A^{\prime n}T_{B}}\left(  \mathcal{C}\right)  $ that is input to the
channel:%
\[
\rho^{A^{\prime n}}\left(  \mathcal{C}\right)  \equiv\text{Tr}_{T_{B}}\left\{
\rho^{A^{\prime n}T_{B}}\left(  \mathcal{C}\right)  \right\}  .
\]

\end{definition}

The above definitions imply the following two results:%
\begin{align*}
\rho^{A^{\prime n}T_{B}}\left(  \mathcal{C}\right)   &  =\mathcal{E}%
^{A_{1}T_{A}\rightarrow A^{\prime n}}\left(  \pi^{A_{1}}\otimes\Phi
^{T_{A}T_{B}}\right)  ,\\
\rho^{A^{\prime n}}\left(  \mathcal{C}\right)   &  =\frac{1}{2^{n\left(
Q+E\right)  }}\sum_{k=1}^{2^{nQ}}\sum_{m=1}^{2^{nE}}\left\vert \phi
_{k,m}\right\rangle \left\langle \phi_{k,m}\right\vert ^{A^{\prime n}}.
\end{align*}

The direct coding theorem for entanglement-assisted quantum communication
gives a method for achieving the multi-letter quantum communication rate and
entanglement consumption rate.

\begin{proposition}
[EAQ Coding Theorem]\label{FP} Consider a quantum channel $\mathcal{N}%
^{A^{\prime}\rightarrow B}$ and its isometric extension $U_{\mathcal{N}%
}^{A^{\prime}\rightarrow BE}$. For any $\epsilon,\delta>0$ and all
sufficiently large $n$, there exists an $\left(  n,\epsilon\right)
$\ entanglement-assisted quantum code defined by isometries $(\mathcal{E}%
,\mathcal{D})$, such that the trace distance between the actual output%
\[
(\mathcal{D}^{B^{n}T_{B}\rightarrow B_{1}B_{E}}\circ U_{\mathcal{N}%
}^{A^{\prime n}\rightarrow B^{n}E^{n}}\circ\mathcal{E}^{A_{1}T_{A}\rightarrow
A^{\prime n}})(\varphi^{\hat{R}A_{1}}\otimes\Phi^{T_{A}T_{B}}),
\]
and the ideal decoupled output%
\begin{equation}
\varphi^{\hat{R}B_{1}}\otimes\rho^{E^{n}B_{E}},
\end{equation}
is no larger than $\epsilon$, for any state $\varphi^{\hat{R}A_{1}}$ with
dimension $2^{nQ}$ in the system $A_{1}$ and any maximally entangled
$\Phi^{T_{A}T_{B}}$ equivalent to $nE$ ebits. The rate of quantum
communication is $Q-\delta=\frac{1}{2}I(A;B)_{\phi}$ provided that the rate of
entanglement consumption is at least $E+\delta={\frac{1}{2}I(A;E)_{\phi}}$.
The entropic quantities are with respect to the following state:%
\begin{equation}
|\phi\rangle^{ABE}\equiv U_{\mathcal{N}}^{A^{\prime}\rightarrow BE}%
|\psi\rangle^{AA^{\prime}}, \label{eq:father-state}%
\end{equation}
where $\left\vert \psi\right\rangle ^{AA^{\prime}}$ is the purification of
some state $\rho^{A^{\prime}}$.
\end{proposition}

\begin{IEEEproof}
See Ref.~\cite{arx2006anura}.
\end{IEEEproof}

\subsection{Random Father Codes}

We cannot say much about the channel input density operator $\rho^{A^{\prime
n}}\left(  \mathcal{C}\right)  $ for a particular EAQ code $\mathcal{C}$. But
we can say something about the expected channel input density operator of a
\textit{random EAQ\ code} $\mathcal{C}$ (where $\mathcal{C}$ itself becomes a
random variable).

\begin{definition}
A \textit{random EAQ code} is an ensemble $\left\{  p_{\mathcal{C}%
},\mathcal{C}\right\}  $ of codes where each code $\mathcal{C}$ occurs with
probability $p_{\mathcal{C}}$. The \textit{expected code density operator}
$\overline{\rho}^{A^{\prime n}T_{B}}$\ is as follows:%
\[
\overline{\rho}^{A^{\prime n}T_{B}}\equiv\mathbb{E}_{\mathcal{C}}\left\{
\rho^{A^{\prime n}T_{B}}\left(  \mathcal{C}\right)  \right\}  .
\]
The \textit{expected channel input density operator} $\overline{\rho
}^{A^{\prime n}}$\ is as follows:%
\[
\overline{\rho}^{A^{\prime n}}\equiv\mathbb{E}_{\mathcal{C}}\left\{
\rho^{A^{\prime n}}\left(  \mathcal{C}\right)  \right\}  .
\]
A random EAQ code is \textquotedblleft$\rho$-like\textquotedblright\ if the
expected channel input density operator is close to a tensor power of some
state $\rho$:%
\begin{equation}
\left\Vert \overline{\rho}^{A^{\prime n}}-\rho^{\otimes n}\right\Vert _{1}%
\leq\epsilon.
\end{equation}

\end{definition}

It follows from the above definition that%
\begin{align*}
\overline{\rho}^{A^{\prime n}T_{B}}  &  =\sum_{\mathcal{C}}p_{\mathcal{C}}%
\rho^{A^{\prime n}T_{B}}\left(  \mathcal{C}\right)  ,\\
\overline{\rho}^{A^{\prime n}}  &  =\text{Tr}_{T_{B}}\left\{  \overline{\rho
}^{A^{\prime n}T_{B}}\right\}  .
\end{align*}

We now state a version of the direct coding theorem that applies to random
father codes. The proof shows that we can produce a random father code with an
expected channel input density operator close to a tensor power state.

\begin{proposition}
\label{thm:random-EA-code}For any $\epsilon,\delta>0$ and all sufficiently
large $n$, there exists a random $\rho^{A^{\prime}}$-like\ EAQ code for a
channel $\mathcal{N}^{A^{\prime}\rightarrow B}$. In particular, the random EAQ
code has quantum rate $\frac{1}{2}I(A;B)_{\phi}-\delta$ and entanglement
consumption rate ${\frac{1}{2}I(A;E)}_{\phi}+\delta$. The entropic quantities
are with respect to the state in (\ref{eq:father-state}) and the state
$\rho^{A^{\prime}}$ is that state's restriction to the system $A^{\prime}$.
\end{proposition}

\begin{IEEEproof}
The proof is in the Appendix \ref{AP_RG0}.
\end{IEEEproof}

\subsection{Associating a Random Father Code with a Classical String}

Suppose that we have an ensemble $\left\{  p\left(  x\right)  ,\rho
_{x}\right\}  _{x\in\mathcal{X}}$ of quantum states. Let $x^{n}\equiv
x_{1}\cdots x_{n}$ denote a classical string generated by the density
$p\left(  x\right)  $ where each symbol $x_{i}\in\mathcal{X}$. Then there is a
density operator $\rho_{x^{n}}$ corresponding to the string $x^{n}$ where%
\[
\rho_{x^{n}}\equiv%
%TCIMACRO{\dbigotimes \limits_{i=1}^{n}}%
%BeginExpansion
{\displaystyle\bigotimes\limits_{i=1}^{n}}
%EndExpansion
\rho_{x_{i}}.
\]
Suppose that we label a random father code by the string $x^{n}$ and let
$\overline{\rho}_{x^{n}}^{A^{\prime n}}$ denote its expected channel input
density operator.

\begin{definition}
A random father code is $\left(  \rho_{x^{n}}\right)  $-like if the expected
channel input density operator $\overline{\rho}_{x^{n}}^{A^{\prime n}}$ is
close to the state $\rho_{x^{n}}$:%
\[
\left\Vert \overline{\rho}_{x^{n}}^{A^{\prime n}}-\rho_{x^{n}}\right\Vert
_{1}\leq\epsilon.
\]

\end{definition}

\begin{proposition}
\label{prop:random-grandfather} Suppose we have an ensemble as above. Consider
a quantum channel $\mathcal{N}^{A^{\prime}\rightarrow B}$ with its isometric
extension $U_{\mathcal{N}}^{A^{\prime}\rightarrow BE}$. Then there exists a
random $\left(  \rho_{x^{n}}\right)  $-like\ entanglement-assisted quantum
code for the channel $\mathcal{N}^{A^{\prime}\rightarrow B}$ for any
$\epsilon,\delta>0$, for all sufficiently large $n$, and for any classical
string $x^{n}$ in the typical set $T_{\delta}^{X^{n}}$ \cite{book1991cover}.
Its quantum communication rate is $I(A;B|X)/2-c^{\prime}\delta$ and its
entanglement consumption rate is $I(A;E|X)/2+c^{\prime\prime}\delta\ $for some
constants $c^{\prime},c^{\prime\prime}$ where the entropic quantities are with
respect to the state in (\ref{eq:maximization-state}) with a trivial system
$E^{\prime}$. The state $\rho_{x^{n}}$ is generated from the restriction of
the ensemble $\{p\left(  x\right)  ,\phi_{x}^{AA^{\prime}}\}_{x\in\mathcal{X}%
}$ to the $A^{\prime}$ system. The states $\phi_{x}^{AA^{\prime}}$ in the
ensemble correspond to the states $\phi_{x}^{AA^{\prime}}$ in
(\ref{eq:maximization-state}).
\end{proposition}

\begin{IEEEproof}
The method of proof involves \textquotedblleft pasting\textquotedblright%
\ random father codes together. The proof is in the Appendix \ref{AP_RG}.
\end{IEEEproof}

\subsection{Construction of a Classically-Enhanced Father Code}

The HSW coding theorem gives an achievable method for sending classical
information over a noisy quantum channel. The crucial property that we exploit
is that it uses a product-state input for sending classical information. This
tensor-product structure is what allows us to \textquotedblleft
piggyback\textquotedblright\ classical information onto father codes.

\begin{proposition}
[HSW Coding Theorem \cite{ieee1998holevo,PhysRevA.56.131}]\label{prop:HSW}%
Consider an input ensemble $\{p\left(  x\right)  ,\rho_{x}^{A^{\prime}}\}$
that gives rise to a classical-quantum state $\sigma^{XB}$ where%
\[
\sigma^{XB}\equiv\sum_{x\in\mathcal{X}}p(x)\left\vert x\right\rangle
\left\langle x\right\vert ^{X}\otimes\mathcal{N}^{A^{\prime}\rightarrow
B}(\rho_{x}^{A^{\prime}}).
\]
Let $C=I(X;B)_{\sigma}-c^{\prime}\delta$ for any $\delta>0$ and for some
constant $c^{\prime}$. Then for all $\epsilon>0$ and for all sufficiently
large $n$, there exists a classical encoding map%
\[
f:\left[  2^{nC}\right]  \rightarrow T_{\delta}^{X^{n}},
\]
and a decoding POVM%
\[
\Lambda^{B^{n}}\equiv(\Lambda_{m}^{B^{n}})_{m\in\lbrack2^{nC}]},
\]
that allows Bob to decode any classical message $m\in\lbrack2^{nC}]$ with high
probability:%
\[
\tr\{\tau_{m}^{B^{n}}\Lambda_{m}^{B^{n}}\}\geq1-\epsilon.
\]
The density operators $\tau_{m}^{B^{n}}$ are the channel outputs%
\begin{equation}
\tau_{m}^{B^{n}}\equiv\mathcal{N}^{A^{\prime n}\rightarrow B^{n}}(\rho
_{f(m)}^{A^{\prime n}}), \label{eq:channel-outputs}%
\end{equation}
and the channel input states $\rho_{x^{n}}^{A^{\prime n}}$ are a tensor
product of states in the ensemble:%
\[
\rho_{x^{n}}^{A^{\prime n}}\equiv{\bigotimes\limits_{i=1}^{n}\ }\rho_{x_{i}%
}^{A^{\prime}}.
\]

\end{proposition}

We are now in a position to prove the direct coding part of the
classically-enhanced father capacity theorem. The proof is similar to that in
Ref.~\cite{cmp2005dev}.

\begin{IEEEproof}
[Direct Coding Theorem] Define the classical message set $\left[
2^{nC}\right]  $, the classical encoding map $f$, the channel output
states$\ \tau_{m}^{B^{n}}$, and the decoding POVM $\Lambda^{B^{n}}$ as in
Proposition \ref{prop:HSW}. Invoking Proposition~\ref{prop:random-grandfather}%
, we know that for each $m\in\lbrack2^{nC}]$, there exists a random
$(\rho_{f\left(  m\right)  }^{A^{\prime n}})$-like\ father code $\mathcal{C}%
_{m}$ whose probability density is $p_{\mathcal{C}_{m}}$. The random father
code $\mathcal{C}_{m}$ has encoding-decoding isometry pairs $(\mathcal{E}%
_{\mathcal{C}_{m}}^{A_{1}T_{A}\rightarrow A^{\prime n}},\mathcal{D}%
_{\mathcal{C}_{m}}^{B_{n}T_{B}\rightarrow B_{1}B_{E}})$ for each of its
realizations. It transmits $n[I(A;B|X)/2-c^{\prime}\delta]$ qubits provided
Alice and Bob share at least $n[I(A;E|X)/2+c^{\prime\prime}\delta]$ ebits. Let
$\mathcal{C}$ denote the \textit{random classically-enhanced father code} that
is the collection of random father codes $\left\{  \mathcal{C}_{m}\right\}
_{m\in\left[  2^{nC}\right]  }$.
We first prove that the expectation of the classical error probability for
message $m$ is small. The expectation is with respect the random father code
$\mathcal{C}_{m}$.\ Let $\tau_{\mathcal{C}_{m}}^{B^{n}}$ denote the
\textit{channel output density operator} corresponding to the father code
$\mathcal{C}_{m}$:%
\[
\tau_{\mathcal{C}_{m}}^{B^{n}}\equiv\text{Tr}_{T_{B}}\left\{  \mathcal{N}%
^{A^{\prime n}\rightarrow B^{n}}\left(  \mathcal{E}_{\mathcal{C}_{m}}%
^{A_{1}T_{A}\rightarrow A^{\prime n}}(\pi^{A_{1}}\otimes\Phi^{T_{A}T_{B}%
})\right)  \right\}  .
\]
Let $\overline{\tau}_{m}^{B^{n}}$ denote the \textit{expected channel output
density operator} of the random father code $\mathcal{C}_{m}$:%
\[
\overline{\tau}_{m}^{B^{n}}\equiv\mathbb{E}_{\mathcal{C}_{m}}\left\{
\tau_{\mathcal{C}_{m}}^{B^{n}}\right\}  =\sum_{\mathcal{C}_{m}}p_{\mathcal{C}%
_{m}}\tau_{\mathcal{C}_{m}}^{B^{n}}.
\]
The following inequality holds%
\[
\left\Vert \overline{\rho}_{f(m)}^{A^{\prime n}}-\rho_{f(m)}^{A^{\prime n}%
}\right\Vert _{1}\leq\left\vert \mathcal{X}\right\vert \epsilon
\]
because the random father code $\mathcal{C}_{m}$ is $(\rho_{f(m)}^{A^{\prime
n}})$-like. Then the expected channel output density operator $\overline{\tau
}_{m}^{B^{n}}$ is close to the tensor product state $\tau_{m}^{B^{n}}$ in
(\ref{eq:channel-outputs}):%
\begin{equation}
\left\Vert \overline{\tau}_{m}^{B^{n}}-\tau_{m}^{B^{n}}\right\Vert _{1}%
\leq\left\vert \mathcal{X}\right\vert \epsilon,
\label{eq:channel-product-closeness}%
\end{equation}
because the trace distance is monotone under the quantum operation
$\mathcal{N}^{A^{\prime n}\rightarrow B^{n}}$. It then follows that the POVM
element $\Lambda_{m}^{B^{n}}$ has a high probability of detecting the expected
channel output density operator $\overline{\tau}_{m}^{B^{n}}$:%
\begin{align}
\tr\{\Lambda_{m}^{B^{n}}\overline{\tau}_{m}^{B^{n}}\}  &  \geq\tr\{\Lambda
_{m}^{B^{n}}\tau_{m}^{B^{n}}\}-\left\Vert \overline{\tau}_{m}^{B^{n}}-\tau
_{m}^{B^{n}}\right\Vert _{1}\nonumber\\
&  \geq1-\epsilon-\left\vert \mathcal{X}\right\vert \epsilon.
\label{eq:average-output-op-good}%
\end{align}
The first inequality follows from the following lemma \cite{thesis2005yard}%
\ that holds for any two quantum states $\rho$ and $\sigma$ and a positive
operator $\Pi$ where $0\leq\Pi\leq I$:%
\[
\text{Tr}\left\{  \Pi\rho\right\}  \geq\text{Tr}\left\{  \Pi\sigma\right\}
-\left\Vert \rho-\sigma\right\Vert _{1}.
\]
The second inequality follows from Proposition~\ref{prop:HSW} and
(\ref{eq:channel-product-closeness}). We define Bob's decoding instrument
$\mathcal{D}_{\mathcal{C}}^{B^{n}T_{B}\rightarrow B_{1}B_{E}\hat{M}}$ for the
random classically-enhanced father code $\mathcal{C}$\ as follows:%
\begin{align*}
&  \mathcal{D}_{\mathcal{C}}^{B^{n}T_{B}\rightarrow B_{1}B_{E}\hat{M}}\left(
\rho^{B^{n}T_{B}}\right) \\
&  \equiv\sum_{m}\mathcal{D}_{\mathcal{C}_{m}}^{B^{n}T_{B}\rightarrow
B_{1}B_{E}}\left(  \sqrt{\Lambda_{m}^{B^{n}}}\rho^{B^{n}T_{B}}\sqrt
{\Lambda_{m}^{B^{n}}}\right)  \otimes\left\vert m\right\rangle \left\langle
m\right\vert ^{\hat{M}},
\end{align*}
where $\mathcal{D}_{\mathcal{C}_{m}}^{B^{n}T_{B}\rightarrow B_{1}B_{E}}$ is
the decoding isometry for the father code $\mathcal{C}_{m}$ and each map
$\mathcal{D}_{\mathcal{C}_{m}}^{B^{n}T_{B}\rightarrow B_{1}B_{E}}%
(\sqrt{\Lambda_{m}^{B^{n}}}\rho\sqrt{\Lambda_{m}^{B^{n}}})$ is trace reducing.
The induced quantum operation corresponding to this instrument is as follows:%
\[
\mathcal{D}_{\mathcal{C}}^{B^{n}T_{B}\rightarrow B_{1}B_{E}}\left(
\rho\right)  =\text{Tr}_{\hat{M}}\left\{  \mathcal{D}_{\mathcal{C}}%
^{B^{n}T_{B}\rightarrow B_{1}B_{E}\hat{M}}\left(  \rho\right)  \right\}  .
\]
Let $p_{e}\left(  \mathcal{C}_{m}\right)  $ denote the classical error
probability for each classical message $m$ of the classically-enhanced father
code $\mathcal{C}$:%
\begin{align*}
p_{e}\left(  \mathcal{C}_{m}\right)   &  \equiv1-\Pr\left\{  M^{\prime
}=m\ |\ M=m\right\} \\
&  =1-\text{Tr}\left\{  \mathcal{D}_{\mathcal{C}_{m}}^{B^{n}T_{B}\rightarrow
B_{1}B_{E}}\left(  \sqrt{\Lambda_{m}^{B^{n}}}\tau_{\mathcal{C}_{m}}^{B^{n}%
}\sqrt{\Lambda_{m}^{B^{n}}}\right)  \right\}  .
\end{align*}
Then by the above definition, (\ref{eq:average-output-op-good}), and the fact
that the trace does not change under the isometry $\mathcal{D}_{\mathcal{C}%
_{m}}^{B^{n}T_{B}\rightarrow B_{1}B_{E}}$, it holds that the expectation of
the classical error probability $p_{e}\left(  \mathcal{C}_{m}\right)  $\ with
respect to the random father code $\mathcal{C}_{m}$ is low:%
\begin{equation}
\mathbb{E}_{\mathcal{C}_{m}}\left\{  p_{e}\left(  \mathcal{C}_{m}\right)
\right\}  \leq\left(  1+|\mathcal{X}|\right)  \epsilon.
\label{eq:classical-error}%
\end{equation}
We now prove that the expectation of the quantum error is small (the
expectation is with respect to the random father code $\mathcal{C}_{m}$).
Input the state $\Phi^{\hat{R}A_{1}}\otimes\Phi^{T_{A}T_{B}}$\ to the encoder
$\mathcal{E}_{\mathcal{C}_{m}}^{A_{1}T_{A}\rightarrow A^{\prime n}}$ followed
by the channel $\mathcal{N}^{A^{\prime n}\rightarrow B^{n}}$.\ The resulting
state is an extension $\Omega_{\mathcal{C}_{m}}^{\hat{R}T_{B}B^{n}}$ of
$\tau_{\mathcal{C}_{m}}^{B^{n}}$:%
\[
\Omega_{\mathcal{C}_{m}}^{\hat{R}T_{B}B^{n}}\equiv\mathcal{N}^{A^{\prime
n}\rightarrow B^{n}}\left(  \mathcal{E}_{\mathcal{C}_{m}}^{A_{1}%
T_{A}\rightarrow A^{\prime n}}(\Phi^{\hat{R}A_{1}}\otimes\Phi^{T_{A}T_{B}})\right)
.
\]
Let $\overline{\Omega}_{m}^{\hat{R}T_{B}B^{n}}$ denote the expectation of
$\Omega_{\mathcal{C}_{m}}^{\hat{R}T_{B}B^{n}}$ with respect to the random code
$\mathcal{C}_{m}$:%
\[
\overline{\Omega}_{m}^{\hat{R}T_{B}B^{n}}\equiv\mathbb{E}_{\mathcal{C}_{m}}\left\{
\Omega_{\mathcal{C}_{m}}^{\hat{R}T_{B}B^{n}}\right\}  .
\]
It follows that $\overline{\Omega}_{m}^{\hat{R}T_{B}B^{n}}$\ is an extension of
$\overline{\tau}_{m}^{B^{n}}$. The following inequality follows from
(\ref{eq:average-output-op-good}):%
\begin{equation}
\tr\{\overline{\Omega}_{m}^{\hat{R}T_{B}B^{n}}\Lambda_{m}^{B^{n}}\}\geq
1-(1+|\mathcal{X}|)\epsilon.
\end{equation}
The above inequality is then sufficient for us to apply a modified version of
the gentle measurement lemma\ (Lemma~\ref{lemma:GM-ensemble} in the
Appendix \ref{sec:gm-ensemble}) so that\ the following inequality holds%
\begin{align}
&  \mathbb{E}_{\mathcal{C}_{m}}\left\{  \left\Vert \sqrt{\Lambda_{m}^{B^{n}}%
}\Omega_{\mathcal{C}_{m}}^{\hat{R}T_{B}B^{n}}\sqrt{\Lambda_{m}^{B^{n}}}%
-\Omega_{\mathcal{C}_{m}}^{\hat{R}T_{B}B^{n}}\right\Vert _{1}\right\} \nonumber\\
&  \leq\sqrt{8(1+|\mathcal{X}|)\epsilon}. \label{gm}%
\end{align}
Monotonicity of the trace distance gives an inequality for the trace-reducing
maps of the quantum decoding instrument:%
\begin{align}
&  \mathbb{E}_{\mathcal{C}_{m}}\left\{  \left\Vert
\begin{array}
[c]{c}%
\mathcal{D}_{\mathcal{C}_{m}}^{B^{n}T_{B}\rightarrow B_{1}B_{E}}\left(
\sqrt{\Lambda_{m}^{B^{n}}}\Omega_{\mathcal{C}_{m}}^{\hat{R}T_{B}B^{n}}\sqrt
{\Lambda_{m}^{B^{n}}}\right)  -\\
\mathcal{D}_{\mathcal{C}_{m}}^{B^{n}T_{B}\rightarrow B_{1}B_{E}}\left(
\Omega_{\mathcal{C}_{m}}^{\hat{R}T_{B}B^{n}}\right)
\end{array}
\right\Vert _{1}\right\} \nonumber\\
&  \leq\sqrt{8(1+|\mathcal{X}|)\epsilon}. \label{eq:epsilon-sqrt-map}%
\end{align}
The following inequality also holds%
\begin{align}
&  \mathbb{E}_{\mathcal{C}_{m}}\left\{  \left\Vert
\begin{array}
[c]{c}%
\mathcal{D}_{\mathcal{C}}^{B^{n}T_{B}\rightarrow B_{1}B_{E}}\left(
\Omega_{\mathcal{C}_{m}}^{\hat{R}T_{B}B^{n}}\right)  -\\
\mathcal{D}_{\mathcal{C}_{m}}^{B^{n}T_{B}\rightarrow B_{1}B_{E}}\left(
\sqrt{\Lambda_{m}^{B^{n}}}\Omega_{\mathcal{C}_{m}}^{\hat{R}T_{B}B^{n}}\sqrt
{\Lambda_{m}^{B^{n}}}\right)
\end{array}
\right\Vert _{1}\right\} \nonumber\\
&  \leq\mathbb{E}_{\mathcal{C}_{m}}\left\{  \sum_{m^{\prime}\neq m}\left\Vert
\mathcal{D}_{\mathcal{C}_{m^{\prime}}}^{B^{n}T_{B}\rightarrow B_{1}B_{E}%
}\left(  \sqrt{\Lambda_{m^{\prime}}^{B^{n}}}\Omega_{\mathcal{C}_{m}}%
^{\hat{R}T_{B}B^{n}}\sqrt{\Lambda_{m^{\prime}}^{B^{n}}}\right)  \right\Vert
_{1}\right\} \nonumber\\
&  =\mathbb{E}_{\mathcal{C}_{m}}\left\{  \sum_{m^{\prime}\neq m}\left\Vert
\sqrt{\Lambda_{m^{\prime}}^{B^{n}}}\Omega_{\mathcal{C}_{m}}^{\hat{R}T_{B}B^{n}}%
\sqrt{\Lambda_{m^{\prime}}^{B^{n}}}\right\Vert _{1}\right\} \nonumber\\
&  =\mathbb{E}_{\mathcal{C}_{m}}\left\{  \sum_{m^{\prime}\neq m}%
\text{Tr}\left\{  \Lambda_{m^{\prime}}^{B^{n}}\Omega_{\mathcal{C}_{m}}%
^{\hat{R}T_{B}B^{n}}\right\}  \right\} \nonumber\\
&  =1-\text{Tr}\left\{  \Lambda_{m}^{B^{n}}\overline{\Omega}_{m}^{\hat{R}T_{B}B^{n}%
}\right\} \nonumber\\
&  \leq(1+|\mathcal{X}|)\epsilon. \label{eq:epsilon-diff-maps}%
\end{align}
The first inequality follows from definitions and the triangle inequality. The
first equality follows because the trace distance is invariant under isometry.
The second equality follows because the operator $\Lambda_{m}^{B^{n}}%
\Omega_{\mathcal{C}_{m}}^{\hat{R}T_{B}B^{n}}$ is positive. The third equality
follows from some algebra, and the second inequality follows from
(\ref{eq:average-output-op-good}). The fidelity of quantum communication for
all classical messages $m$ and codes $\mathcal{C}_{m}$ is high%
\[
F\left(  \mathcal{D}_{\mathcal{C}_{m}}^{B^{n}T_{B}\rightarrow B_{1}B_{E}%
}\left(  \Omega_{\mathcal{C}_{m}}^{\hat{R}T_{B}B^{n}}\right)  ,\Phi^{\hat{R}B_{1}}\right)
\geq1-\epsilon
\]
because each code $\mathcal{C}_{m}$\ in the random father code is good for
quantum communication. It then follows that%
\begin{equation}
\mathbb{E}_{\mathcal{C}_{m}}\left\{  \left\Vert \mathcal{D}_{\mathcal{C}_{m}%
}^{B^{n}T_{B}\rightarrow B_{1}B_{E}}\left(  \Omega_{\mathcal{C}_{m}}%
^{\hat{R}T_{B}B^{n}}\right)  -\Phi^{\hat{R}B_{1}}\right\Vert _{1}\right\}  \leq
2\sqrt{\epsilon} \label{eq:good-q-comm}%
\end{equation}
because of the relation between the trace distance and fidelity
\cite{thesis2005yard}. Application of the triangle inequality to
(\ref{eq:good-q-comm}), (\ref{eq:epsilon-diff-maps}), and
(\ref{eq:epsilon-sqrt-map}) gives the following bound on the expected quantum
error%
\begin{equation}
\mathbb{E}_{\mathcal{C}_{m}}\left\{  q_{e}\left(  \mathcal{C}_{m}\right)
\right\}  \leq\epsilon^{\prime} \label{eq:quantum-error}%
\end{equation}
where%
\[
\epsilon^{\prime}\equiv(1+|\mathcal{X}|)\epsilon+\sqrt{8(1+|\mathcal{X}%
|)\epsilon}+2\sqrt{\epsilon},
\]
and where we define the quantum error $q_{e}\left(  \mathcal{C}_{m}\right)  $
of the code $\mathcal{C}_{m}$ as follows:%
\[
q_{e}\left(  \mathcal{C}_{m}\right)  \equiv\left\Vert \mathcal{D}%
_{\mathcal{C}}^{B^{n}T_{B}\rightarrow B_{1}B_{E}}\left(  \Omega_{\mathcal{C}%
_{m}}^{\hat{R}T_{B}B^{n}}\right)  -\Phi^{\hat{R}B_{1}}\right\Vert _{1}.
\]
The above random classically-enhanced father code relies on Alice and Bob
having access to a source of common randomness. We now show that they can
eliminate the need for common randomness and select a good
classically-enhanced father code $\mathcal{C}$ that has a low quantum error
$q_{e}\left(  \mathcal{C}_{m}\right)  $ and low classical error $p_{e}\left(
\mathcal{C}_{m}\right)  $ for all classical messages $m$ in a large subset of
$\left[  2^{nC}\right]  $. By the bounds in (\ref{eq:classical-error}) and
(\ref{eq:quantum-error}), the following bound holds for the expectation of the
averaged summed error probabilities:%
\[
\mathbb{E}_{\mathcal{C}_{m}}\left\{  \frac{1}{2^{nC}}\sum_{m}p_{e}\left(
\mathcal{C}_{m}\right)  +q_{e}\left(  \mathcal{C}_{m}\right)  \right\}
\leq\epsilon^{\prime}+(1+|\mathcal{X}|)\epsilon.
\]
If the above bound holds for the expectation over all random codes, it follows
that there exists a particular classically-enhanced father code $\mathcal{C}%
=\left\{  \mathcal{C}_{m}\right\}  _{m\in\left[  2^{nC}\right]  }$ with the
following bound on its averaged summed error probabilities:%
\[
\frac{1}{2^{nC}}\sum_{m}p_{e}\left(  \mathcal{C}_{m}\right)  +q_{e}\left(
\mathcal{C}_{m}\right)  \leq\epsilon^{\prime}+(1+|\mathcal{X}|)\epsilon.
\]
We fix the code $\mathcal{C}$ and expurgate the worst half of the father
codes---those father codes with classical messages $m$ that have the highest
value of $p_{e}\left(  \mathcal{C}_{m}\right)  +q_{e}\left(  \mathcal{C}%
_{m}\right)  $. This derandomization and expurgation yields a
classically-enhanced father code that has each classical error $p_{e}\left(
\mathcal{C}_{m}\right)  $ and each quantum error $q_{e}\left(  \mathcal{C}%
_{m}\right)  $ upper bounded by $2\left(  \epsilon^{\prime}+(1+|\mathcal{X}%
|)\epsilon\right)  $ for the remaining classical messages $m$. This
expurgation decreases the classical rate by a negligible factor of $\frac
{1}{n}$.
\end{IEEEproof}

Note that the above proof is a scheme for entanglement transmission. This task
is equivalent to the task of subspace transmission (quantum communication) by
the methods in Ref.~\cite{ieee2000barnum}.

\subsection{Child Protocols}

\label{sec:children}We detail five protocols that are children of the
classically-enhanced father protocol in the sense of Ref.~\cite{arx2005dev}.
Recall the classically-enhanced father resource inequality in (\ref{eq:CEFR}).
Recall the three respective unit resource inequalities for teleportation,
super-dense coding, and entanglement distribution:%
\begin{align}
2\left[  c\rightarrow c\right]  +\left[  qq\right]   &  \geq\left[
q\rightarrow q\right]  ,\label{TP}\\
\left[  qq\right]  +\left[  q\rightarrow q\right]   &  \geq2\left[
c\rightarrow c\right]  ,\label{SD}\\
\left[  q\rightarrow q\right]   &  \geq\left[  qq\right]  . \label{ED}%
\end{align}

We can first append entanglement distribution to the classically-enhanced
father resource inequality. This appending gives rise to the
classically-enhanced quantum communication protocol in Ref.~\cite{cmp2005dev}.
The development proceeds as follows:%
\begin{align*}
&  \langle\mathcal{N}^{A^{\prime}\rightarrow B}\rangle+\frac{1}{2}I\left(
A;E|X\right)  \left[  qq\right] \\
&  \geq\frac{1}{2}I\left(  A;B|X\right)  \left[  q\rightarrow q\right]
+I\left(  X;B\right)  \left[  c\rightarrow c\right] \\
&  =\left(  \frac{1}{2}I\left(  A;E|X\right)  +I\left(  A\rangle BX\right)
\right)  \left[  q\rightarrow q\right]  +I\left(  X;B\right)  \left[
c\rightarrow c\right] \\
&  \geq\frac{1}{2}I\left(  A;E|X\right)  \left[  qq\right]  +I\left(  A\rangle
BX\right)  \left[  q\rightarrow q\right]  +I\left(  X;B\right)  \left[
c\rightarrow c\right]  ,
\end{align*}
where the first inequality is the classically-enhanced father resource
inequality, the first equality exploits the identity in
(\ref{eq:coherent-identity}), and the last inequality follows from
entanglement distribution. By the cancellation lemma (Lemma~4.6 of
Ref.~\cite{arx2005dev}), the following resource inequality holds%
\begin{equation}
\langle\mathcal{N}^{A^{\prime}\rightarrow B}\rangle+o\left[  qq\right]  \geq
I\left(  A\rangle BX\right)  \left[  q\rightarrow q\right]  +I\left(
X;B\right)  \left[  c\rightarrow c\right]  , \label{eq:DS}%
\end{equation}
where $o\left[  qq\right]  $ represents a sublinear amount of entanglement.
The above resource inequality is equivalent to the classically-enhanced
quantum communication protocol in Ref.~\cite{cmp2005dev} (modulo the sublinear
entanglement). Combining the above resource inequality further with
entanglement distribution gives the classically-enhanced entanglement
generation protocol from Ref.~\cite{cmp2005dev}:%
\[
\langle\mathcal{N}^{A^{\prime}\rightarrow B}\rangle+o\left[  qq\right]  \geq
I\left(  A\rangle BX\right)  \left[  qq\right]  +I\left(  X;B\right)  \left[
c\rightarrow c\right]  .
\]

We can combine the classically-enhanced father protocol with super-dense
coding and entanglement distribution. Let CEF-SD-ED denote the resulting
protocol.\ The development proceeds by first using qubits at a rate $\frac
{1}{2}H\left(  A|X\right)  $ for entanglement distribution:%
\begin{align*}
&  \langle\mathcal{N}^{A^{\prime}\rightarrow B}\rangle+\frac{1}{2}I\left(
A;E|X\right)  \left[  qq\right] \\
&  \geq\frac{1}{2}I\left(  A;B|X\right)  \left[  q\rightarrow q\right]
+I\left(  X;B\right)  \left[  c\rightarrow c\right] \\
&  =\left(  \frac{1}{2}H\left(  A|X\right)  +\frac{1}{2}I\left(  A\rangle
BX\right)  \right)  \left[  q\rightarrow q\right]  +I\left(  X;B\right)
\left[  c\rightarrow c\right] \\
&  \geq\frac{1}{2}H\left(  A|X\right)  \left[  qq\right]  +\frac{1}{2}I\left(
A\rangle BX\right)  \left[  q\rightarrow q\right]  +I\left(  X;B\right)
\left[  c\rightarrow c\right]
\end{align*}
After this step, the above protocol is equivalent to the following one:%
\[
\langle\mathcal{N}\rangle+o\left[  qq\right]  \geq\frac{1}{2}I\left(  A\rangle
BX\right)  ( \left[  qq\right]  + \left[  q\rightarrow q\right] ) +I\left(
X;B\right)  \left[  c\rightarrow c\right]  ,
\]
so that it has generated entanglement at a net rate of $\frac{1}{2}I\left(
A\rangle BX\right)  $ ebits. We can then further combine with super-dense
coding to achieve the protocol CEF-SD-ED:%
\[
\langle\mathcal{N}^{A^{\prime}\rightarrow B}\rangle+o\left[  qq\right]  \geq
I\left(  A\rangle BX\right)  \left[  c\rightarrow c\right]  +I\left(
X;B\right)  \left[  c\rightarrow c\right]  .
\]

We can combine the classically-enhanced father protocol with super-dense
coding to get Shor's entanglement-assisted classical (EAC)\ communication
protocol \cite{arx2004shor}:%
\begin{align}
&  \langle\mathcal{N}^{A^{\prime}\rightarrow B}\rangle+H\left(  A|X\right)
\left[  qq\right] \nonumber\\
&  =\langle\mathcal{N}^{A^{\prime}\rightarrow B}\rangle+\frac{1}{2}I\left(
A;E|X\right)  \left[  qq\right]  +\frac{1}{2}I\left(  A;B|X\right)  \left[
qq\right] \nonumber\\
&  \geq\frac{1}{2}I\left(  A;B|X\right)  \left[  q\rightarrow q\right]
+I\left(  X;B\right)  \left[  c\rightarrow c\right] \nonumber\\
&  +\frac{1}{2}I\left(  A;B|X\right)  \left[  qq\right] \nonumber\\
&  \geq I\left(  X;B\right)  \left[  c\rightarrow c\right]  +I\left(
A;B|X\right)  \left[  c\rightarrow c\right] \nonumber\\
&  =I\left(  AX;B\right)  \left[  c\rightarrow c\right]  . \label{unEAC}%
\end{align}
The first equality uses the identity in (\ref{eq:entropy-identity}). The first
inequality uses the classically-enhanced father resource inequality. The
second inequality uses super-dense coding, and the last equality uses the
chain-rule identity in (\ref{eq:chain-rule}). The above rates are the same as
those in Refs.~\cite{arx2004shor,arx2005dev}.

Teleportation is the last unit resource inequality with which we can combine
the classically-enhanced father protocol. Let CEF-TP (classically-enhanced
father combined with teleportation) denote the resulting protocol. Consider
that the classically-enhanced father protocol generates classical
communication at a rate $I\left(  X;B\right)  $. Alice and Bob can teleport
quantum information if they have an extra $I\left(  X;B\right)  /2$ ebits of
entanglement. The development proceeds as follows:%
\begin{align*}
&  \langle\mathcal{N}^{A^{\prime}\rightarrow B}\rangle+\frac{1}{2}I\left(
A;E|X\right)  \left[  qq\right]  +\frac{1}{2}I\left(  X;B\right)  \left[
qq\right] \\
&  \geq\frac{1}{2}I\left(  A;B|X\right)  \left[  q\rightarrow q\right]
+I\left(  X;B\right)  \left[  c\rightarrow c\right]  +\frac{1}{2}I\left(
X;B\right)  \left[  qq\right] \\
&  \geq\frac{1}{2}I\left(  A;B|X\right)  \left[  q\rightarrow q\right]
+\frac{1}{2}I\left(  X;B\right)  \left[  q\rightarrow q\right] \\
&  =\frac{1}{2}I\left(  AX;B\right)  \left[  q\rightarrow q\right]  .
\end{align*}
We apply teleportation to get the second inequality and the chain rule
in\ (\ref{eq:chain-rule}) to get the last equality. We can rewrite the above
protocol as follows:%
\begin{multline*}
\langle\mathcal{N}^{A^{\prime}\rightarrow B}\rangle+\frac{1}{2}\left(
I\left(  A;E|X\right)  +I\left(  X;B\right)  \right)  \left[  qq\right] \\
\geq\frac{1}{2}I\left(  AX;B\right)  \left[  q\rightarrow q\right]  .
\end{multline*}
This protocol is the same as the father protocol if random variable $X$ has a
degenerate distribution.

\section{Single-Letter Examples}

Theorem~\ref{gf}\ is a general theorem that determines the capacity region of
any entanglement-assisted channel for classical and quantum communication.
Unfortunately, the theorem is of a multi-letter nature, implying that it is an
intractable problem to compute the capacity region corresponding to an
arbitrary channel.

In the forthcoming subsections, we provide several examples of channels for
which we can exactly compute their corresponding capacity regions. The first
example is the trivial completely depolarizing channel (the channel that
replaces the input state with the maximally mixed state). We find this example
interesting despite its triviality because it coincides with our results in
Ref.~\cite{HW09}. The second example is the quantum erasure channel
\cite{GBP97}. The advantage of the quantum erasure channel is that we can
apply simple reasoning to determine the outer bound of its corresponding
capacity region. We then show that the inner bound corresponding to the
achievable region of this channel matches the outer bound. Thus, we know the
full capacity region for the quantum erasure channel. The final channel that
we single-letterize is the qubit dephasing channel. Perhaps surprisingly, we
are able to do so by arguing that the Devetak-Shor CQ region and the Shor CE
region each single-letterize.

\subsection{The Completely Depolarizing Channel}

The first single-letter example that we consider is the completely
depolarizing channel. This channel simply replaces the input state with the
maximally mixed state. Therefore, no classical or quantum information can
traverse it, even with the help of entanglement.

\begin{corollary}
\label{cor:complete-dep}The following set of inequalities specifies the
entanglement-assisted capacity of the completely depolarizing channel:%
\begin{align*}
C+2Q  &  \leq0,\\
Q  &  \leq E,\\
C+Q  &  \leq E.
\end{align*}

\end{corollary}

\begin{IEEEproof}
The proof follows by considering that the mutual information $I(AX;B)$ and the Holevo information
$I(X;B)$ in
Theorem~\ref{gf}\ vanish for any $k$-qudit state transmitted through the
completely depolarizing channel and the coherent information is either negative or zero for any input state. Then the inequalities (\ref{gf1}-\ref{gf3})
there become the respective inequalities above.
\end{IEEEproof}

One should observe that the region is actually trivial (it is empty) because
$C+2Q\leq0$. Nevertheless, we still find the inequalities in
Corollary~\ref{cor:complete-dep}\ interesting because they coincide with those
that we found in Ref.~\cite{HW09}\ for the \textquotedblleft unit resource
capacity\textquotedblright\ region\footnote{The unit resource capacity region
is the set of rates that are achievable without the aid of a noisy resource.}
(modulo a different sign convention with the entanglement rate $E$). The proof
techniques in Ref.~\cite{HW09} involve \textit{reductio ad absurdum} arguments
that show how points outside the region conflict with physical law, whereas
the arguments in the converse proof of Theorem~\ref{gf}\ are information
theoretic. One should expect that the set of inequalities in
Corollary~\ref{cor:complete-dep}\ coincide with those for the unit resource
capacity region because having access to the completely depolarizing channel
is equivalent to having no quantum channel at all---Bob can actually simulate
this resource locally merely by preparing the maximally mixed state in his laboratory.

\subsection{The Quantum Erasure Channel}

The quantum erasure channel is perhaps one of the simplest noisy quantum
channels \cite{GBP97}, because it has a simple specification and its known
transmission capacities admit simple formulas \cite{PhysRevLett.78.3217}. A
quantum erasure channel passes the input state along to the environment and
gives Bob an erasure state $\left\vert e\right\rangle $ with probability
$\epsilon$. It passes the input state along to Bob and gives the environment
an erasure state $\left\vert e\right\rangle $\ with probability $1-\epsilon$.
It induces the following map on a density operator $\rho^{A^{\prime}}$:%
\[
\rho^{A^{\prime}}\rightarrow\left(  1-\epsilon\right)  \rho^{B}+\epsilon
\left\vert e\right\rangle \left\langle e\right\vert ^{B},
\]
and its isometric extension acts as follows:%
\[
\left\vert \psi\right\rangle ^{AA^{\prime}}\rightarrow\sqrt{1-\epsilon
}\left\vert \psi\right\rangle ^{AB}\left\vert e\right\rangle ^{E}%
+\sqrt{\epsilon}\left\vert \psi\right\rangle ^{AE}\left\vert e\right\rangle
^{B},
\]
where $\left\vert \psi\right\rangle ^{AA^{\prime}}$\ is some purification of
$\rho^{A^{\prime}}$.%
%TCIMACRO{\TeXButton{B}{\begin{table}[tbp] \centering}}%
%BeginExpansion
\begin{table}[tbp] \centering
%EndExpansion%
\begin{tabular}
[c]{l|l}\hline\hline
\textbf{Capacity} & Rate Triple $\mathbf{(}C,Q,E\mathbf{)}$\\\hline\hline
Entanglement-assisted classical capacity (EAC) & $\left(  2\left(
1-\epsilon\right)  ,0,1\right)  $\\\hline
Quantum capacity (LSD) & $\left(  0,1-2\epsilon,0\right)  $\\\hline
Classical capacity (HSW) & $\left(  1-\epsilon,0,0\right)  $\\\hline
Entanglement-assisted quantum capacity (EAQ) & $\left(  0,1-\epsilon
,\epsilon\right)  $\\\hline\hline
\end{tabular}
\caption{The left column gives a particular type of capacity for the quantum erasure channel,
and the right column gives the corresponding optimal rate triple.}\label{tbl:known-points}%
%TCIMACRO{\TeXButton{E}{\end{table}}}%
%BeginExpansion
\end{table}%
%EndExpansion

Table~\ref{tbl:known-points}\ lists the known optimal transmission capacities
for the quantum erasure channel. Bennett \textit{et al}. determined the
classical capacity of the quantum erasure channel with an intuitive argument
(the outer bound exploits the Holevo bound \cite{book2000mikeandike} and the
inner bound uses an encoding with orthogonal states), and they determined its
quantum capacity with a different intuitive argument (the well-known
no-cloning argument combined with linear interpolation for the outer bound and
one-way random hashing for the inner bound \cite{PhysRevLett.78.3217}). The
optimality of the classical rate $2\left(  1-\epsilon\right)  $\ and the
quantum rate $1-\epsilon$ of an entanglement-assisted quantum erasure channel
follows from the arguments in Ref.~\cite{PhysRevLett.83.3081}. The optimality
of the respective entanglement consumption rates follows from our forthcoming
arguments. Finally, note that we can obtain the quantum capacity result by
combining the father protocol (entanglement-assisted quantum
communication)\ with entanglement distribution at a rate $\epsilon$, and we
can obtain entanglement-assisted classical communication from
entanglement-assisted quantum communication by consuming all of its quantum
communication at rate $1-\epsilon$ with super-dense coding.

Corollary~\ref{cor:erasure-region} below shows that the CQE capacity region of
a quantum erasure channel admits a simple characterization in terms of three
inequalities. We prove the converse by intuitive reasoning that one would
perhaps expect to be able to apply to the quantum erasure channel, given
earlier intuitive reasoning that authors have applied to this channel. We
prove the direct coding theorem by giving an explicit ensemble that reaches
all of the bounds in the inequalities in Corollary~\ref{cor:erasure-region}.
The result is that time-sharing\footnote{Time-sharing is a simple method of
combining coding strategies \cite{book1991cover}. As an example, consider the
case of time-sharing a channel between an $\left(  n,Q_{1},\epsilon\right)  $
quantum code and another $\left(  n,Q_{2},\epsilon\right)  $ quantum code. For
any $\lambda$ where $0<\lambda<1$, the sender uses the first code for a
fraction $\lambda$ of the channel uses and uses the other code for a fraction
$\left(  1-\lambda\right)  $ of the channel uses. This time-sharing strategy
produces a quantum code with rate $\lambda Q_{1}+\left(  1-\lambda\right)
Q_{2}$ and error at most $2\epsilon$. Time-sharing immediately gives that the
convex hull of any set of achievable points is an achievable region.} between
the four protocols in Table~\ref{tbl:known-points}\ is the optimal coding strategy.

\begin{corollary}
\label{cor:erasure-region}Suppose a quantum erasure channel has an erasure
probability $\epsilon$. The following set of inequalities specifies the
capacity region of this entanglement-assisted channel for transmitting
classical and quantum information:%
\begin{align}
C+2Q  &  \leq2\left(  1-\epsilon\right)  ,\label{eq:erase-1}\\
\frac{1-2\epsilon}{1-\epsilon}C+Q  &  \leq E+\left(  1-2\epsilon\right)
,\label{eq:erase-2}\\
C+\left(  1+\epsilon\right)  Q  &  \leq\left(  1-\epsilon\right)  \left(
1+E\right)  . \label{eq:erase-3}%
\end{align}

\end{corollary}

\begin{IEEEproof}
We first prove the converse. The first bound in (\ref{eq:erase-1})\ holds
because the sum rate $C+2Q$ can never exceed $2\left(  1-\epsilon\right)  $.
Otherwise, one could beat the entanglement-assisted classical capacity by
dense coding or one could beat the entanglement-assisted quantum capacity by teleportation.
We next consider the second bound in (\ref{eq:erase-2}). We first prove that
time-sharing between the HSW\ point and the LSD\ point is an optimal
strategy\footnote{Devetak and Shor stated (but did not explicitly prove) that
time-sharing between HSW\ and LSD\ is optimal for the erasure channel
\cite{cmp2005dev}.} and then show that this result implies the bound in
(\ref{eq:erase-2}). Consider a scheme of quantum error correction for an
erasure channel with erasure parameter $\epsilon$. If Alice transmits $n$
qubits, then Bob receives $n(1-\epsilon)$ of these and the environment
receives $n\epsilon$ of them (for the case of large $n$). From these
$n(1-\epsilon)$ physical qubits, Bob can perform a decoding to obtain
$n(1-2\epsilon)$ logical qubits, by the quantum capacity result for the
erasure channel. This implies an optimal \textquotedblleft decoding
ratio\textquotedblright\ of $n(1-2\epsilon)$ decoded qubits for the
$n(1-\epsilon)$ received qubits: $(1-2\epsilon)/(1-\epsilon)$. Now let us
consider a Devetak-Shor-like code for the erasure channel. Suppose that Alice
can achieve the rate triple $(\lambda(1-\epsilon),(1-\lambda)(1-2\epsilon
)+\delta,0)$ where $\delta$ is some small positive number (so that this rate
triple represents any point that beats the time-sharing limit). Now if Alice
transmits $n$ qubits, Bob receives $n(1-\epsilon)$ of them and the environment
again receives $n\epsilon$ of them. But this time, Bob performs measurements
on $n\lambda(1-\epsilon)$ of them in order to obtain the classical
information. Thus, these qubits are no longer available for decoding quantum
information because the measurements completely dephase them. This leaves
$n(1-\epsilon)-n\lambda(1-\epsilon)=n(1-\lambda)(1-\epsilon)$ qubits available
for decoding the quantum information. If Bob could decode $n((1-\lambda
)(1-2\epsilon)+\delta)$ logical qubits, this would contradict the optimality
of the above \textquotedblleft decoding ratio\textquotedblright\ because
$n((1-\lambda)(1-2\epsilon)+\delta)/(n(1-\lambda)(1-\epsilon))=(1-2\epsilon
)/(1-\epsilon)+\delta/(1-\lambda)(1-\epsilon)$ is greater than the optimal
decoding ratio $(1-2\epsilon)/(1-\epsilon)$. Therefore, he must only be able
to decode $n(1-\lambda)(1-2\epsilon)$ logical qubits. This proves that
time-sharing between HSW\ and LSD\ is an optimal strategy for the quantum
erasure channel.
Now, the capacity region excludes any point lying above the $CQ$-plane with
which we can combine with entanglement distribution to reach a point on the
$CQ$-plane outside the Devetak-Shor time-sharing bound (otherwise, we would be
able to beat the time-sharing bound between HSW\ and LSD by combining this
point with entanglement distribution). In particular, this means that
achievable points cannot be outside the plane containing the vector connecting
LSD\ to HSW\ and the vector of entanglement distribution. It is
straightforward to calculate the equation for this plane. The vector
connecting LSD\ to HSW\ is%
\[
\left(  0,1-2\epsilon,0\right)  -\left(  1-\epsilon,0,0\right)  =\left(
-\left(  1-\epsilon\right)  ,1-2\epsilon,0\right)  .
\]
The vector of entanglement distribution is $\left(  0,-1,-1\right)  $. A
normal vector for the plane containing the two vectors is%
\[
\left(  -\frac{1-2\epsilon}{1-\epsilon},-1,1\right)  .
\]
Then the equation for the plane is%
\[
-\frac{1-2\epsilon}{1-\epsilon}\left(  C-\left(  1-\epsilon\right)  \right)
-Q+E=0,
\]
implying that achievable points must obey the bound in (\ref{eq:erase-2})
because they cannot lie outside this plane.
The above argument also shows that the EAQ rate triple $\left(  0,1-\epsilon
,\epsilon\right)  $ is optimal (in particular, that the entanglement
consumption rate is optimal) because it lies at the intersection of the two
bounds in (\ref{eq:erase-1})\ and (\ref{eq:erase-2}).
We now prove the last bound in (\ref{eq:erase-3}) in three steps.\ We first
prove that the entanglement consumption rate of the EAC\ protocol is optimal.
We then prove that time-sharing between EAC\ and HSW\ is optimal, and finally
rule out all points outside a plane containing the vector connecting EAC\ to
HSW and the vector of super-dense coding.
Consider the EAC\ rate triple $\left(  2\left(  1-\epsilon\right)
,0,1\right)  $. The entanglement consumption rate of one ebit per channel use
is optimal, i.e., one cannot achieve the classical rate of $2\left(
1-\epsilon\right)  $ with less than one ebit per channel use. The state that
achieves capacity is the maximally entangled state $|\Phi^{+}\rangle$. The
minimum amount of entanglement that this capacity-achieving state requires is
$H\left(  A\right)  =1$ ebit (we give a more detailed proof in Appendix~\ref{sec:EAC-EA-rate}).
%, according to the characterization in Ref.~\cite{arx2005dev}.
Thus, no lower amount of entanglement could suffice for achieving the maximal
classical rate.
We now prove that time-sharing between EAC\ and HSW\ is optimal by an argument
similar to the argument for our other time-sharing bound. Consider a scheme of
entanglement-assisted classical communication for an erasure channel with
erasure parameter $\epsilon$. If Alice transmits $n$ qubits (that could
potentially be entangled with $n$ qubits of Bob's), then Bob receives
$n(1-\epsilon)$ of these and the environment receives $n\epsilon$ of them (for
the case of large $n$). From these $n(1-\epsilon)$ physical qubits (and his
halves of the ebits), Bob can perform a decoding to obtain $n2(1-\epsilon)$
classical bits, by the entanglement-assisted classical capacity result for the
erasure channel. This implies an optimal \textquotedblleft EA\ decoding
ratio\textquotedblright\ of $n2(1-\epsilon)$ decoded bits for the
$n(1-\epsilon)$ received qubits: $2(1-\epsilon)/(1-\epsilon)$. Now let us
consider a Shor-like code\footnote{\textquotedblleft
Shor-like\textquotedblright\ in the sense of Ref.~\cite{arx2004shor}.} for the
erasure channel. Suppose that Alice can achieve the rate triple $(\lambda
(1-\epsilon)+(1-\lambda)2(1-\epsilon)+\delta,0,1-\lambda)$ where $\delta$ is
some small positive number (so that this rate triple represents any point that
beats the time-sharing limit). Now if Alice transmits $n$ qubits, then Bob
receives $n(1-\epsilon)$ of them and the environment again receives
$n\epsilon$ of them. But this time, Bob performs some measurement on
$n\lambda(1-\epsilon)$ of them in order to obtain some of the classical
information. Thus, these qubits are no longer available for decoding any more
classical information because they have already been decoded. This leaves
$n(1-\epsilon)-n\lambda(1-\epsilon)=n(1-\lambda)(1-\epsilon)$ qubits available
for decoding the extra classical information. If Bob could decode
$n((1-\lambda)2(1-\epsilon)+\delta)$ extra classical bits, this would
contradict the optimality of the above \textquotedblleft EA\ decoding
ratio\textquotedblright\ because $n((1-\lambda)2(1-\epsilon)+\delta
)/(n(1-\lambda)(1-\epsilon))=2(1-\epsilon)/(1-\epsilon)+\delta/(1-\lambda
)(1-\epsilon)$ is greater than the optimal decoding ratio $2(1-\epsilon
)/(1-\epsilon)$. Therefore, he must only be able to decode $n(1-\lambda
)2(1-\epsilon)$ classical bits. This proves that time-sharing between HSW\ and
EAC\ is an optimal strategy for the quantum erasure channel.
Now, the capacity region excludes any point lying to the right of the
$CE$-plane with which we can combine with super-dense coding to reach a point
on the $CE$-plane outside the time-sharing bound (otherwise, we would be able
to beat the time-sharing bound between HSW\ and EAC by combining this point
with super-dense coding). In particular, this means that achievable points
cannot be outside the plane containing the vector connecting HSW\ to EAC and
the vector of super-dense coding. It is straightforward to calculate the
equation for this plane. Consider that the vector between EAC\ and HSW\ is%
\[
\left(  2\left(  1-\epsilon\right)  ,0,1\right)  -\left(  1-\epsilon
,0,0\right)  =\left(  1-\epsilon,0,1\right)  .
\]
The vector of dense coding is $\left(  2,-1,1\right)  $. A\ normal vector for
this plane is
\[
\left(  -1,-\left(  1+\epsilon\right)  ,1-\epsilon\right)  .
\]
The equation for the plane is%
\[
-\left(  C-\left(  1-\epsilon\right)  \right)  -\left(  1+\epsilon\right)
Q+\left(  1-\epsilon\right)  E=0,
\]
implying that achievable points must obey the bound in (\ref{eq:erase-3})
because they cannot lie outside this plane. We have now completed the proof of
the outer bound.
We prove the direct coding theorem. The simple way to prove it follows simply
by time-sharing between the four protocols HSW, LSD, EAQ, and EAC, but it is
interesting to explore a particular ensemble of states of the form
(\ref{eq:maximization-state})\ in Theorem~\ref{gf} that achieves the capacity.
We consider transmitting the $A^{\prime}$ system of the following
classical-quantum state through the channel:%
\begin{equation}
\sigma^{XAA^{\prime}}\equiv\frac{1}{2}\left(  \left\vert 0\right\rangle
\left\langle 0\right\vert ^{X}\otimes\psi_{0}^{AA^{\prime}}+\left\vert
1\right\rangle \left\langle 1\right\vert ^{X}\otimes\psi_{1}^{AA^{\prime}%
}\right)  , \label{eq:mu-ensemble-cq-state}%
\end{equation}
where%
\begin{align*}
\left\vert \psi_{0}\right\rangle ^{AA^{\prime}}  &  \equiv\sqrt{\mu}\left\vert
00\right\rangle ^{AA^{\prime}}+\sqrt{1-\mu}\left\vert 11\right\rangle
^{AA^{\prime}},\\
\left\vert \psi_{1}\right\rangle ^{AA^{\prime}}  &  \equiv\sqrt{1-\mu
}\left\vert 00\right\rangle ^{AA^{\prime}}+\sqrt{\mu}\left\vert
11\right\rangle ^{AA^{\prime}},
\end{align*}
and $\mu\in\left[  0,\frac{1}{2}\right]  $. This classical-quantum state is a
purified version of the ensemble considered in Ref.~\cite{cmp2005dev}.
We can evaluate various relevant entropic quantities for this state:%
\begin{align*}
H\left(  B\right)  _{\sigma}  &  =1-\epsilon+H_{2}\left(  \epsilon\right)  ,\\
H\left(  A\right)  _{\sigma}  &  =H_{2}\left(  \mu\right)  ,\\
H\left(  A|X\right)  _{\sigma}  &  =H_{2}\left(  \mu\right)  ,\\
H\left(  B|X\right)  _{\sigma}  &  =\left(  1-\epsilon\right)  H_{2}\left(
\mu\right)  +H_{2}\left(  \epsilon\right)  ,\\
H\left(  E|X\right)  _{\sigma}  &  =\epsilon H_{2}\left(  \mu\right)
+H_{2}\left(  \epsilon\right)  ,
\end{align*}
where the state $\sigma$ is the state resulting from sending the $A^{\prime}$
system through the erasure channel. It then follows that%
\begin{align*}
I\left(  X;B\right)  _{\sigma}  &  =\left(  1-\epsilon\right)  \left(
1-H_{2}\left(  \mu\right)  \right)  ,\\
I\left(  A\rangle BX\right)  _{\sigma}  &  =\left(  1-2\epsilon\right)
H_{2}\left(  \mu\right)  ,\\
\frac{1}{2}I\left(  A;B|X\right)  _{\sigma}  &  =\left(  1-\epsilon\right)
H_{2}\left(  \mu\right)  ,\\
\frac{1}{2}I\left(  A;E|X\right)  _{\sigma}  &  =\left(  1-\epsilon\right)
H_{2}\left(  \mu\right)  ,\\
I\left(  AX;B\right)   &  =\left(  1+H_{2}\left(  \mu\right)  \right)  \left(
1-\epsilon\right)  .
\end{align*}
A quick glance over the above information quantities reveals that exploiting
coding strategies such as the classically-enhanced father protocol gives no
improvement over time-sharing because $H_{2}\left( \mu\right) $ varies between
zero and one as $\mu$ varies between zero and $1/2$ (the classically-enhanced
father protocol gives exactly the same performance as time-sharing, as does
the classically-enhanced quantum communication strategy of Devetak and Shor
\cite{cmp2005dev}). Thus, the region obtained as the union of the one-shot,
one-state regions is indeed equivalent to the outer bound given above.
Figure~\ref{fig:erasure-region} plots this region for a quantum erasure
channel with erasure parameter $\epsilon=1/4$, demonstrating that this region
is equivalent to the outer bound.%
%TCIMACRO{\FRAME{ftbpFU}{3.4411in}{1.9009in}{0pt}{\Qcb{(Color online) The
%capacity region of the quantum erasure channel with erasure parameter
%$\epsilon=1/4$. Planes I, II, and III correspond to the respective bounds in
%(\ref{eq:erase-1}-\ref{eq:erase-3}). The optimal strategy is to time-share
%between classical coding (HSW), quantum coding (LSD), entanglement-assisted
%quantum coding (EAQ), and entanglement-assisted classical coding (EAC). The
%classically-enhanced father protocol does not give any improvement over
%time-sharing for a quantum erasure channel.}}{\Qlb{fig:erasure-region}%
%}{erasure-region.pdf}{\special{ language "Scientific Word";  type "GRAPHIC";
%width 3.4411in;  height 1.9009in;  depth 0pt;  original-width 11.2642in;
%original-height 6.1808in;  cropleft "0";  croptop "1";  cropright "1";
%cropbottom "0";  filename 'erasure-region.pdf';file-properties "XNPEU";}} }%
%BeginExpansion
\begin{figure}
[ptb]
\begin{center}
\includegraphics[
natheight=6.180800in,
natwidth=11.264200in,
width=3.4411in
]%
{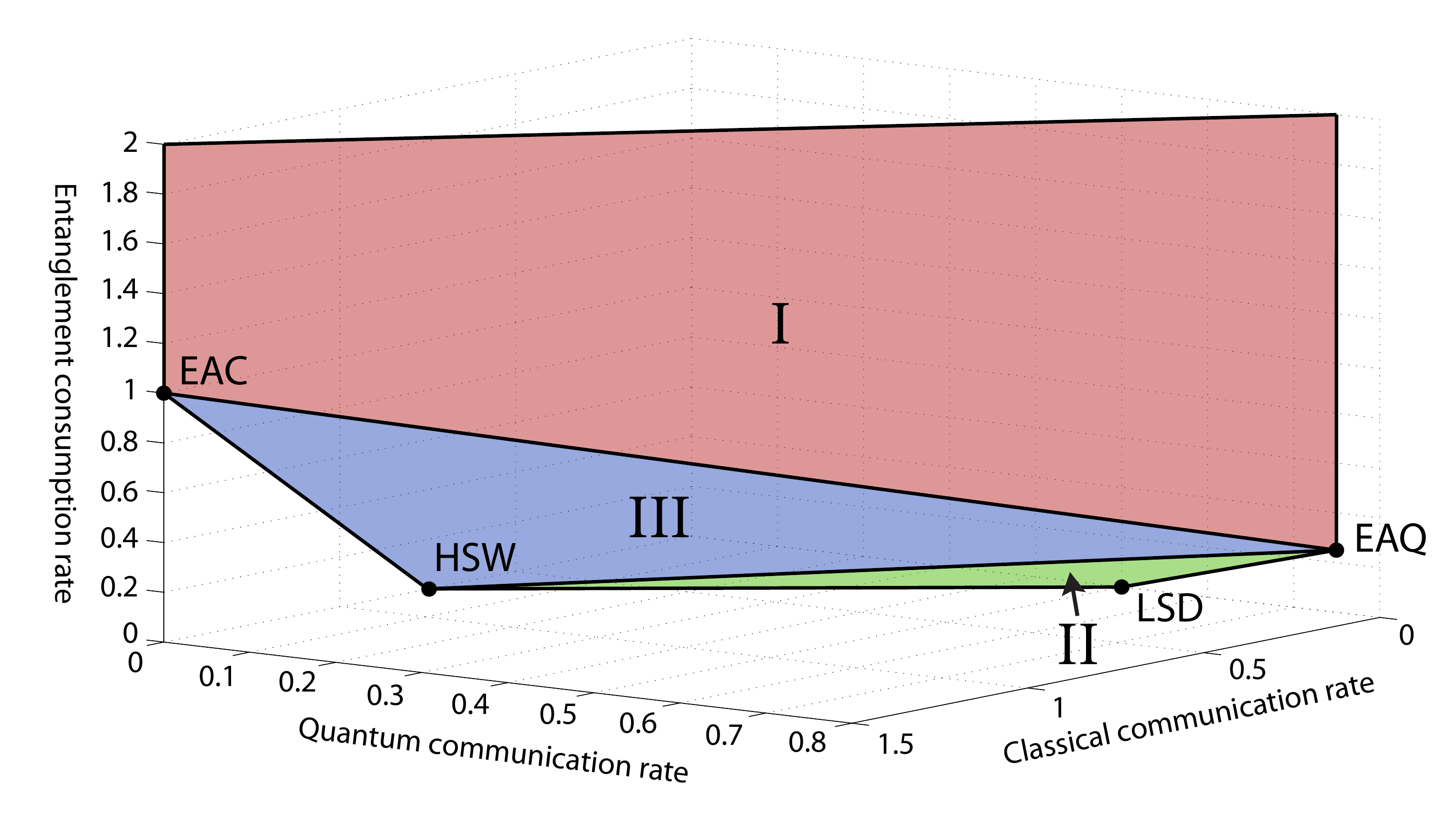}%
\caption{(Color online) The capacity region of the quantum erasure channel
with erasure parameter $\epsilon=1/4$. Planes I, II, and III correspond to the
respective bounds in (\ref{eq:erase-1}-\ref{eq:erase-3}). The optimal strategy
is to time-share between classical coding (HSW), quantum coding (LSD),
entanglement-assisted quantum coding (EAQ), and entanglement-assisted
classical coding (EAC). The classically-enhanced father protocol does not give
any improvement over time-sharing for a quantum erasure channel.}%
\label{fig:erasure-region}%
\end{center}
\end{figure}
%EndExpansion
\end{IEEEproof}

The following corollary applies to the noiseless qubit channel by simply
plugging in $\epsilon=0$.

\begin{corollary}
The following set of inequalities specifies the entanglement-assisted capacity
of the noiseless qubit channel for transmitting classical and quantum
information:%
\begin{align*}
C+2Q  &  \leq2,\\
C+Q  &  \leq E+1.
\end{align*}

\end{corollary}

\subsection{The Qubit Dephasing Channel}

In this section, we show that we can compute the full capacity region of a
qubit dephasing channel and plot it in Figure~\ref{fig:dephasing-region}\ for
a channel with dephasing parameter $p=0.2$. We show also that the
classically-enhanced father protocol can beat time-sharing for a qubit
dephasing channel (the example is an extension of the argument in
Ref.~\cite{cmp2005dev}).

\subsubsection{Single-Letterization}

We first show that the classically-enhanced father trade-off curve is optimal
in the sense that it lies along the boundary of the capacity region for the
qubit dephasing channel. A surprisingly simple argument proves this result by
resorting to the result of Devetak and Shor in Ref.~\cite{cmp2005dev}. There,
they showed that the following trade-off curve in the $CQ$-plane is optimal:%
\begin{equation}
\left\{  \left(  C_{\text{CQ}}\left(  \mu\right)  ,Q_{\text{CQ}}\left(
\mu\right)  ,0\right)  :0\leq\mu\leq1/2\right\}  , \label{eq:DS-bound}%
\end{equation}
where%
\begin{align*}
C_{\text{CQ}}\left(  \mu\right)   &  \equiv1-H_{2}\left(  \mu\right)  ,\\
Q_{\text{CQ}}\left(  \mu\right)   &  \equiv H_{2}\left(  \mu\right)
-H_{2}\left(  g\left(  p,\mu\right)  \right)  ,\\
g\left(  p,\mu\right)   &  \equiv\frac{1}{2}+\frac{1}{2}\sqrt{1-16\cdot
\frac{p}{2}\left(  1-\frac{p}{2}\right)  \mu\left(  1-\mu\right)  }.
\end{align*}
Now, consider the surface formed by the following set of points:%
\begin{equation}
\left\{  \left(  C_{\text{CQ}}\left(  \mu\right)  ,Q_{\text{CQ}}\left(
\mu\right)  +E,E\right)  :0\leq\mu\leq1/2,\ \ E\geq0\right\}  .
\label{eq:DS-surface}%
\end{equation}
This surface is an outer bound for the capacity region (if it were not so, one
could combine points outside this surface with entanglement distribution and
beat the optimal bound in (\ref{eq:DS-bound}) for the Devetak-Shor case).

Now consider sending the $\mu$-parametrized ensemble in
(\ref{eq:mu-ensemble-cq-state}), where $\mu\in\left[  0,1/2\right]  $, through
the qubit dephasing channel with dephasing parameter $p$. It is
straightforward to show that the various entropic quantities in the
classically-enhanced father protocol are as follows for the $\mu$-parametrized
ensemble:%
\begin{align*}
C_{\text{CEF}}\left(  \mu\right)   &  \equiv I\left(  X;B\right)  _{\sigma
}=1-H_{2}\left(  \mu\right)  ,\\
Q_{\text{CEF}}\left(  \mu\right)   &  \equiv\frac{1}{2}I\left(  A;B|X\right)
=H_{2}\left(  \mu\right)  -\frac{1}{2}H_{2}\left(  g\left(  p,\mu\right)
\right)  ,\\
E_{\text{CEF}}\left(  \mu\right)   &  \equiv\frac{1}{2}I\left(  A;E|X\right)
=\frac{1}{2}H_{2}\left(  g\left(  p,\mu\right)  \right)  .
\end{align*}
Thus, the following set of points contains all points along the
classically-enhanced father trade-off curve:%
\[
\left\{  \left(  C_{\text{CEF}}\left(  \mu\right)  ,Q_{\text{CEF}}\left(
\mu\right)  ,E_{\text{CEF}}\left(  \mu\right)  \right)  :0\leq\mu
\leq1/2\right\}  .
\]
All points along the classically-enhanced father lie along the boundary
because they are of the form in (\ref{eq:DS-surface}) with $E=H_{2}\left(
g\left(  p,\mu\right)  \right)  /2$. This proves that the points along the
classically-enhanced father trade-off curve are optimal. One can also achieve
any point along the surface in (\ref{eq:DS-surface}) with entanglement
consumption below the classically-enhanced father by combining the
classically-enhanced father with entanglement distribution.

We now outline the proof that Shor's trade-off curve for entanglement-assisted
classical communication single-letterizes for the qubit dephasing channel
(full details appear in Ref.~\cite{BHTW10}---the argument complements the
argument in Appendix~B of Ref.~\cite{cmp2005dev}). Any point along Shor's
trade-off curve achieves a classical communication rate of $I\left(
AX;B^{n}\right)  $ at an entanglement consumption rate of $H\left(
A|X\right)  $ \cite{arx2005dev}. Therefore, to determine a point along the
trade-off curve, we would like to maximize the classical communication rate
while minimizing the entanglement consumption rate. To do so, we can define
the following function%
\[
f_{\lambda}\left(  \mathcal{N}^{\otimes n}\right)  \equiv\max_{\sigma}\left(
I\left(  AX;B^{n}\right)  -\lambda H\left(  A|X\right)  \right)  ,
\]
where $\lambda>0$ and the maximization is over all states of the form
(\ref{eq:maximization-state}), with the exception that the $E^{\prime}$ system
is not necessary for Shor's trade-off curve \cite{arx2005dev}. By a sequence
of arguments similar to those in Appendix B of Ref.~\cite{cmp2005dev}, we can
show that%
\[
f_{\lambda}\left(  \mathcal{N}^{\otimes n}\right)  \leq nh_{\lambda}\left(
\mathcal{N}\right)  ,
\]
where%
\[
h_{\lambda}\left(  \mathcal{N}\right)  \equiv\max_{\sigma_{\mu}}\left(
H\left(  Y\right)  +\left(  1-\lambda\right)  H\left(  A|X\right)  -H\left(
E|X\right)  \right)  ,
\]
$Y$ is the completely dephased version of $B$, and $\sigma_{\mu}$ is a state
that arises after sending the $A^{\prime}$ system of a state of the form in
(\ref{eq:mu-ensemble-cq-state}) through a \textit{single use} of the qubit
dephasing channel. This then shows that the region single-letterizes and that
states of the form in (\ref{eq:mu-ensemble-cq-state}) give rise to optimal
points that lie along Shor's trade-off curve. Shor's trade-off curve in the
CE-plane has the following form:%
\begin{equation}
\left\{  \left(  C_{\text{CE}}\left(  \mu\right)  ,0,E_{\text{CE}}\left(
\mu\right)  \right)  :0\leq\mu\leq1/2\right\}  , \label{eq:shor-curve}%
\end{equation}
where%
\begin{align*}
C_{\text{CE}}\left(  \mu\right)   &  \equiv1+H_{2}\left(  \mu\right)
-H_{2}\left(  g\left(  p,\mu\right)  \right)  ,\\
E_{\text{CE}}\left(  \mu\right)   &  \equiv H_{2}\left(  \mu\right)  .
\end{align*}

We can now exploit Shor's trade-off curve to outline a bounding surface in the
CQE space (just as we did before with the Devetak-Shor curve and entanglement
distribution). Consider the surface formed by the following set of points:%
\begin{equation}
\left\{  \left(  C_{\text{CE}}\left(  \mu\right)  -2E,E,E_{\text{CE}}\left(
\mu\right)  -E\right)  :0\leq\mu\leq1/2,\ \ E\geq0\right\}  .
\label{eq:shor-surface}%
\end{equation}
This surface is an outer bound for the capacity region (if it were not so, one
could combine points outside this surface with super-dense coding and beat the
optimal bound in (\ref{eq:shor-curve})). Interestingly, this surface
intersects the surface in (\ref{eq:DS-surface}) at exactly the
classically-enhanced father trade-off curve.

\begin{figure}
[ptb]
\begin{center}
\includegraphics[
natheight=5.138700in,
natwidth=6.625300in,
width=3.5405in
]%
{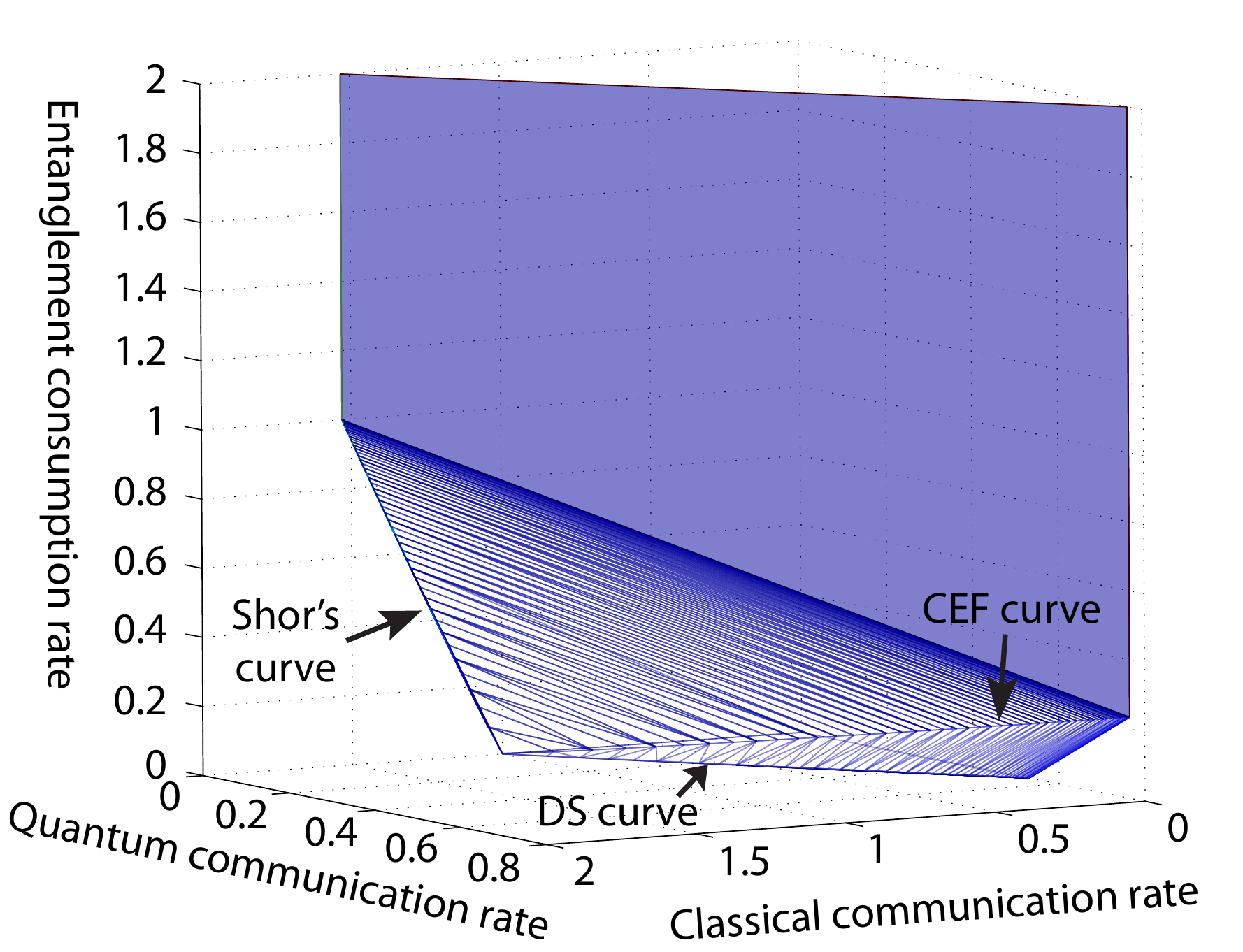}%
\caption{(Color online)\ The above figure plots the full capacity region for
the qubit dephasing channel with dephasing parameter $p=0.2$. It outlines
Shor's trade-off curve, the Devetak-Shor (DS) trade-off curve, and the
classically-enhanced father (CEF) trade-off curve. The surface between Shor's
curve and the CEF\ curve is that in (\ref{eq:shor-surface}). The surface
between the CEF curve and the DS\ curve is that in (\ref{eq:DS-surface}).
Finally, (\ref{eq:solid-plane}) specifies the solid plane. This region is a
union of regions formed by translating the unit resource capacity region from Ref.~\cite{HW09}
along the classically-enhanced father trade-off curve. This point
is perhaps more clear in Ref.~\cite{HW09} where we plot the full triple trade-off.}%
\label{fig:dephasing-region}%
\end{center}
\end{figure}

We can finally outline the full capacity region by combining the two surfaces
in (\ref{eq:DS-surface}) and (\ref{eq:shor-surface}) with the bound:%
\begin{equation}
C+2Q\leq2-H_{2}\left(  g\left(  p,1/2\right)  \right)  .
\label{eq:solid-plane}%
\end{equation}
The above bound is the largest that the entanglement-assisted classical
capacity can be and therefore bounds the sum rate $C+2Q$ as we have argued
previously. The intersection of these three surfaces forms a single-letter
bound for the capacity region, and all points on the boundary are achievable
by combining the classically-enhanced father trade-off curve with entanglement
distribution, super-dense coding, or the wasting of entanglement.
Figure~\ref{fig:dephasing-region}\ plots the full capacity region.

%We remark that this proof technique shows that isometric encodings are optimal
%(so that the $E^{\prime}$ register is not necessary), at least for the case of
%the qubit dephasing channel. It might be straightforward to extend this idea
%to show that isometric encodings are optimal for an arbitrary channel, but we
%leave this to future work.%
%

%TCIMACRO{\FRAME{ftbpFU}{3.5405in}{2.7527in}{0pt}{\Qcb{(Color online)\ The
%above figure plots the full capacity region for the qubit dephasing channel
%with dephasing parameter $p=0.2$. It outlines Shor's trade-off curve, the
%Devetak-Shor (DS) trade-off curve, and the classically-enhanced father (CEF)
%trade-off curve. The surface between Shor's curve and the CEF\ curve is that
%in (\ref{eq:shor-surface}). The surface between the CEF curve and the
%DS\ curve is that in (\ref{eq:DS-surface}). Finally, (\ref{eq:solid-plane})
%specifies the solid plane. This region is a
%union of regions formed by translating the unit resource capacity region from Ref.~\cite{HW09}
%along the classically-enhanced father trade-off curve. This point
%is perhaps more clear in Ref.~\cite{HW09} where we plot the full triple trade-off.}}{\Qlb{fig:dephasing-region}}{dephasing-region.pdf}%
%width 3.5405in;  height 2.7527in;  depth 0pt;  original-width 6.6253in;
%original-height 5.1387in;  cropleft "0";  croptop "1";  cropright "1";
%cropbottom "0";  filename 'dephasing-region.pdf';file-properties "XNPEU";}}}%
%BeginExpansion
%EndExpansion

\subsubsection{The Classically-Enhanced Father Protocol can beat Time-Sharing}%

%TCIMACRO{\FRAME{ftbpFU}{6.5518in}{3.5276in}{0pt}{\Qcb{(Color online)\ (a) The
%figure on the left displays the points achievable by time-sharing between
%entanglement-assisted quantum coding and classical coding on the solid red
%line, and it displays the points achievable with the classically-enhanced
%father protocol on the dotted blue line. The channel for which we are coding
%is the qubit dephasing channel with dephasing parameter $p=0.2$. The figure
%demonstrates that one can achieve more quantum communication with less
%entanglement consumption, while having the same rate of classical
%communication, by employing the classically-enhanced father protocol instead
%of a time-sharing strategy. (b) The figure on the right makes the previous
%statement precise, by showing the difference between quantum communication and
%entanglement consumption for achievable points on the classically-enhanced
%father trade-off curve that attain the same rate of classical communication as
%a time-sharing strategy.}}{\Qlb{fig:dephasing-CEF-curve}}{ts-vs-cef.pdf}%
%width 6.5518in;  height 3.5276in;  depth 0pt;  original-width 12.583in;
%original-height 6.7499in;  cropleft "0";  croptop "1";  cropright "1";
%cropbottom "0";  filename 'TS-vs-CEF.pdf';file-properties "XNPEU";}}}%
%BeginExpansion
\begin{figure*}
[ptb]
\begin{center}
\includegraphics[
natheight=6.749900in,
natwidth=12.583000in,
width=6.5518in
]%
{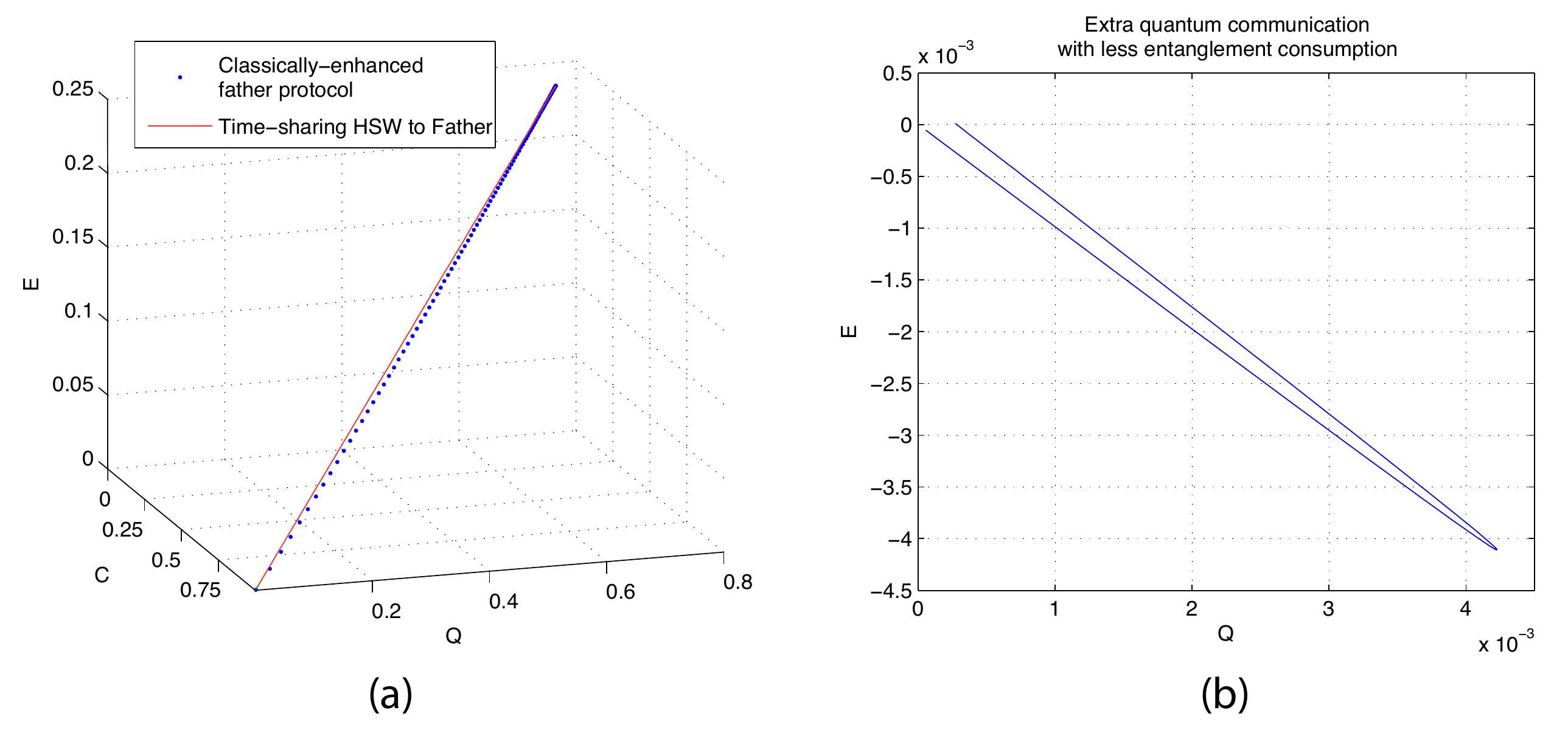}%
\caption{(Color online)\ (a) The figure on the left displays the points
achievable by time-sharing between entanglement-assisted quantum coding and
classical coding on the solid red line, and it displays the points achievable
with the classically-enhanced father protocol on the dotted blue line. The
channel for which we are coding is the qubit dephasing channel with dephasing
parameter $p=0.2$. The figure demonstrates that one can achieve more quantum
communication with less entanglement consumption, while having the same rate
of classical communication, by employing the classically-enhanced father
protocol instead of a time-sharing strategy. (b) The figure on the right makes
the previous statement precise, by showing the difference between quantum
communication and entanglement consumption for achievable points on the
classically-enhanced father trade-off curve that attain the same rate of
classical communication as a time-sharing strategy.}%
\label{fig:dephasing-CEF-curve}%
\end{center}
\end{figure*}
%EndExpansion

An important question for entanglement-assisted classical-quantum coding is
whether a time-sharing strategy is optimal for all channels or if the
classically-enhanced father protocol can give an improvement over
time-sharing. There are three time-sharing strategies that one could employ in
EACQ\ coding. In all three strategies, we suppose that the sender and receiver
share some finite amount of entanglement $E$. The three strategies are as follows:

\begin{enumerate}
\item Use an entanglement-assisted quantum code with rate triple $\left(
0,Q_{1},E_{1}\right)  $ and an HSW\ code with rate triple $\left(
C_{2},0,0\right)  $. If $E=\lambda E_{1}$, then time-sharing produces an EACQ
code with rate triple $\left(  \left(  1-\lambda\right)  C_{2},\lambda
Q_{1},E\right)  $.

\item Use an entanglement-assisted classical code with rate triple $\left(
C_{1},0,E_{1}\right)  $ and a quantum channel code with rate triple $\left(
0,Q_{2},0\right)  $. If $E=\lambda E_{1}$, then time-sharing produces an EACQ
code with rate triple $\left(  \lambda C_{1},\left(  1-\lambda\right)
Q_{2},E\right)  $.

\item Use an entanglement-assisted quantum code with rate triple $\left(
0,Q_{1},E_{1}\right)  $ and an entanglement-assisted classical code with rate
triple $\left(  C_{2},0,E_{2}\right)  $. If $E=\lambda E_{1}+\left(
1-\lambda\right)  E_{2}$, then time-sharing produces an EACQ code with rate
triple $\left(  \left(  1-\lambda\right)  C_{2},\lambda Q_{1},E\right)  $.
\end{enumerate}

We should compare the classically-enhanced father protocol to the first
time-sharing strategy because the two points EAQ and HSW are special cases of
it. For the second time-sharing strategy, it is clear that this strategy is
not optimal because the line connecting EAC\ to LSD\ is strictly inside the
capacity region. For the third time-sharing strategy, time-sharing is the
optimal strategy. If it were not (in the sense that one could achieve a higher
quantum or classical rate than a point along the time-sharing bound), then one
could beat the bound in (\ref{gf1}) by combining this protocol with either
teleportation or super-dense coding.

We now consider the first case for the qubit dephasing channel and show that
the classically-enhanced father protocol can beat a time-sharing strategy.
Consider the qubit dephasing channel with dephasing parameter $p=0.2$. The
classical capacity of this channel is one bit per channel use, and the
entanglement-assisted quantum capacity is about 0.7655 qubits per channel use
while using about 0.2345 ebits per channel use. The solid red line in
Figure~\ref{fig:dephasing-CEF-curve}(a)\ corresponds to the time-sharing line
between these two optimal points. The blue dotted line in
Figure~\ref{fig:dephasing-CEF-curve}(a)\ corresponds to the various points
along the classically-enhanced father protocol. In comparing the time-sharing
line to the classically-enhanced father trade-off curve, we see that the
classically-enhanced father protocol achieves more quantum communication for
less entanglement consumption for any point along the time-sharing line that
achieves the same amount of classical communication.
Figure~\ref{fig:dephasing-CEF-curve}(b) makes this statement precise by
comparing the difference in quantum communication and entanglement consumption
for all points along the trade-off curve that achieve the same amount of
classical communication as a time-sharing point.

\section{Conclusion}

We have proven the entanglement-assisted classical and quantum capacity
theorem. This theorem determines the ultimate rates at which a noisy quantum
channel can communicate both classical and quantum information reliably, while
consuming entanglement to do so. The coding strategy exploits a new
entanglement-assisted classical-quantum coding strategy, the
\textit{classically-enhanced father protocol}, and the unit protocols of
teleportation, super-dense coding, and entanglement distribution. Several
protocols in the family tree of quantum Shannon theory are now child protocols
of the classically-enhanced father. We also have provided example channels
whose corresponding CQE\ capacity regions single-letterize, so that we can
actually determine the region for these channels, and we have shown that
classically-enhanced father protocol beats a time-sharing strategy for the
case of a qubit dephasing channel. We discuss follow-up work and several open
problems in what follows.

\subsection{The Full Triple Trade-off}

The present article addresses only one octant of the channel coding
scenario---the octant where we consume entanglement and generate classical and
quantum communication. We characterize the full triple trade-off region in
Ref.~\cite{HW09}, where we show that the classically-enhanced father protocol
combined with the unit resource protocols in (\ref{TP}-\ref{ED}) achieves the
full capacity region for all octants.

\subsection{The Structure of Classically-Enhanced Father Codes}

\label{sec:optimal-codes}In Ref.~\cite{arx2008wildeUQCC}, one of the authors
constructed a classically-enhanced father code that uses only ancilla qubits
for encoding classical information. In Ref.~\cite{kremsky:012341}, the other
author constructed a classically-enhanced father code that uses both ancilla
qubits and ebits for encoding classical information. One might think that
using ebits in addition to ancilla qubits for encoding classical information
could improve performance and it was unclear which coding structure might
perform better.

The structure of our classically-enhanced father protocol actually gives a
hint for constructing classically-enhanced father codes that achieve the rates
in Theorem~\ref{gf}. Consider the protocol in the proof of the direct coding
part of Theorem~\ref{gf}. Bob decodes the classical information by measuring
the channel outputs only. He does not need to measure his half of the
entanglement to decode the classical information. This decoding implies that
he is not using the entanglement for sending classical information---if he
were, he would need to measure his half of the entanglement as well. This
observation lends creedence to the conjecture that it is sufficient to encode
classical information into ancilla qubits when attempting to construct codes
that achieve the trade-off rate triple in Theorem~\ref{gf}.

\subsection{Other Issues}

Another issue remains with the \textquotedblleft pasting\textquotedblright%
\ proof technique. It relies on the assumption that the channel is IID and
thus does not apply in a straightforward way to channels with memory. Many
proof techniques in quantum Shannon theory rely on a \textquotedblleft
one-shot\textquotedblright\ lemma applied to the IID\ case. The usefulness of
this method of proof is that the one-shot result can apply to more general
scenarios such as channels that have memory. So it may be useful to develop a
one-shot result for the code pasting technique.

\section{Acknowledgements}

The authors thank Kamil Br\'{a}dler, Igor Devetak, Patrick Hayden, Dave
Touchette, and Andreas Winter for useful discussions. The authors thank
Chung-Hsien Chou and the National Center for Theoretical Science (South) for
hosting M.-H.H. as a visitor and thank Martin R\"{o}tteler and
NEC\ Laboratories America for hosting M.M.W.\ as a visitor.
M.M.W.\ acknowledges support from the National Research Foundation \& Ministry
of Education, Singapore. The authors acknowledge support from the MDEIE
(Qu\'{e}bec) PSR-SIIRI international collaboration grant.

\appendices

\section{Proof of Proposition ~\ref{thm:random-EA-code}}

\label{AP_RG0}The proof of Proposition~\ref{thm:random-EA-code} is an
extension of the development in Appendix~D of Ref.~\cite{ieee2005dev}.

\begin{IEEEproof}
[Proposition~\ref{thm:random-EA-code}] Consider an arbitrary density operator
$\rho^{A^{\prime}}$ whose spectral decomposition is as follows:%
\[
\rho^{A^{\prime}}=\sum_{x\in\mathcal{X}}p\left(  x\right)  \left\vert
x\right\rangle \left\langle x\right\vert ^{A^{\prime}}.
\]
The $n^{th}$ extension of the above state as a tensor power state is as
follows:
\[
\rho^{A^{\prime n}}\equiv(\rho^{A^{\prime}})^{\otimes n}=\sum_{x^{n}%
\in\mathcal{X}^{n}}p^{n}\left(  x^{n}\right)  \left\vert x^{n}\right\rangle
\left\langle x^{n}\right\vert ^{A^{\prime n}}.
\]
We define the pruned distribution $p^{\prime n}$ as follows:%
\[
p^{\prime n}\left(  x^{n}\right)  \equiv\left\{
\begin{array}
[c]{ccc}%
p^{n}\left(  x^{n}\right)  /\sum_{x^{n}\in T_{\delta}^{X^{n}}}p^{n}\left(
x^{n}\right)  & : & x^{n}\in T_{\delta}^{X^{n}}\\
0 & : & \text{else},
\end{array}
\right.
\]
where $T_{\delta}^{X^{n}}$ denotes the $\delta$-typical set of sequences with
length $n$. Let $\widetilde{\rho}^{A^{\prime n}}$ denote the following
\textquotedblleft pruned state\textquotedblright:%
\begin{equation}
\widetilde{\rho}^{A^{\prime n}}\equiv\sum_{x^{n}\in T_{\delta}^{X^{n}}%
}p^{\prime n}\left(  x^{n}\right)  \left\vert x^{n}\right\rangle \left\langle
x^{n}\right\vert ^{A^{\prime n}}. \label{eq:pruned-state}%
\end{equation}
For any $\epsilon>0$ and sufficiently large $n$, the state $\rho^{A^{\prime
n}}$ is close to $\widetilde{\rho}^{A^{\prime n}}$\ by the gentle measurement
lemma \cite{itit1999winter}\ and the typical subspace theorem
\cite{book2000mikeandike}:%
\[
\left\Vert \rho^{A^{\prime n}}-\widetilde{\rho}^{A^{\prime n}}\right\Vert
_{1}\leq2\epsilon.
\]
For any density operator $\rho^{A^{\prime}}$, it is possible to construct an
entanglement-assisted quantum code that achieves the quantum communication
rate and entanglement consumption rate in Proposition~\ref{thm:random-EA-code}%
. Ref.~\cite{arx2006anura} provides group-theoretical and other clever
arguments to show how to achieve the rates in
Proposition~\ref{thm:random-EA-code}.
Another method for achieving the rates in Proposition~\ref{thm:random-EA-code}
is to exploit the connection between quantum privacy and quantum coherence in
constructing quantum codes \cite{ieee2005dev,PhysRevLett.93.080501}. Indeed,
in Ref.~\cite{hsieh:042306}, one of the current authors showed how to
construct secret-key-assisted private classical codes for a quantum channel.
Using the methods of \cite{ieee2005dev,PhysRevLett.93.080501}, it is possible
to make \textquotedblleft coherent\textquotedblright\ versions, i.e.,
entanglement-assisted quantum codes, of these secret-key-assisted private
classical codes. Let $\left[  k\right]  $ denote a set of size $\sim2^{nQ}$
and let $\left[  m\right]  $ denote a set of size $\sim2^{nE}$. Let $U_{k,m}$
denote $\sim2^{n\left(  Q+E\right)  }$ random variables that we choose
according to the pruned distribution $p^{\prime n}\left(  x^{n}\right)  $. The
realizations $u_{k,m}$ of the random variables $U_{k,m}$ are sequences in
$\mathcal{X}^{n}$ and are the basis for constructing an entanglement-assisted
quantum code $\mathcal{C}$ whose codewords are as follows%
\[
\mathcal{C}=\{\left\vert \phi_{k}\right\rangle ^{A^{n}T_{B}}\}_{k}.
\]
The entanglement-assisted quantum codewords $\left\vert \phi_{k}\right\rangle
^{A^{n}T_{B}}$ in $\mathcal{C}$ are as follows%
\[
\left\vert \phi_{k}\right\rangle ^{A^{n}T_{B}}\equiv\frac{1}{\sqrt{2^{nE}}%
}\sum_{m=1}^{2^{nE}}|\phi_{u_{k,m}}\rangle^{A^{\prime n}}\left\vert
m\right\rangle ^{T_{B}},
\]
where%
\[
|\phi_{u_{k,m}}\rangle^{A^{\prime n}}\equiv|u_{k,m}\rangle^{A^{\prime n}}.
\]
We then expurgate this code to improve its performance and this expurgation
has a minimal impact on the rate of the code. After expurgation, the code
forms a good entanglement-assisted quantum code, resulting in failure with
probability $\epsilon+10\sqrt[4]{\epsilon}$ by the arguments in
Refs.~\cite{ieee2005dev,PhysRevLett.93.080501}.
Suppose that we choose a particular entanglement-assisted quantum code
$\mathcal{C}$ according to the above prescription. Its code density operator
is%
\[
\rho^{A^{\prime n}T_{B}}(\mathcal{C})=\frac{1}{2^{nQ}}\sum_{k=1}^{2^{nQ}%
}\left\vert \phi_{k}\right\rangle \left\langle \phi_{k}\right\vert ^{A^{\prime
n}T_{B}},
\]
and its input code density operator is%
\begin{align*}
\rho^{A^{\prime n}}(\mathcal{C})  &  =\text{Tr}_{T_{B}}\left\{  \rho
^{A^{\prime n}T_{B}}(\mathcal{C})\right\} \\
&  =\frac{1}{2^{n\left(  Q+E\right)  }}\sum_{m=1}^{2^{nE}}\sum_{k=1}^{2^{nQ}%
}|\phi_{u_{k,m}}\rangle\langle\phi_{u_{k,m}}|^{A^{\prime n}}.
\end{align*}
Suppose we now consider the entanglement-assisted code chosen according to the
above prescription as a \textit{random} code $\mathcal{C}$ (where
$\mathcal{C}$ is now a random variable). Let $\rho^{\prime A^{\prime n}%
}\left(  \mathcal{C}\right)  $ be the channel input density operator for the
random code before expurgation and $\rho^{A^{\prime n}}\left(  \mathcal{C}%
\right)  $ its channel input density operator after expurgation:%
\begin{align*}
\rho^{\prime A^{\prime n}}\left(  \mathcal{C}\right)   &  \equiv\frac
{1}{2^{n\left(  Q^{\prime}+E^{\prime}\right)  }}\sum_{k=1}^{2^{nQ^{\prime}}%
}\sum_{m=1}^{2^{nE^{\prime}}}|\phi_{U_{k,m}}\rangle\langle\phi_{U_{k,m}%
}|^{A^{\prime n}},\\
\rho^{A^{\prime n}}\left(  \mathcal{C}\right)   &  \equiv\frac{1}{2^{n\left(
Q+E\right)  }}\sum_{k=1}^{2^{nQ}}\sum_{m=1}^{2^{nE}}|\phi_{U_{k,m}}%
\rangle\langle\phi_{U_{k,m}}|^{A^{\prime n}},
\end{align*}
where the primed rates are the rates before expurgation and the unprimed rates
are those after expurgation (they are slightly different but identical for
large $n$). Let $\overline{\rho}^{\prime A^{\prime n}}$ and $\overline{\rho
}^{A^{\prime n}}$ denote the expectation of the above channel input density
operators:%
\begin{align*}
\overline{\rho}^{\prime A^{\prime n}}  &  \equiv\mathbb{E}_{\mathcal{C}%
}\left\{  \rho^{\prime A^{\prime n}}\left(  \mathcal{C}\right)  \right\}  ,\\
\overline{\rho}^{A^{\prime n}}  &  \equiv\mathbb{E}_{\mathcal{C}}\left\{
\rho^{A^{\prime n}}\left(  \mathcal{C}\right)  \right\}  .
\end{align*}
Choosing our code in the particular way that we did leads to an interesting
consequence. The expectation of the density operator corresponding to Alice's
restricted codeword $|\phi_{U_{k,m}}\rangle^{A^{\prime n}}$\ is equal to the
pruned state in\ (\ref{eq:pruned-state}):%
\[
\mathbb{E}_{\mathcal{C}}\left\{  |\phi_{U_{k,m}}\rangle\langle\phi_{U_{k,m}%
}|^{A^{\prime n}}\right\}  =\sum_{x^{n}}p^{\prime n}\left(  x^{n}\right)
\left\vert \phi_{x^{n}}\right\rangle \left\langle \phi_{x^{n}}\right\vert
^{A^{\prime n}},
\]
because we choose the codewords $|\phi_{U_{k,m}}\rangle$ randomly according to
the pruned distribution $p^{\prime n}\left(  x^{n}\right)  $. Then the
expected channel input density operator $\overline{\rho}^{\prime A^{\prime n}%
}$\ is as follows:%
\begin{align}
\overline{\rho}^{\prime A^{\prime n}}  &  =\mathbb{E}_{\mathcal{C}}\left\{
\rho^{\prime A^{\prime n}}\left(  \mathcal{C}\right)  \right\} \\
&  =\frac{1}{2^{n\left(  Q^{\prime}+E^{\prime}\right)  }}\sum_{k=1}%
^{2^{nQ^{\prime}}}\sum_{m=1}^{2^{nE^{\prime}}}\mathbb{E}_{\mathcal{C}}\left\{
|\phi_{U_{k,m}}\rangle\langle\phi_{U_{k,m}}|^{A^{\prime n}}\right\} \\
&  =\sum_{x^{n}}p^{\prime n}\left\vert \phi_{x^{n}}\right\rangle \left\langle
\phi_{x^{n}}\right\vert ^{A^{\prime n}}.
\end{align}
Then we know that the following inequality holds for $\overline{\rho}^{\prime
A^{\prime n}}$ and the tensor power state $\rho^{A^{\prime n}}$%
\begin{equation}
\left\Vert \overline{\rho}^{\prime A^{\prime n}}-\rho^{A^{\prime n}%
}\right\Vert _{1}\leq2\epsilon\label{eq:typical-close}%
\end{equation}
by the typical subspace theorem and the gentle measurement lemma. The
expurgation of any entanglement-assisted code $\mathcal{C}$ has a minimal
effect on the resulting channel input density operator \cite{ieee2005dev}:%
\[
\left\Vert \rho^{\prime A^{\prime n}}\left(  \mathcal{C}\right)
-\rho^{A^{\prime n}}\left(  \mathcal{C}\right)  \right\Vert _{1}\leq
4\sqrt[4]{\epsilon}.
\]
The above inequality implies that the following one holds for the expected
channel input density operators $\overline{\rho}^{\prime A^{\prime n}}$ and
$\overline{\rho}^{A^{\prime n}}$%
\begin{equation}
\left\Vert \overline{\rho}^{\prime A^{\prime n}}-\overline{\rho}^{A^{\prime
n}}\right\Vert _{1}\leq4\sqrt[4]{\epsilon}, \label{eq:expected-close}%
\end{equation}
because the trace distance is convex. The following inequality holds%
\begin{equation}
\left\Vert \overline{\rho}^{A^{\prime n}}-\rho^{A^{\prime n}}\right\Vert
_{1}\leq2\epsilon+4\sqrt[4]{\epsilon},
\end{equation}
by applying the triangle inequality to (\ref{eq:typical-close}) and
(\ref{eq:expected-close}). Therefore, the random entanglement-assisted quantum
code is $\rho$-like.
\end{IEEEproof}

\section{Proof of Proposition~\ref{prop:random-grandfather}}

\label{AP_RG} We now prove Proposition~\ref{prop:random-grandfather} that
applies to a random father code that has an associated classical string.

\begin{IEEEproof}
[Proposition~\ref{prop:random-grandfather}] The proof is similar to the proof
of Proposition~5 in Ref.~\cite{cmp2005dev}. Suppose that we have an ensemble
$\{p_{x},\rho_{x}^{A^{\prime}}\}$ where each density operator $\rho
_{x}^{A^{\prime}}$ has a purification $\psi_{x}^{AA^{\prime}}$ and state
$\phi_{x}^{ABE}=U_{\mathcal{N}}^{A^{\prime}\rightarrow BE}(\psi_{x}%
^{AA^{\prime}})$ arising from the channel $\mathcal{N}^{A^{\prime}\rightarrow
B}$. By Proposition \ref{thm:random-EA-code}, for sufficiently large $n$ and
for all $x\in\mathcal{X}$, there exists a random $\rho_{x}^{A^{\prime}}$-like
entanglement-assisted $(n[p_{x}-\delta],\epsilon)$ code of quantum rate
$Q_{x}=I(A;B)_{\phi_{x}}/2-\delta$ and entanglement consumption rate
$E_{x}=I\left(  A;E\right)  _{\phi_{x}}/2+\delta$. Its expected channel input
density operator $\overline{\rho}_{x}^{A^{\prime n\left[  p_{x}-\delta\right]
}}$ is close to a tensor power of the state $\rho_{x}^{A^{\prime}}$:%
\[
\left\Vert \overline{\rho}_{x}^{A^{\prime n\left[  p_{x}-\delta\right]  }%
}-\rho_{x}^{\otimes n[p_{x}-\delta]}\right\Vert _{1}\leq\epsilon.
\]
The code's quantum rate is $Q_{x}=\frac{1}{2}I(A;B)_{\phi_{x}}-\delta$ because
it transmits $n\left[  p_{x}-\delta\right]  Q_{x}$ qubits for $n\left[
p_{x}-\delta\right]  $ uses of the channel. The code's entanglement
consumption rate is $E_{x}=\frac{1}{2}I(A;E)_{\phi_{x}}+\delta$ because it
consumes at least $n\left[  p_{x}-\delta\right]  E_{x}$ ebits for $n\left[
p_{x}-\delta\right]  $ uses of the channel. We produce an $(n-|\mathcal{X}%
|\delta,|\mathcal{X}|\epsilon)$ entanglement-assisted code with expected
channel input density operator%
\[
\overline{\rho}^{A^{\prime n\left(  1-|\mathcal{X}|\delta\right)  }%
}=\bigotimes\limits_{x}\overline{\rho}_{x}^{A^{\prime n\left[  p_{x}%
-\delta\right]  }}%
\]
by \textquotedblleft pasting\textquotedblright\ $|\mathcal{X}|$ of these codes
together (one for each $x$). Applying the triangle inequality $|\mathcal{X}|$
times, the expected channel input density operator $\overline{\rho}^{A^{\prime
n\left(  1-|\mathcal{X}|\delta\right)  }}$ of the pasted code is close to a
pasting of the tensor power states $\{\rho_{x}^{\otimes n[p_{x}-\delta]}%
\}_{x}$:%
\begin{equation}
\left\Vert \overline{\rho}^{A^{\prime n\left(  1-|\mathcal{X}|\delta\right)
}}-\bigotimes\limits_{x}\rho_{x}^{\otimes n[p_{x}-\delta]}\right\Vert _{1}%
\leq|\mathcal{X}|\epsilon.
\end{equation}
Consider the classical sequence $x^{n}$. Let random variable $X$ have the
probability distribution $p$ and define the typical set
\[
T_{\delta}^{X^{n}}=\{x^{n}:\forall x\ \ |n_{x}-np_{x}|\leq\delta n\},
\]
where $n_{x}\equiv N(x|x^{n})$ is the number of occurrences of the symbol $x$
in $x^{n}$. If $x^{n}$ lies in the typical set $T_{\delta}^{X^{n}}$, then we
can construct a conditional permutation operation that permutes the elements
of the input sequence as follows \cite{krovi:012321}:%
\[
x^{n}\rightarrow\underbrace{x_{1}\cdots x_{1}}_{n\left[  p_{x_{1}}%
-\delta\right]  }\underbrace{x_{2}\cdots x_{2}}_{n\left[  p_{x_{2}}%
-\delta\right]  }\cdots\underbrace{x_{\left\vert \mathcal{X}\right\vert
}\cdots x_{\left\vert \mathcal{X}\right\vert }}_{n\left[  p_{x_{\left\vert
\mathcal{X}\right\vert }}-\delta\right]  }x_{g}%
\]
where $x_{g}$ (for \textquotedblleft$x$ garbage\textquotedblright) denotes the
remaining $n\left\vert \mathcal{X}\right\vert \delta$\ symbols in $x^{n}$. The
density operator $\rho_{x^{n}}$ corresponds to the input sequence $x^{n}$. We
can construct a conditional permutation unitary that acts on the density
operator $\rho_{x^{n}}$ and changes the ordering of the state $\rho_{x^{n}}$
as follows:%
\[
\rho_{x^{n}}\rightarrow\bigotimes\limits_{x}\rho_{x}^{n\left[  p_{x}%
-\delta\right]  }\otimes\rho_{x_{g}}%
\]
where $\dim\left(  \rho_{x_{g}}\right)  \leq n\left\vert \mathcal{X}%
\right\vert \delta\log d_{A^{\prime}}$. We modify the random
entanglement-assisted code slightly by inserting $|\mathcal{X}|\delta$
\textquotedblleft garbage states\textquotedblright\ with density operator
$\rho_{x_{g}}$\ and define the expected channel input density operator
$\overline{\rho}^{A^{\prime n}}$\ for the full code as follows:%
\[
\overline{\rho}^{A^{\prime n}}\equiv\overline{\rho}^{A^{\prime n\left(
1-|\mathcal{X}|\delta\right)  }}\otimes\rho_{x_{g}}.
\]
Then the expected channel input density operator $\overline{\rho}^{A^{\prime
n}}$ is close to the permuted version of $\rho_{x^{n}}$%
\[
\left\Vert \overline{\rho}^{A^{\prime n}}-\bigotimes\limits_{x}\rho
_{x}^{n\left[  p_{x}-\delta\right]  }\otimes\rho_{x_{g}}\right\Vert _{1}%
\leq|\mathcal{X}|\epsilon.
\]
The quantum rate $Q$ for the random \textquotedblleft pasted\textquotedblright%
\ father code is as follows:%
\begin{align*}
Q  &  =\frac{\sum_{x}nQ_{x}\left[  p_{x}-\delta\right]  }{n}\\
&  =\sum_{x}Q_{x}\left[  p_{x}-\delta\right] \\
&  =\sum_{x}p_{x}\left(  \frac{I(A;B)_{\phi_{x}}}{2}-\delta\right)  -\delta
Q_{x}\\
&  =\frac{I(A;B|X)}{2}-c^{\prime}\delta,
\end{align*}
where%
\[
c^{\prime}\equiv1+\sum_{x}Q_{x}.
\]
The entanglement consumption rate $E$ is as follows:%
\begin{align*}
E  &  =\frac{\sum_{x}nE_{x}[p_{x}-\delta]}{n}\\
&  =\sum_{x}E_{x}[p_{x}-\delta]\\
&  =\sum_{x}p_{x}\left(  \frac{I(A;E)_{\phi_{x}}}{2}-\delta\right)  -\delta
E_{x}\\
&  =\frac{I(A;E|X)}{2}-c^{\prime\prime}\delta,
\end{align*}
where%
\[
c^{\prime\prime}\equiv1+\sum_{x}E_{x}.
\]
A permutation relates the states $\rho_{x^{n}}$ and $\bigotimes\limits_{x}%
\rho_{x}^{n\left[  p_{x}-\delta\right]  }\otimes\rho_{x_{g}}$. Therefore,
there exists an $(n,|\mathcal{X}|\epsilon)$ random entanglement-asissted code
of the same quantum communication rate and entanglement consumption rate with
an expected channel input density operator $\overline{\rho}^{\prime A^{\prime
n}}$ that is close to the tensor power state $\rho_{x^{n}}$:%
\[
\left\Vert \overline{\rho}^{\prime A^{\prime n}}-\rho_{x^{n}}\right\Vert
_{1}\leq|\mathcal{X}|\epsilon,
\]
because the action of the IID\ channel $\mathcal{N}^{\otimes n}$\ is invariant
under permutations of the input Hilbert spaces.
\end{IEEEproof}

\section{Gentle Measurement for Ensembles}

\begin{lemma}
[Gentle Measurement for Ensembles]\label{lemma:GM-ensemble}Let $\left\{
p_{x},\rho_{x}\right\}  $ be an ensemble with average $\overline{\rho}%
\equiv\sum_{x}p_{x}\rho_{x}$. Given a positive operator $X$ with $X\leq I$ and
Tr$\left\{  \overline{\rho}X\right\}  \geq1-\epsilon$ where $\epsilon\leq1$,
then%
\[
\sum_{x}p_{x}\left\Vert \rho_{x}-\sqrt{X}\rho_{x}\sqrt{X}\right\Vert _{1}%
\leq\sqrt{8\epsilon}.
\]

\end{lemma}

\begin{IEEEproof}
We can apply the same steps in the proof of the gentle measurement lemma
\cite{thesis1999winter}\ to get the following inequality:%
\[
\left\Vert \rho_{x}-\sqrt{X}\rho_{x}\sqrt{X}\right\Vert _{1}^{2}\leq8\left(
1-\text{Tr}\left\{  \rho_{x}X\right\}  \right)  .
\]
Summing over both sides produces the following inequality:%
\begin{align*}
\sum_{x}p_{x}\left\Vert \rho_{x}-\sqrt{X}\rho_{x}\sqrt{X}\right\Vert _{1}^{2}
&  \leq8\left(  1-\text{Tr}\left\{  \rho X\right\}  \right) \\
&  \leq8\epsilon.
\end{align*}
Taking the square root of the above inequality gives the following one:%
\[
\sqrt{\sum_{x}p_{x}\left\Vert \rho_{x}-\sqrt{X}\rho_{x}\sqrt{X}\right\Vert
_{1}^{2}}\leq\sqrt{8\epsilon}.
\]
Concavity of the square root implies then implies the result:%
\[
\sum_{x}p_{x}\sqrt{\left\Vert \rho_{x}-\sqrt{X}\rho_{x}\sqrt{X}\right\Vert
_{1}^{2}}\leq\sqrt{8\epsilon}.
\]
\label{sec:gm-ensemble}
\end{IEEEproof}

\section{Entanglement Consumption Rate of the EAC Classical Capacity}

\label{sec:EAC-EA-rate}

We prove that the entanglement consumption rate corresponding to the maximal
EAC rate is one ebit. Consider a general qubit density operator $\rho
^{A^{\prime}}$ that Alice can input to the erasure channel. Let $\psi^{A
A^{\prime}}$ denote the purification of $\rho^{A^{\prime}}$. Suppose that
$\rho$ has the spectral decomposition $\rho= p \vert\phi_{0} \rangle
\langle\phi_{0} \vert+ (1-p) \vert\phi_{1} \rangle\langle\phi_{1} \vert$ for
some orthonormal states $\vert\phi_{0} \rangle, \vert\phi_{1} \rangle$. After
Alice transmits this density operator through an erasure channel with erasure
parameter $\epsilon$, Bob has the following state:
\[
\sigma^{B} \equiv(1 -\epsilon) \rho+ \epsilon\vert e \rangle\langle e \vert,
\]
and Eve has
\[
\sigma^{E} \equiv\epsilon\rho+ (1 -\epsilon) \vert e \rangle\langle e \vert,
\]
where $\vert e \rangle$ is an erasure state. The entropies $H(A)$, $H(B)$, and
$H(E)$ are as follows:
\begin{align}
H(A)  &  = H_{2}(p),\nonumber\\
H(B)  &  = (1-\epsilon) H_{2}(p) + H_{2}(\epsilon),\nonumber\\
H(E)  &  = \epsilon H_{2}(p) + H_{2}(\epsilon),\nonumber
\end{align}
and the mutual information $I(A;B)$ is as follows:
\[
I(A;B) = H(A) + H(B) - H(E) = 2 (1-\epsilon) H_{2}(p).
\]
This quantity is maximized only when $p=\frac{1}{2}$, implying that the
entanglement consumed for this state is exactly one ebit because $H(A) =
H_{2}(p)$. Thus, Alice and Bob cannot consume entanglement at a lower rate
than this amount in order to achieve the EAC capacity.

\section{Isometric Encodings suffice in the CQE Theorem}

\label{sec:isometric-encoders}We prove that it is only necessary to consider
isometric encodings for achieving points in the CQE capacity region. Our
argument follows the technique of Ref.~\cite{arx2005dev}, by showing that a
protocol can only improve upon measuring the environment of a non-isometric encoder.

Suppose that we exploit the following state that results from a non-isometric
encoder, rather than the state in (\ref{eq:maximization-state}):
\begin{equation}
\widetilde{\sigma}^{XABEE^{\prime}}\equiv\sum_{x}p(x)\left\vert x\right\rangle
\left\langle x\right\vert ^{X}\otimes U_{\mathcal{N}}^{A^{\prime}\rightarrow
BE}(\phi_{x}^{AA^{\prime}E^{\prime}}). \label{eq:non-isometric-state}%
\end{equation}
The inequalities in (\ref{gf1}-\ref{gf3}) for the CQE capacity region involve
the mutual information $I(AX;B)_{\widetilde{\sigma}}$, the Holevo information
$I(X;B)_{\widetilde{\sigma}}$, and the coherent information $I(A\rangle
BX)_{\widetilde{\sigma}}$. As we show below, each of these entropic quantities
can only improve if Alice measures the system $E^{\prime}$. This improvement
then implies that it is only necessary to consider isometric encodings in the
CQE capacity theorem.

Suppose that Alice sends the system $E^{\prime}$ through a completely
dephasing channel $\Delta^{E^{\prime}\to Y}$ to obtain a classical variable
$Y$ (this simulates a measurement). Let $\overline{\sigma}^{XYABE}$ denote
this later state, a state of the form:
\begin{equation}
\overline{\sigma}^{XYABE}\equiv\sum_{x}p(x,y)\left\vert x\right\rangle
\left\langle x\right\vert ^{X}\otimes\left\vert y\right\rangle \left\langle
y\right\vert ^{Y}\otimes U_{\mathcal{N}}^{A^{\prime}\rightarrow BE}(\psi
_{x,y}^{AA^{\prime}}). \label{eq:measured-state}%
\end{equation}
This state is actually a state of the form in (\ref{eq:maximization-state}) if
we subsume the classical variables $X$ and $Y$ into one classical variable.

The following three inequalities each follow from an application of the
quantum data processing inequality (or, equivalently, strong subadditivity):
\begin{align}
I(X;B)_{\widetilde{\sigma}}  &  = I(X;B)_{\overline{\sigma}} \leq
I(XY;B)_{\overline{\sigma}},\\
I(AX;B)_{\widetilde{\sigma}}  &  = I(AX;B)_{\overline{\sigma}} \leq
I(AXY;B)_{\overline{\sigma}}\\
I(A\rangle BX)_{\widetilde{\sigma}}  &  = I(A\rangle BX)_{\overline{\sigma}}
\leq I(A\rangle BXY)_{\overline{\sigma}}%
\end{align}
Each of these inequalities proves the desired result for the respective Holevo
information, mutual information, and coherent information.

\bibliographystyle{IEEEtran}
\bibliography{Ref}

\end{document}